\documentclass[reqno,12pt]{article}
\usepackage{amsfonts,amssymb,
amsmath,amscd, bm, mathrsfs, 
mathrsfs,delarray,subfigure}
\usepackage{hyperref,cite}
\usepackage{xypic}
\usepackage[dvips]{color}
\usepackage[dvips]{graphicx}

\DeclareMathOperator{\ad}{ad}
\DeclareMathOperator{\Ad}{Ad}
\DeclareMathOperator{\vol}{vol}

\DeclareMathOperator{\id}{id}

\DeclareMathOperator{\Conn}{Conn}
\DeclareMathOperator{\Curv}{Curv}
\DeclareMathOperator{\Map}{Map}

\DeclareMathOperator{\Tor}{Tor}

\DeclareMathOperator{\Gau}{Gau}
\DeclareMathOperator{\OGau}{OGau}

\DeclareMathOperator{\End}{End}
\DeclareMathOperator{\Aut}{Aut}
\DeclareMathOperator{\OAut}{OAut}

\DeclareMathOperator{\tr}{tr}

\DeclareMathOperator{\PD}{PD}

\DeclareMathOperator{\UU}{U}

\DeclareMathOperator{\SU}{SU}

\DeclareMathOperator{\CS}{CS}
\DeclareMathOperator{\cs}{cs}
\DeclareMathOperator{\cc}{c}

\DeclareMathOperator{\Vol}{Vol}
\DeclareMathOperator{\Det}{Det}

\numberwithin{equation}{subsection} 
\numberwithin{subsection}{section} 

\newcommand{\sss}{{\hbox{$\sum$}}}

\font\sansserif=cmss12
\font\scriptsansserif=cmss12 at 7 truept
\font\scriptscriptsansserif=cmss10 at 5 truept
\def\sans{\fam=14}
\newcommand{\mathsans}[1]{{{\sans #1}}}

\font\euler=eusm10 at 12.8 truept
\font\scripteuler=eusm7
\font\scriptscripteuler=eusm5 
\def\eul{\fam=12}
\newcommand{\matheul}[1]{{{\eul #1}}}

\newcommand{\bfdot}{{\boldsymbol{\,\cdot\,}}}

\newcommand{\ul}[1]{{\underline{#1}}{}}

\begin{document}

\hrule\vskip.5cm
\hbox to 14.5 truecm{December 2015 \hfil DIFA 15}
\vskip.5cm\hrule
\vskip.7cm
\centerline{\textcolor{blue}{\bf A LIE BASED 4--DIMENSIONAL}}   
\centerline{\textcolor{blue}{\bf HIGHER CHERN--SIMONS THEORY}}   
\vskip.2cm
\centerline{by}
\vskip.2cm
\centerline{\bf Roberto Zucchini}
\centerline{\it Dipartimento di Fisica ed Astronomia, Universit\`a di Bologna}
\centerline{\it V. Irnerio 46, I-40126 Bologna, Italy}
\centerline{\it I.N.F.N., sezione di Bologna, Italy}
\centerline{\it E--mail: zucchinir@bo.infn.it}
\vskip.7cm
\hrule
\vskip.6cm
\centerline{\bf Abstract} 
\par\noindent
We present and study a model of 4--dimensional higher Chern-Simons theory, 
special Chern--Simons (SCS) theory, instances of which have appeared in the string literature, 
whose symmetry is encoded in a skeletal semistrict Lie 2--algebra constructed from a compact 
Lie group with non discrete center. The field content of SCS theory consists of a Lie valued
2--connection coupled to a background closed 3--form. SCS theory enjoys a large  
gauge and gauge for gauge symmetry organized in an infinite dimensional strict Lie 2--group. 
The partition function of SCS theory is simply related to that of a topological gauge theory 
localizing on flat connections with degree 3 second characteristic class determined by 
the background 3--form. Finally, SCS theory is related to a 3--dimensional special gauge theory 
whose 2--connection space has a natural symplectic structure with respect to which the 1--gauge 
transformation action is Hamiltonian, the 2--curvature map acting as moment map.
\vfil\eject

~~

{\it To Raymond Stora (1930 - 2015), in memoriam}

\vfil\eject

\tableofcontents

\vfil\eject


\section{\normalsize \textcolor{blue}{Introduction}}\label{sec:intro}

\hspace{.5cm} 
{\it Higher gauge theory} is an extension of ordinary gauge theory 
where gauge potentials and their gauge curvatures are higher degree forms.
It is believed that higher gauge theory describes 
the dynamics of the higher--dimensional extended objects
thought to be the basic building blocks of fundamental interactions,
such as strings and branes. 
See ref. \cite{Baez:2010ya} and references therein. 

\vfil
Higher gauge theory, in its Abelian variant, originated in 
supergravity. Subsequently, it 
turned out to be relevant 
in string theory \cite{Polchinski:1998rr,Becker:2007zj,Johnson:2003gi}, in particular in the study of 
$D$-- and $M$--branes, and in quantum gravity \cite{Baez:1999sr,Rovelli:2004tv},
especially in loop and spin foam models. 
Presently, the interest in higher gauge theory stems from the hope that 
it may eventually provide a Lagrangian formulation of the 
$N=(2, 0)$ $6$--dimensional superconformal field theory 
describing the effective dynamics of $M5$--branes \cite{Fiorenza:2012tb,
Fiorenza:2015gla,Palmer:2012ya,Saemann:2013pca,Lavau:2014iva}.

\vfil
Higher gauge theory intersects many areas of contemporary mathematics, 
primarily the theory of higher
algebraic structures, such as 2--categories, 2--groups \cite{Baez5,Baez:2003fs} 
and strong homotopy Lie or $L_\infty$ algebras \cite{Lada:1992wc,Lada:1994mn}, 
and higher geometrical structures, such as gerbes \cite{Brylinski:1993ab,Breen:2001ie}.
An illustration of these topics and their relationship to
fundamental physics is provided in \cite{Schreiber2011,Gruetzmann:2014ica,Sharpe:2015mja}. 

\vfil
Quite early in the history of the subject, it was realized that 
higher gauge theory should be built as a categorification of
ordinary gauge theory by codifying the higher gauge symmetry into the 
algebraic structures yielded by the categorification of ordinary 
groups, that is weak or coherent $2$--groups 
\cite{Baez:1998,Baez:2002jn,Baez:2004in,Baez:2005qu}.
In the initial stages, most studies on the subject were limited to the case where 
the structure $2$--group was strict. More recently, the investigation of higher gauge theory 
with non strict structure $2$--group was undertaken in the very general 
context of $\infty$--Lie theory in refs. \cite{Sati:2008eg,Fiorenza2011,Fiorenza:2011jr}. 
An alternative approach to the problem was followed in refs. \cite{Ritter:2013wpa,Jurco:2014mva}.
\vfil

\subsection{\normalsize \textcolor{blue}{The scope and the plan of this paper}}\label{subsec:scope}

\vspace{.4mm}
\hspace{.5cm} 
Chern--Simons theory is a Schwarz type $3$--dimensional topological field theory
first formulated in 1989 by E. Witten in ref. \cite{Witten:1988hf}. (See. ref. 
\cite{Marino:2005sj} for a recent review of the subject and extensive referencing). 
Witten was able to show that many topological knot and link invariants discovered 
by topologists in the 1980, such as the Jones and HOMFLY polynomials, could be
expressed as correlation functions of gauge theoretic Wilson loop operators in Chern--Simons theory. 
Witten also proved that the Chern--Simons partition function is a 
topological invariant of the base $3$--manifold. 
Intimate relationships to the $2$--dimensional WZW model and the $A$ and $B$ type topological sigma models 
were also found in the subsequent years \cite{Elitzur:1989nr,Witten:1992fb}.

\vspace{.4mm}
The present paper is a further modest step in our project of building a model of 
$4$--dimensional non strict higher Chern--Simons gauge theory applicable to the study of 
$4$--dimensional topology just as the ordinary Chern--Simons theory is in $3$ dimensions. 
Our goal is eventually obtaining a field theoretic expression of $2$--knot
and link invariants of $4$--manifolds and unveiling $3$-dimensional
higher analogs of WZW theory. Although there is no guarantee
that this endeavour will eventually succeed, it may be worthy to explore this possibility. 

\vspace{.4mm}
The version of non strict higher gauge theory we employ, called {\it semistrict}, 
was first formulated by the author in ref. \cite{Zucchini:2011aa} and 
further developed in ref. \cite{Soncini:2014ara}. 
As it is not widely known, we review it in some detail in sect. \ref{sec:highcs}. 

\vspace{.4mm}
In the $4$--dimensional higher Chern--Simons model considered in refs. 
\cite{Zucchini:2011aa,Soncini:2014ara}, symmetry is encoded in a balanced 
semistrict Lie $2$--algebra $\mathfrak{v}$ equipped with an invariant non singular 
bilinear form. At this level of generality, only the canonical 
quantization of the model appears to be possible. Moreover, 
the higher gauge theoretic framework, although general enough, 
is limited by the lack of a computational scheme for higher holonomies \pagebreak 
as efficient as that available in strict higher gauge theory. 

In this paper, we employ a special choice $\mathfrak{v}_k(\mathfrak{g})$ of
the balanced Lie $2$--algebra $\mathfrak{v}$ built from a compact connected Lie group 
$G$ whose the Lie algebra $\mathfrak{g}$ has a non trivial center $\mathfrak{z}(\mathfrak{g})$
and is equipped with an invariant symmetric non singular bilinear form $(\cdot,\cdot)$
and a choice of an element $k\in\mathfrak{z}(\mathfrak{g})$ such that $(k,k)\not=0$. 
Although $\mathfrak{v}_k(\mathfrak{g})$ is semistrict, the familiar Lie theoretic techniques 
are still available and allow one to carry out many explicit computations. 
$\mathfrak{v}_k(\mathfrak{g})$ turns out to be skeletal: its boundary map $\partial$ vanishes. 
Since every Lie 2--algebra is equivalent to a skeletal Lie 2--algebra, we are covering 
here a broad range of prototypical examples. 

A special $G$--gauge theory is a semistrict higher gauge theory whose symmetry is encoded 
in the semistrict Lie 2--algebra $\mathfrak{v}_k(\mathfrak{g})$. 
A special $G$--2--connection is a $\mathfrak{v}_k(\mathfrak{g})$--2--connection,
a pair of a $\mathfrak{g}$--valued 1-- and 2--form fields $\omega$, $\varOmega_\omega$.
Special $G$--1--gauge transformations act on special 2--connections and are related by 
special $G$--2-- gauge transformations. Together, they form an infinite dimensional strict Lie 
2--group $\Gau(N,G)$, the special gauge transformation 2--group. 
See sect. \ref{sec:specgau} for a thorough exposition of these matters. 

Just as ordinary gauge theory can be framed geometrically in the theory of principal $G$--bundles,
the geometry of special gauge theory is naturally described by the theory of
special $G$--2--bundles expounded in sect. \ref{sec:specgau}. Very roughly speaking, 
a special $G$--2--bundle $Q$ is specified by smooth $G$--valued trivialization matching data $\gamma_{ij}$ 
and constant $Z(G)$--valued trivialization matching compatibility data $K_{ijk}$ with respect to a suitable
open covering $U_i$ of the base manifold $N$. The date $K_{ijk}$ define a flat $Z(G)$--gerbe $B$ 
constituting the obstruction to the data $\gamma_{ij}$ defining a principal $G$--bundle,
\begin{equation}
\gamma_{ij}\gamma_{jk}\gamma_{ki}=K_{ijk}.
\label{intspec2cs8}
\end{equation}
The $\gamma_{ij}$ in turn are local 1--gauge transformations which, together with other 
appended trivialization matching data, describe how the local trivializing 2--connection data $\omega_i$,
$\varOmega_i$ fit globally. 

Special $G$--2--Chern--Simons theory is the semistrict higher Chern--Simons theory 
associated with the Lie 2--algebra $\mathfrak{v}_k(\mathfrak{g})$ in the framework of refs. 
\cite{Zucchini:2011aa,Soncini:2014ara}. The field content of the model consists of a special 
$G$--2--connection $\omega$, $\varOmega_\omega$ coupled to a background closed 3--form field $H$. 
The action of the model reads 
\begin{align}
\overline{\CS}_2(\omega,\varOmega_\omega;H)&
=\kappa_2\int_N\Big\{\Big(d\omega+\frac{1}{2}[\omega,\omega],\varOmega_\omega\Big)
\vphantom{\Big]}
\label{intspec2cs3}
\\
&\hspace{4cm}+(\omega,k)\Big[-\frac{1}{6}(\omega,[\omega,\omega])+8\pi^2H\Big]\Big\}.
\vphantom{\Big]}
\nonumber 
\end{align}
Because of the assortment of its constitutive elements, 
special 2--Chern--Simons theory turns out to be quite rich. 
Its field equations take the form
\begin{subequations}
\label{intspec2cs4-5}
\begin{align}
&d\omega+\frac{1}{2}[\omega,\omega]=0,
\vphantom{\Big]}
\label{intspec2cs4}
\\
&d\varOmega_\omega+[\omega,\varOmega_\omega]-\frac{1}{6~}(\omega,[\omega,\omega])k
+\frac{1}{2}(\omega,k)[\omega,\omega]=-8\pi^2Hk,
\vphantom{\Big]}
\label{intspec2cs5}
\end{align}
\end{subequations}
which, in the case where $H=0$, reduce to the flatness conditions of the 2--con\-nection $\omega,\varOmega_\omega$. 
Further, it enjoys full special $G$--1--gauge invariance. The theory is many respects an ordinary gauge theory 
with gauge group $G$ and gauge field $\omega$ extended by including a 2--form field $\varOmega_\omega$
acting as the $B$ field of a $BF$ theory. It is not however a $BF$ theory as it may naively appear, since 
$\varOmega_\omega$ transforms inhomogeneously under gauge transformations, as required for the 2--form 
component of a 2--connection. 

Special 2--Chern--Simons theory is related to a 3--dimensional special gauge theory 
whose 2--connection space has a natural symplectic structure with respect to which 
the 1--gauge transformation action is Hamiltonian, with the 2--curvature map roughly acting as moment map, 
as ordinary Chern-Simons theory. Further, the partition function of the model is that of an ordinary gauge 
theory localizing on flat connections with prescribed degree 3 second \pagebreak  
characteristic class depending on the background 3--form. 
See sect. \ref{sec:skhighcs} for an in depth analysis of these points. 

\subsection{\normalsize \textcolor{blue}{Physical origin of special 2--Chern--Simons theory}}\label{subsec:motivation}

\hspace{.5cm} 
Specific instances of the special 2--Chern--Simons theory studied in this paper 
have appeared in disguised form in the physical literature 
(see e. g. \cite{Shiu:2015xda} and references therein). 
In string based cosmology, axions and their coupling to gauge fields  
play an important role. In a simplified version that fits best our purposes, 
the 4--dimensional effective theory describing the axion--gauge system can be formulated 
as follows. The space--time manifold $N$ is endowed with a background metric $g$, that 
we assume here to have Euclidean signature. The field content of the theory comprises
an axion field $\vartheta$, an $\UU(1)$ gauge field $A$ with gauge curvature $F=dA$ 
and an $\SU(n)$ gauge field $B$ with gauge curvature $G=dB+[B,B]/2$. 
The Lagrangian has the form
\begin{align}
&\mathcal{L}=\frac{1}{2e_A{}^2}F*F+\frac{m_A{}^2}{2}(d\vartheta-A)*(d\vartheta-A)
\vphantom{\Big]}
\label{mot1}
\\
&\hspace{4cm}+\frac{1}{2e_B{}^2}\tr(G*G)+\frac{i\vartheta}{8\pi^2}\tr(GG),
\vphantom{\Big]}
\nonumber
\end{align}
where $m_A$ is the Stueckelberg mass of $A$ and $e_A$ and $e_B$ are the gauge coupling constants. 

The model enjoys full $\SU(n)$ gauge invariance: 
the Lagrangian $\mathcal{L}$ 
is invariant under any $\SU(n)$ gauge transformation $\gamma$ 
\begin{equation}
{}^\gamma B=\gamma B\gamma^{-1}-d\gamma\gamma^{-1}. \vphantom{\Bigg]}
\label{mot2}
\end{equation}
If $\vartheta$ did not couple to $B$ via its Pontryagin density, $\mathcal{L}$  would also be invariant 
under any $\UU(1)$ gauge transformation $\alpha$ 
\begin{align}
&{}^\alpha A=A+d\alpha,
\vphantom{\Big]}
\label{mot3}
\\
&{}^\alpha\vartheta=\vartheta+\alpha.
\vphantom{\Big]}
\label{mot4}
\end{align}
The presence of the coupling breaks this symmetry. \pagebreak 
However, the Boltzmann exponential $\exp\big(-\int_N\mathcal{L}\big)$ is still invariant under 
a residual $\UU(1)$ gauge symmetry 
\begin{align}
&{}^nA=A,
\vphantom{\Big]}
\label{mot5}
\\
&{}^n\vartheta=\vartheta+2\pi n,
\vphantom{\Big]}
\label{mot6}
\end{align}
where $n$ is an integer. $\vartheta$ is therefore an $S^1$-valued field. 

\vspace{.33mm}
In the presence of chiral fermions coupling to the $\UU(1)$ gauge 
field $A$ and transforming under appropriate representations
of the $\UU(1)$ and $\SU(n)$ gauge groups, full $\UU(1)$ gauge invariance can be 
restored by including certain Chern--Simons and fermionic coupling terms
\cite{Aldazabal:2002py,Andrianopoli:2004sv,Anastasopoulos:2006cz,DeRydt:2007vg}. 
The Lagrangian takes the form 
\begin{align}
&\mathcal{L}'=\frac{1}{2e_A{}^2}F*F+\frac{m_A{}^2}{2}(d\vartheta-A)*(d\vartheta-A)
\vphantom{\Big]}
\label{mot7}
\\
&\hspace{3cm}+\frac{1}{2e_B{}^2}\tr(G*G)+\frac{i\vartheta}{8\pi^2}\tr(GG)+i\lambda A(\cs_{1B}+*J/\lambda), 
\vphantom{\Big]}
\nonumber
\end{align}
where $\cs_{1B}$ is the Chern--Simons 3--form of $B$
\begin{equation}
\cs_{1B}=\frac{1}{8\pi^2}\tr\Big(BdB+\frac{2}{3}BBB\Big),
\label{mot8}
\end{equation}
$J$ is the fermion current coupled to $A$ 
and $\lambda$ is the coefficient measuring the strength of the 
anomalous violation of the conservation equation of $J$
\begin{equation}
d*J=\frac{\lambda}{8\pi^2}\tr(GG).
\label{mot9}
\end{equation}
$\lambda$ is determined by the representations of the fermions. 
If $\lambda=1/2$, the $\UU(1)$ gauge symmetry \eqref{mot3}, \eqref{mot4} is recovered. 
This has however a price: the Chern--Simons form $\cs_{1B}$ explicitly breaks 
$\SU(n)$ invariance. Indeed, under a $\SU(n)$ gauge transformation \eqref{mot2}, 
\begin{equation}
{}^\gamma\cs_{1B}=\cs_{1B}+w(\gamma)-\frac{1}{8\pi^2}\tr(B\gamma^{-1}d\gamma),
\label{mot10}
\end{equation}
where $w(\gamma)$ is the winding number density of $\gamma$,
\begin{equation}
w(\gamma)=\frac{1}{24\pi^2}\tr(\gamma^{-1}d\gamma\gamma^{-1}d\gamma\gamma^{-1}d\gamma).
\label{mot11}
\end{equation}
The $\UU(1)$ invariance 
allows us to provide $A$ with a mass equal to $m_A$ 
through the Stueckelberg mechanism. 

\vspace{.33mm}
We now replace the Lagrangian $\mathcal{L}'$ by an equivalent Lagrangian $\mathcal{L}''$ 
containing an $\UU(1)$ auxiliary 2--form field $U$ and an $\SU(n)$ auxiliary 2--form field $V$,
\begin{align}
&\mathcal{L}''=\frac{\lambda^2e_A{}^2}{2}U*U+i\lambda UF
+\frac{m_A{}^2}{2}(d\vartheta-A)*(d\vartheta-A)
\vphantom{\Big]}
\label{mot12}
\\
&\hspace{1.3cm}+\frac{\lambda^2e_B{}^2}{2}\tr(V*V)+i\lambda\tr(VG)
+\frac{i\vartheta}{8\pi^2}\tr(GG)+i\lambda A(\cs_{1B}+*J/\lambda).
\vphantom{\Big]}
\nonumber
\end{align}
$\mathcal{L}''$ contains a purely topological portion
\begin{equation}
\mathcal{L}_{\mathrm{CS}}=i\lambda\big[\tr(VG)+A(\cs_{1B}+\tilde J+dU)\big],
\label{mot13}
\end{equation}
where $\tilde J=*J/\lambda$ is treated as background 3--form field. This is an instance of the 
special 2--Chern--Simons action studied in this paper.
The gauge group $G$ is here $\UU(n)$ having the former groups $\UU(1)$ and $\SU(n)$ 
as its center and adjoint group (after modding the $\mathbb{Z}_n$
center). The bilinear form is defined through the trace $\tr$ on the fundamental
representation of $\UU(n)$ and central element $k$ is just $i1_n/n^{1/2}$, say. The 2--connection components 
are given by
\begin{subequations}
\label{intspec2cs6-7}
\begin{align}
&\omega=Ak+B,
\vphantom{\Big]}
\label{intspec2cs6}
\\
&\varOmega_\omega=8\pi^2[Uk+V+AB].
\vphantom{\Big]}
\label{intspec2cs7}
\end{align}
\end{subequations}
The background 3--form $H=\tilde J$.

\subsection{\normalsize \textcolor{blue}{Mathematical ramifications}}\label{subsec:math}

\vspace{.33mm}
\hspace{.5cm} 
Special 2--Chern--Simons theory has ramifications also in the differential topology
of principal bundles. Consider a 
principal $G_s$--bundle $P$ on a fourfold $N$, where $G_s$ is some compact connected Lie group. 
Then, $P$ is characterized topologically by its 2nd Chern class, which is an integer cohomology 
class $C_2\in H^4(M,\mathbb{Z})$ \cite{Chern:1946}.
The image of $C_2$ in the real cohomology $H^4(M,\mathbb{R})$ \pagebreak 
is represented by the closed differential forms $\cc_2$ given by 
\begin{equation}
\cc_2=\frac{1}{8\pi^2}(F_s,F_s)_s
\label{math1}
\end{equation}
where $F_s$ is the gauge curvature of any connection $A_s$ of $P$,
\begin{equation}
F_s=dA_s+\frac{1}{2}[A_s,A_s]
\label{math2}
\end{equation}
and $(\cdot,\cdot)_s$ is a suitably normalized invariant symmetric non singular 
bilinear form on $\mathfrak{g}_s$. It is indeed a standard result of differential geometry that 
\begin{equation}
d\cc_2=0
\label{math3}
\end{equation}
and that $\cc_2$ is independent from the choice of $A_s$ up to exact terms. 

If the principal bundle $P$ is flat, $\cc_2$ is exact. Indeed, one has
\begin{equation}
\cc_2=d\cs_1,
\label{math4}
\end{equation}
where $\cs_1$ is the Chern--Simons 3--form
\begin{equation}
\cs_1=\frac{1}{8\pi^2}\Big(A_s,dA_s+\frac{1}{3}[A_s,A_s]\Big)_s
\label{math5}
\end{equation}
\!\!
\cite{Chern:1974}. A remarkable property of $\cs_1$ is that it is itself closed
when one restricts to flat connections $A_s$, as $\cc_2=0~\Rightarrow~d\cs_1=0$. 
In that case, $\cs_1$ defines a class $CS_1\in H^3(N,\mathbb{R})$ in real cohomology. 
However, this class does not characterize the flat bundle $P$ as it is gauge dependent:
under a gauge transformation $\gamma$, 
\begin{equation}
{}^\gamma\cs_1=\cs_1+w(\gamma)+\text{exact term},
\label{math6}
\end{equation}
where $w(\gamma)$ is a closed 3--form representing a cohomology class 
of $H^3(N,\mathbb{R})$ in the image of $H^3(N,\mathbb{Z})$, 
\begin{equation}
w(\gamma)=\frac{1}{48\pi^2}(\gamma^{-1}d\gamma,[\gamma^{-1}d\gamma,\gamma^{-1}d\gamma])_s.
\label{math7}
\end{equation}
Thus, $\cs_1$ defines rather a class $CS_1\in H^3(N,\mathbb{R}/\mathbb{Z})$ in real mod integer
cohomology. This is the secondary characteristic class of $P$ \cite{Peterson:1962}. 

$\vphantom{\dot{\dot{\dot{x}}}}$A basic problem of the theory is the determination of the actual value the 
secondary characteristic class $CS_1$ can take. Special 2--Chern--Simons
theory furnishes a potential answer. 
Extend the group $G_s$ by a central $\UU(1)$ factor
to the group $G=\UU(1)\times G_s$. Extend correspondingly the invariant form $(\cdot,\cdot)_s$
of $\mathfrak{g}_s$ to one $(\cdot,\cdot)$ on $\mathfrak{g}=\mathfrak{u}(1)\oplus\mathfrak{g}_s$. 
Then, roughly speaking, the partition function of the 
theory with background 3--form $H$ depends only on the class $[H]$ of $H$ in real mod integer cohomology 
and localizes on the flat gauge fields $A_s$ such that $CS_1=-[H]$ (see sect. \ref{subsec:funct2cs} 
for a more precise statement).

\subsection{\normalsize \textcolor{blue}{Outlook}}\label{subsec:outlook}

\vspace{.33mm}

\hspace{.5cm}
The motivation that prompted the author to write the present paper was the quest to 
construct concrete models of 4--dimensional higher Chern--Simons theory applicable to the 
study of the topology of surface knots based on the general framework  
of refs. \cite{Zucchini:2011aa,Soncini:2014ara}. While this has produced an example 
of a genuinely semistrict higher gauge theory amenable by the familiar methods of quantum field 
theory, special 2--Chern--Simons theory, adding to the presently rather short list of such examples,
the  problem of devising an efficient computational scheme for surface holonomies is unfortunately still 
open. The methods currently available, which are sufficiently concrete to be usable in practice, 
work mostly for strict higher gauge theories (see \cite{Soncini:2014zra,Zucchini:2015wba,Zucchini:2015xba}
and references therein). We hope to come back to this in future work. 

\vspace{.33mm}
The elementary special $G$--2--bundle theory developed in this paper 
is closely related to that describing t'Hooft's magnetic flux in ordinary 
gauge theory \cite{thooft:1979}, opening the possibility of using it to study various non perturbative
aspects of gauge theory. This will require however a deeper understanding
of the role of the central element $k$, which presently is still a bit mysterious. 

\vspace{.33mm}
A similar topological framework has been considered also in refs. 
\cite{Kapustin:2014gua,Gaiotto:2014kfa} in the analysis of the 
coupling of ordinary 
and topological quantum field theories and their higher degree form global symmetries. 
These symmetries however are by their nature Abelian.
In special 2--Chern--Simons theory such restriction apparently does not emerge.
It is possible that there is a deeper relationship of the constructions of \cite{Kapustin:2014gua,Gaiotto:2014kfa}
to ours, but we have not been able to elucidate this point satisfactorily. 

\vspace{.5cm}

\noindent
\textcolor{blue}{Dedication.} 
This paper is dedicated to Raymond Stora, who passed away on July 20-th 2015.
A deeply open minded and curious man and a sophisticated field theorist, 
he certainly would have enjoyed discussing with the author about the present 
work pointing out its (to be sure numerous) imperfections and providing
invaluable suggestions for correcting them. The theoretical community 
will miss him as a man and a scientist. 

\vspace{.5cm}

\noindent
\textcolor{blue}{Acknowledgements.} 
The author thanks V. Mathai, P. Ritter and D. Stevenson for useful discussions.
We acknowledge financial support from INFN Research Agency
under the provisions of the agreement between Bologna University and INFN. 
He also thanks the Erwin Schroedinger Institute of Vienna, where part of this work was done,  
for hospitality during the 2015 ESI Program on "Higher structures in string theory 
and quantum field theory''.

\vfil\eject

\section{\normalsize \textcolor{blue}{Semistrict higher Chern--Simons theory}}\label{sec:highcs}

\hspace{.5cm} 
In this section, we review the formulation of higher Chern--Simons theory of refs.
\cite{Zucchini:2011aa,Soncini:2014ara}. The reader already familiar with the content of 
these works can skip directly to section \ref{sec:specgau}. 

In semistrict higher gauge theory, symmetry is codified in a Lie $2$--algebra $\mathfrak{v}$. 
In semistrict highder Chern--Simons theory, the symmetry Lie 2--algebra $\mathfrak{v}$ 
is required to be balanced and equipped with an invariant bilinear form 
$(\cdot,\cdot)$. Our definitions and conventions of Lie 2--algebra theory are 
collected in apps. \ref{sec:linfty}, \ref{sec:linfty}.



\subsection{\normalsize \textcolor{blue}{Semistrict higher gauge symmetry}}\label{subsec:highgausym}

\hspace{.5cm}
In this subsection, we review semistrict higher gauge symmetry and its 2--group structure. 

\vspace{2.3mm}
 
{\it Orthogonal semistrict higher gauge transformations} 

\vspace{2.3mm}

The set $\OGau_1(N,\mathfrak{v})$ of {\it orthogonal 1--gauge transformations} consists
of all quadruples $(g,\sigma_g,\varSigma_g,\tau_g)$ with 
$g\in\Map(M,\OAut_1(\mathfrak{v}))$ (cf. app. \ref{sec:linftyauto}, eqs. \eqref{linfty4}, \eqref{linfty5}), 
$\sigma_g\in\Omega^1(N,\mathfrak{v}_0)$, $\varSigma_g\in\Omega^2(N,\mathfrak{v}_1)$,
and $\tau_g\in\Omega^1(M,\mathfrak{aut}_1(\mathfrak{v}))$ obeying the higher Maurer--Cartan equations 
\begin{subequations}
\label{1linfdglob}
\begin{align}
&d\sigma_g+\frac{1}{2}[\sigma_g,\sigma_g]-\partial\varSigma_g=0,
\vphantom{\Big]}
\label{1linfdgloba}
\\
&d\varSigma_g+[\sigma_g,\varSigma_g]-\frac{1}{6}[\sigma_g,\sigma_g,\sigma_g]=0,
\vphantom{\Big]}
\label{1linfdglobb}
\\
&d\tau_g(\pi)+[\sigma_g,\tau_g(\pi)]-[\pi,\varSigma_g]+\frac{1}{2}[\sigma_g,\sigma_g,\pi]
\vphantom{\Big]}
\label{2linfdglob}
\\
&\qquad\qquad\qquad\qquad\qquad\qquad
+\tau_g([\sigma_g,\pi]+\partial\tau_g(\pi))=0
\vphantom{\Big]}
\nonumber
\end{align}
\end{subequations}
and the orthogonality condition 
\begin{equation}
(x,\tau_g(y))+(y,\tau_g(x))=0
\label{orthogauge1}
\end{equation}
for $x,y\in\mathfrak{v}_0$. It is further required that 
$g$, $\sigma_g$, $\varSigma_g$, $\tau_g$ satisfy the relations.
\begin{subequations}
\label{3linfdglob}
\begin{align}
&g_0{}^{-1}dg_0(\pi)-[\sigma_g,\pi]-\partial\tau_g(\pi)=0,
\vphantom{\Big]}
\label{3linfdgloba}
\\
&g_1{}^{-1}dg_1(\varPi)-[\sigma_g,\varPi]-\tau_g(\partial \varPi)=0,
\vphantom{\Big]} 
\label{3linfdglobb}
\\
&g_1{}^{-1}(dg_2(\pi,\pi)-2g_2(g_0{}^{-1}dg_0(\pi),\pi))
\vphantom{\Big]}
\label{3linfdglobc}
\\
&\qquad\qquad -[\sigma_g,\pi,\pi]-\tau_g([\pi,\pi])-2[\pi,\tau_g(\pi)]=0.
\vphantom{\Big]}
\nonumber
\end{align}
\end{subequations} 
Notice that our notation does not imply that $\sigma_g$, $\varSigma_g$, $\tau_g$
are determined by $g$, but only that they are the partners of $g$ in the 
1--gauge transformation. 

In semistrict higher gauge theory, one has in addition gauge for gauge symmetry. 
Let $g,h\in\OGau_1(M,\mathfrak{v})$ be orthogonal $1$--gauge transformations. 
The set $\OGau_2(N,\mathfrak{v})(g,h)$ of {\it orthogonal 2--gauge transformations} $g\Rightarrow h$ 
consists of the pairs $(F,A_F)$ with $F\in\Map(M,\OAut_2(\mathfrak{v}))(g,h)$, 
where $\Map(M,\OAut_2(\mathfrak{v}))(g,h)$ is the space of 
sections of the fiber bundle $\bigcup_{m\in M}\OAut_2(\mathfrak{v})(g(m),h(m))\rightarrow M$ 
(cf. app. \ref{sec:linftyauto}) and $A_F\in\Omega^1(M,\mathfrak{v}_1)$. 
It is required that $F$, $A_F$ obey the relations
\begin{subequations}
\label{0linfdglob}
\begin{align}
&\sigma_g-\sigma_h=\partial A_F, 
\vphantom{\Big]}
\label{0linfdgloba}
\\
&\varSigma_g-\varSigma_h=dA_F+[\sigma_h,A_F]+\frac{1}{2}[\partial A_F,A_F], \hspace{2.6cm}
\vphantom{\Big]}
\label{0linfdglobb}
\\
&\tau_g(\pi)-\tau_h(\pi)=-[\pi,A_F]+g_1{}^{-1}\big(dF(\pi)-F([\sigma_h,\pi]+\partial\tau_h(\pi))\big). 
\vphantom{\Big]}
\label{0linfdglobc}
\end{align}
\end{subequations}
Notice that our notation does not imply that $A_F$
is determined by $F$, but only that it is the partner of $F$ in the 
2--gauge transformation. 
We shall denote the set of all $2$--gauge transformations 
by $\OGau_2(M,\mathfrak{v})$.

\vspace{2.3mm}

{\it Orthogonal semistrict higher gauge transformation $2$--group}

\vspace{2.3mm}

$\OGau(M,\mathfrak{v})$ is an infinite dimensional strict Lie $2$--group, the {\it orthogonal 
gauge transformation $2$--group} of the theory. 
Analogously to the ordinary case, by this statement we mean simply that 
$\OGau(M,\mathfrak{v})$ is a strict $2$--group and that there is a natural definition of
$1$-- and $2$--cells infinitesimally close to the $1$-- and $2$--identities respectively and of Lie 
$2$--algebra brackets thereof 
by formal linearization of finite cells and their properly defined 
finite higher commutators in a neighborhood of the identities
such that the resulting infinitesimal cells constitute an infinite dimensional 
strict Lie $2$--algebra, as it will be detailed momentarily below.
The composition and inversion laws and the unit of the $1$--gauge transformations are 
\begin{subequations}
\label{4linfdglob}
\begin{align}
&h\diamond g=h\circ g, 
\vphantom{\Big]}
\label{4linfdglobz}
\\
&\sigma_{h\,\diamond \,g}
=\sigma_g+ g_0{}^{-1}(\sigma_h),
\vphantom{\Big]}
\label{4linfdgloba}
\\
&\varSigma_{h\,\diamond \,g}
=\varSigma_g+ g_1{}^{-1}\Big(\varSigma_h
+\frac{1}{2} g_2(g_0{}^{-1}(\sigma_h),g_0{}^{-1}(\sigma_h))\Big)-\tau_g(g_0{}^{-1}(\sigma_h)),
\vphantom{\Big]}
\label{4linfdglobb}
\\
&\tau_{h\,\diamond \,g}(\pi)
=\tau_g(\pi)+ g_1{}^{-1}\big(\tau_h(g_0(\pi))-g_2(g_0{}^{-1}(\sigma_h),\pi)\big),
\vphantom{\Big]}
\label{4linfdglobc}
\\
&g^{-1_\diamond}=g^{-1_\circ},
\vphantom{\Big]}
\label{6linfdglobz}
\\
&\sigma_{g^{-1_\diamond}}=-g_0(\sigma_g),
\vphantom{\Big]}
\label{6linfdgloba}
\\
&\varSigma_{g^{-1_\diamond}}=- g_1(\varSigma_g+\tau_g(\sigma_g))-\frac{1}{2} g_2(\sigma_g,\sigma_g),
\vphantom{\Big]}
\label{6linfdglobb}
\\
&\tau_{g^{-1_\diamond}}(\pi)=- g_1(\tau_g( g_0{}^{-1}(\pi)))- g_2(\sigma_g, g_0{}^{-1}(\pi)),
\vphantom{\Big]}
\label{6linfdglobc}
\\
&i=\id,    
\vphantom{\Big]}
\label{5linfdglobz}
\\
&\sigma_i=0,
\vphantom{\Big]}
\label{5linfdgloba}
\\
&\varSigma_i=0,
\vphantom{\Big]}
\label{5linfdglobb}
\\
&\tau_i(\pi)=0, \hspace{10.1cm}
\vphantom{\Big]}
\label{5linfdglobc}
\end{align}
\end{subequations}
where $g,h\in\OGau_1(N,\mathfrak{v})$. The horizontal and vertical composition 
and inversion laws and the units of the $2$--gauge transformations are defined by 
\begin{subequations}
\begin{align}
&G\diamond F=G\circ F,   
\vphantom{\Big]}
\label{50linfdgloba}
\\
&A_{G\,\diamond\, F}=A_F+h^{-1}{}_1(A_G)-g_1{}^{-1}Fh_0{}^{-1}(\sigma_k),
\vphantom{\Big]}
\label{50linfdglobb}
\\
&F^{-1_\diamond}=F^{-1_\circ},
\vphantom{\Big]}
\label{50linfdglobc}
\\
&A_{F^{-1_\diamond }}=-g_1(A_F)-F(\sigma_h), 
\vphantom{\Big]}
\label{50linfdglobd}
\end{align}
\begin{align}
&K\bullet H=K\bfdot H,
\vphantom{\Big]}
\label{50linfdglobe}
\\
&A_{K\,\bullet\, H}=A_H+A_K, \hspace{3.5cm}
\vphantom{\Big]}
\label{50linfdglobf}
\\
&H^{-1_\bullet}=H^{-1_\bfdot},
\vphantom{\Big]}
\label{50linfdglobg}
\\
&A_{H^{-1_\bullet}}=-A_H,
\vphantom{\Big]}
\label{50linfdglobh}
\\
&I_g=\mathrm{Id}_g, 
\vphantom{\Big]}
\label{50linfdglobi}
\\
&A_{I_g}=0, 
\vphantom{\Big]}
\label{50linfdglobj}
\end{align}
\end{subequations}  
where $g,h,k,l\in\OGau_1(M,\mathfrak{v})$ and $F,G,H,K\in\OGau_2(M,\mathfrak{v})$, 
with $F:g\Rightarrow h$, $G:k\Rightarrow l$ and $H,K$ composable. 
In \eqref{4linfdglobz}, \eqref{6linfdglobz}, \eqref{5linfdglobz},
the composition, inversion and unit in the right hand side are those of $\Aut_1(\mathfrak{v})$
thought of as holding pointwise on $M$ (cf. eqs. \eqref{mor3tlinalga}--\eqref{mor3tlinalgc},
\eqref{mor3/2tlinalgd}--\eqref{mor3/2tlinalgf},
\eqref{mor3tlinalgg}--\eqref{mor3tlinalgi}). 
In \eqref{50linfdgloba}, \eqref{50linfdglobc}, \eqref{50linfdglobe}, \eqref{50linfdglobg},
\eqref{50linfdglobi},
the horizontal and vertical compositions and inversions and the units in the right hand side 
are those of $\Aut_2(\mathfrak{v})$ thought of as holding pointwise on $M$ (cf. eqs. 
\eqref{mor4tlinalga}, \eqref{mor4/1tlinalgb},
\eqref{mor4tlinalgb}, \eqref{mor4/1tlinalgd}, \eqref{mor4tlinalgc}).

\vspace{2.3mm}

{\it Infinitesimal orthogonal semistrict higher gauge transformations}

\vspace{2.3mm}

{\it Infinitesimal higher $1$--gauge transformations} are $1$--gauge transformations
in linearized form as in the ordinary case. Their form can be easily realised expanding \eqref{1linfdglob}, 
\eqref{orthogauge1} around the unit transformation $i$ to first order.
The set $\mathfrak{ogau}_0(N,\mathfrak{v})$ of orthogonal infinitesimal higher $1$--gauge transformations
consists of the quadruples $(u,\dot\sigma_u,\dot\varSigma_u,\dot\tau_u)$ with 
$u\in\Map(M,\mathfrak{oaut}_0(\mathfrak{v}))$ (cf. app. \ref{sec:linftyauto})
$\dot\sigma_u\in\Omega^1(N,\mathfrak{v}_0)$, $\dot\varSigma_u\in\Omega^2(N,\mathfrak{v}_1)$,
and $\dot \tau_u\in\Omega^1(M,\mathfrak{aut}_1(\mathfrak{v}))$ satisfying the 
linearized higher Maurer--Cartan equations 
\begin{subequations}
\label{1inflinfdglob}
\begin{align}
&d\dot\sigma_u-\partial\dot\varSigma_u=0,
\vphantom{\Big]}
\label{1inflinfdgloba}
\\
&d\dot\varSigma_u=0
\vphantom{\Big]}
\label{1inflinfdglobb}
\end{align}
\begin{align}
&d\dot\tau_u(\pi)-[\pi,\dot\varSigma_u]=0
\vphantom{\Big]}
\label{2inflinfdglob}
\end{align}
\end{subequations}

and the linearized orthogonality condition  
\begin{equation}
(x,\dot\tau_u(y))+(y,\dot\tau_u(x))=0.
\label{orthogauge2}
\end{equation}
$u$, $\dot\sigma_u$, $\dot\varSigma_u$, $\dot\tau_u$ 
are further required to satisfy the relations
stemming from \eqref{3linfdglob} upon linearization. 
With $u=(u_0,u_1,u_2)$ (cf. app. \ref{sec:linftyauto}), these read
\begin{subequations}
\label{3inflinfdglob}
\begin{align}
&du_0(\pi)-[\dot\sigma_u,\pi]-\partial\dot\tau_u(\pi)=0,
\vphantom{\Big]}
\label{3inflinfdgloba}
\\
&du_1(\varPi)-[\dot\sigma_u,\varPi]-\dot\tau_u(\partial \varPi)=0,
\vphantom{\Big]} 
\label{3inflinfdglobb}
\\
&du_2(\pi,\pi) -[\dot\sigma_u,\pi,\pi]-\dot\tau_u([\pi,\pi])-2[\pi,\dot\tau_u(\pi)]=0.
\vphantom{\Big]}
\label{3inflinfdglobc}
\end{align}
\end{subequations}
As usual, our notation means only that 
$\dot\sigma_u$, $\dot\varSigma_u$, $\dot\tau_u$ are the partners of $u$ in the 
gauge transformation. 

The gauge for gauge symmetry of semistrict higher gauge theory also has an infinitesimal version.
{\it Infinitesimal higher 2--gauge transformations} 
are 2--gauge transformation in linearized form obtained by 
expanding \eqref{0linfdglob} around the unit transformation $I_i$ to first order. 
The set $\mathfrak{ogau}_1(N,\mathfrak{v})$ of {\it infinitesimal orthogonal higher $2$--gauge transformations} 
consists of the pairs $(Q,\dot A_Q)$ with $Q\in\Map(M,\mathfrak{oaut}_1(\mathfrak{v}))$
and $\dot A_Q\in\Omega^1(M,\mathfrak{v}_1)$. 
There are no further restrictions these objects must obey.
Our notation means that $\dot A_Q$ is the partner of $Q$ in the 
gauge transformation, 

\vspace{2mm}

{\it Infinitesimal orthogonal semistrict higher gauge transformation Lie $2$--algebra}

\vspace{2mm}

$\mathfrak{ogau}(M,\mathfrak{v})$ is an infinite dimensional strict Lie $2$--algebra,
in fact that of the gauge transformation Lie $2$--group $\OGau(M,\mathfrak{v})$. 
The boundary map and the brackets of $\mathfrak{ogau}(M,\mathfrak{v})$ are given by the expressions
\begin{subequations}
\label{4inflinfdglob}
\begin{align}
&[u,v]_\diamond =[u,v]_\circ,
\vphantom{\Big]}
\label{4inflinfdglobz}
\\
&\dot\sigma_{[u,v]_\diamond }=u_0(\dot\sigma_v)-v_0(\dot\sigma_u),
\vphantom{\Big]}
\label{4inflinfdgloba}
\end{align}
\begin{align}
&\dot\varSigma_{[u,v]_\diamond }=u_1(\dot\varSigma_v)-v_1(\dot\varSigma_u)
+\dot\tau_u(\dot\sigma_v)-\dot\tau_v(\dot\sigma_u),
\vphantom{\Big]}
\label{4inflinfdglobb}
\\
&\dot\tau_{[u,v]_\diamond }(\pi)
=u_1\dot\tau_v(\pi)-v_1\dot\tau_u(\pi)+\dot\tau_uv_0(\pi)
\vphantom{\Big]}
\label{4inflinfdglobc}
\\
&\hspace{5cm}-\dot\tau_vu_0(\pi)
+u_2(\dot\sigma_v,\pi)-v_2(\dot\sigma_u,\pi),
\vphantom{\Big]}
\nonumber
\\
&[u,v,w]_\diamond =[u,v,w]_\circ =0,
\vphantom{\Big]}
\label{5inflinfdglobb}
\end{align}
\end{subequations} %
where $u,v,w\in\mathfrak{ogau}_0(N,\mathfrak{v})$ and 
\begin{subequations}
\begin{align}
&\partial_\diamond Q=\partial_\circ Q, \hspace{7cm}
\vphantom{\Big]}
\label{6inflinfdglobb}
\\
&\dot\sigma_{\partial_\diamond Q}=-\partial \dot A_Q,
\vphantom{\Big]}
\label{6inflinfdglobc}
\\
&\dot\varSigma_{\partial_\diamond Q}=-d \dot A_Q,
\vphantom{\Big]}
\label{5inflinfdglobz}
\\
&\dot\tau_{\partial_\diamond Q}(\pi)=[\pi,\dot A_Q]-dQ(\pi),
\vphantom{\Big]}
\label{5inflinfdgloba}
\\
&\hspace{4cm}-\dot\tau_vu_0(\pi)
+u_2(\dot\sigma_v,\pi)-v_2(\dot\sigma_u,\pi),
\vphantom{\Big]}
\nonumber
\\
&[u,Q]_\diamond =[u,Q]_\circ ,
\vphantom{\Big]}
\label{6inflinfdglobz}
\\
&\dot A_{[u,Q]_\diamond }=u_1(\dot A_Q)-P(\dot\sigma_u), \hspace{3.75cm}
\vphantom{\Big]}
\label{6inflinfdgloba}
\end{align}
\end{subequations} 
where $u\in\mathfrak{ogau}_0(N,\mathfrak{v})$ and 
and $Q\in \mathfrak{ogau}_1(M,\mathfrak{v})$. 
In \eqref{6inflinfdglobb}, \eqref{4inflinfdglobz}, \eqref{6inflinfdglobz}, \eqref{5inflinfdglobb},
the boundary and the brackets in the right hand side are those of $\mathfrak{oaut}(\mathfrak{v})$
thought of as holding pointwise on $M$ (cf. eqs. \eqref{mor7tlinalgx}--\eqref{mor7tlinalgz},
\eqref{mor7tlinalga}--\eqref{mor7tlinalgc},
\eqref{mor7tlinalgv}, \eqref{mor7tlinalgw}). 


\subsection{\normalsize \textcolor{blue}{Semistrict higher Chern--Simons theory}}
\label{subsec:2tchern}

\hspace{.5cm}
In this subsect, we review the formulation of semistrict higher Chern--Simons theory
originally proposed in ref. \cite{Zucchini:2011aa} and further developed in ref. \cite{Soncini:2014ara}.
The reader is advised to consult these papers for motivation and background information.

In semistrict higher gauge theory with structure Lie $2$--algebra $\mathfrak{v}$
(cf. app. \ref{sec:linfty}), fields group in {\it field doublets} 
$(\phi,\varPhi_\phi)\in\Omega^m(M,\mathfrak{v}_0[n])\times \Omega^{m+1}(M,\mathfrak{v}_1[n])$
with $m,n\in\mathbb{Z}$, $-1\leq m\leq d$. When $m=-1$, the first component of the doublet vanishes. 
When $m=d$, the second one does. Such doublets are characterized by the form/ghost bidegree
$(m,n)$. In the following, as a rule, we shall denote the first component of a field doublet $(\phi,\varPhi_\phi)$,
$\phi$, in lower case and the second component, $\varPhi_\phi$, as the upper case form of the first component
with a suffix $\phi$ attached to indicate that $\varPhi_\phi$ is the partner of $\phi$ in the doublet. 
This allows us to conveniently denote and identify the doublet $(\phi,\varPhi_\phi)$ simply as $\phi$ 
in all those instances where the listing of all the components is not strictly necessary. 

\vspace{2mm}
 
{\it Connection doublets and their curvature doublets}

\vspace{2mm} 

In semistrict higher gauge theory, there is a distinguished field doublet $\omega$ 
of bidegree $(1,0)$, we shall call the theory's {\it $\mathfrak{v}$--2--connection} or simply {\it 2--connection}. 
Associated with it is another field doublet $f$ 
of bidegree $(2,0)$, called the {\it 2--curvature} of the 2--connection,  defined by the expressions
\begin{subequations}
\label{fFcurv}
\begin{align}
&f=d\omega+\frac{1}{2}[\omega,\omega]-\partial\varOmega_\omega,
\vphantom{\Big]}
\label{fcurv}
\\
&F_f=d\varOmega_\omega+[\omega,\varOmega_\omega]-\frac{1}{6}[\omega,\omega,\omega].
\vphantom{\Big]}
\label{Fcurv}
\end{align}
\end{subequations}
$f$ satisfies the {\it 2--Bianchi identities}
\begin{subequations}
\label{fFBianchi}
\begin{align}
&df+[\omega,f]+\partial F_f=0,
\vphantom{\Big]}
\label{fBianchi}
\\
&dF_f+[\omega,F_f]-[f,\varOmega_\omega]+\frac{1}{2}[\omega,\omega,f]=0
\vphantom{\Big]}
\label{FBianchi}
\end{align}
\end{subequations}
analogous to  the Bianchi identity of ordinary gauge theory. 
The 2--connection $\omega$ is said flat when its 2--curvature $f$ vanishes. 
We shall denote the space of 2--connections by $\Conn_2(N,\mathfrak{v})$.

The definition \eqref{fFcurv} of 2--curvature we have given is
designed in such a way that the flatness condition of a 2--connection 
has the same form as the Chevalley--Eilenberg differential relation \eqref{2tlinalgQ} 
of $\mathfrak{v}$, just as in ordinary gauge theory. 

In refs. \cite{Zucchini:2011aa}  a consistent definition of \pagebreak 
$\OGau_1(N,\mathfrak{v})$--action on the 2--connection space $\Conn_2(N,\mathfrak{v})$ has been worked out.
For $g\in\OGau_1(N,\mathfrak{v})$, 
the gauge transform ${}^g\omega$ of a 2--connection $\omega\in\Conn_2(N,\mathfrak{v})$ is 
\begin{subequations}
\label{7linfdglob}
\begin{align}
&{}^g\omega=g_0(\omega-\sigma_g), 
\vphantom{\Big]}
\label{7linfdgloba}
\\
&{}^g\varOmega_\omega=g_1(\varOmega_\omega-\varSigma_g+\tau_g(\omega-\sigma_g))
-\frac{1}{2}g_2(\omega-\sigma_g,\omega-\sigma_g).
\vphantom{\Big]}
\label{7linfdglobb}
\end{align}
\end{subequations}
Correspondingly, the gauge transform of the 2--curvature $f$ of $\omega$ is 
\begin{subequations}
\label{8linfdglob}
\begin{align}
&{}^gf=g_0(f), 
\vphantom{\Big]}
\label{8linfdgloba}
\\
&{}^gF_f=g_1(F_f-\tau_g(f))+g_2(\omega-\sigma_g,f).
\vphantom{\Big]}
\label{8linfdglobb}
\end{align}
\end{subequations}  

Turning to the Lie $2$--algebra $\mathfrak{ogau}(M,\mathfrak{v})$ of $\OGau(M,\mathfrak{v})$, we can 
express \eqref{7linfdglob} in infinitesimal form (cf. subsect. \ref{subsec:highgausym}). 
For an infinitesimal $1$--gauge transformation
$u\in\mathfrak{ogau}_0(M,\mathfrak{v})$, the 1--gauge variation $\delta_u\omega$  
of a 2--connection $\omega\in\Conn_2(N,\mathfrak{v})$ reads
\begin{subequations}
\label{7inflinfdglob}
\begin{align}
&\delta_u\omega=u_0(\omega)-\dot\sigma_u, 
\vphantom{\Big]}
\label{7inflinfdgloba}
\\
&\delta_u\varOmega_\omega=u_1(\varOmega_\omega)
-\dot\varSigma_u+\dot\tau_u(\omega)-\frac{1}{2}u_2(\omega,\omega).
\vphantom{\Big]}
\label{7inflinfdglobb}
\end{align}
\end{subequations}
The 1--gauge variation $\delta_uf$ of the 2--curvature $f$ of $\omega$ reads correspondingly as 
follows \hphantom{xxxxxxxxxxxxxxx}
\begin{subequations}
\label{8inflinfdglob}
\begin{align}
&\delta_uf=u_0(f), \hspace{4.3cm}
\vphantom{\Big]}
\label{8inflinfdgloba}
\\
&\delta_uF_f=u_1(F_f)-\dot\tau_u(f)+u_2(\omega,f).
\vphantom{\Big]}
\label{8inflinfdglobb}
\end{align}
\end{subequations}  

\vspace{2.5mm}
 
{\it Semistrict higher Chern--Simons theory} 

\vspace{2.5mm} 

We now introduce the {\it semistrict higher Chern--Simons theory}, 
on which this paper is based. The model's basic algebraic datum 
is a balanced Lie $2$--algebra $\mathfrak{v}$ endowed with an 
invariant form $(\cdot,\cdot)$ (cf. app. \ref{sec:linfty}).
The topological background is a compact oriented $4$--fold $N$. 
The field content consists in a $\mathfrak{v}$--2--connection 
$\omega\in\Conn_2(N,\mathfrak{v})$. The action functional is given by \pagebreak 
\begin{equation}
\CS_2(\omega)=\kappa_2\int_N\bigg[\frac{1}{2}(2f+\partial\varOmega_\omega,\varOmega_\omega)
-\frac{1}{24}(\omega, [\omega,\omega,\omega])\bigg],
\label{2tchern1}
\end{equation}
where $f$ is 2--form component of the 2--connection's 2--curvature given explicitly by \eqref{fcurv}.
The action $\CS_2(\omega)$ is designed so that its associated classical 
field equations are the flatness condition of the 2--connection $\omega$, 
\begin{subequations}
\label{2tchern2}
\begin{align}
&f=0,
\vphantom{\Big]}
\label{2tchern2a}
\\
&F_f=0
\vphantom{\Big]}
\label{2tchern2b}
\end{align}
\end{subequations}
(cf. eqs. \eqref{fFcurv}). For this reason, by its analogy to the standard Chern--Simons theory
and as implied by its given name, 
the present model can be legitimately considered a semistrict higher Chern--Simons theory.

Let $X$ be any manifold. In semistrict gauge theory, analogously to ordinary gauge theory, 
the de Rham complex $\Omega^*(X)$ includes the special subcomplex $\Omega_{\mathfrak{v}}{}^*(X)$
formed by those forms that are polynomial in the components of one or more 2--connections
and their differentials. In turn, 
$\Omega_{\mathfrak{v}}{}^*(X)$ contains the subcomplex $\Omega_{\mathfrak{v}\mathrm{inv}}{}^*(X)$ 
constituted by those elements which are invariant under the action \eqref{7linfdglob} of the orthogonal $1$--gauge 
transformation group $\OGau_1(X,\mathfrak{v})$ on $\Conn_2(N,\mathfrak{v})$.
For any $\mathfrak{v}$--2--connection $\omega\in\Conn_2(N,\mathfrak{v})$, a form $\mathcal{L}_2\in\Omega^4(X)$ 
\begin{equation}
\mathcal{L}_2=\frac{1}{2}(2f+\partial\varOmega_\omega,\varOmega_\omega)
-\frac{1}{24}(\omega, [\omega,\omega,\omega]).
\label{h2tchern1}
\end{equation}
of the same form as the Lagrangian density of the $\CS_2$ action is given. 
While clearly $\mathcal{L}_2\in\Omega_{\mathfrak{v}}{}^4(X)$, in general 
$\mathcal{L}_2\not\in\Omega_{\mathfrak{v}\mathrm{inv}}{}^4(X)$, as 
\begin{align}
&{}^g\mathcal{L}_2=\mathcal{L}_2-\frac{1}{4}(\sigma_g,d\varSigma_g)
-d\bigg[\frac{1}{2}(\sigma_g,\varSigma_g)
\vphantom{\Big]}
\label{h2tchern2}
\\
&\hspace{2cm}
+\frac{1}{6}(\omega-\sigma_g, g_1{}^{-1}g_2(\omega-\sigma_g,\omega-\sigma_g)
+6\varSigma_g-3\tau_g(\omega-\sigma_g))\bigg].
\vphantom{\Big]}
\nonumber
\end{align}
for $g\in\OGau_1(X,\mathfrak{v})$. Similarly to standard gauge theory, one has 
\begin{equation}
d\mathcal{L}_2=\mathcal{C}_2,
\label{h2tchern3}
\end{equation}
where $\mathcal{C}_2\in\Omega^5(X)$ is the 2--curvature bilinear 
\begin{equation}
\mathcal{C}_2=(f,F_f).
\label{h2tchern4}
\end{equation}
Hence, $\mathcal{C}_2\in\Omega_{\mathfrak{v}}{}^5(X)$. However, 
unlike $\mathcal{L}_2$, $\mathcal{C}_2$ is invariant under the action of 
$\OGau_1(X,\mathfrak{v})$ on $\Conn_2(N,\mathfrak{g})$, 
\begin{align}
{}^g\mathcal{C}_2=\mathcal{C}_2,
\label{h2tchern6}
\end{align}
so that $\mathcal{C}_2\in\Omega_{\mathfrak{v}\mathrm{inv}}{}^5(X)$. 
From \eqref{h2tchern2} and \eqref{h2tchern3}, it follows so that $\mathcal{C}_2$, 
while exact in the complex $\Omega_{\mathfrak{v}}{}^*(X)$, 
is generally only closed in the $\OGau_1(X,\mathfrak{v})$--invariant complex 
$\Omega_{\mathfrak{v}\mathrm{inv}}{}^*(X)$. 
It thus defines a class $[\mathcal{C}_2]_{\mathrm{inv}}\in H_{\mathfrak{v}\mathrm{inv}}{}^5(X)$. 
Further, the variation $\delta\mathcal{C}_2$ of $\mathcal{C}_2$ under 
arbitrary variations  $\delta\omega$, $\delta\varOmega_\omega$ of $\omega$, $\varOmega_\omega$
is given by 
\begin{equation}
\delta\mathcal{C}_2=d\big[(\delta\omega,F_f)+(f,\delta\varOmega_\omega)\big].
\label{h2tchern7}
\end{equation}
where the $5$--form in the right hand side is $\OGau_1(X,\mathfrak{v})$ invariant 
\begin{equation}
({}^g\delta\omega,{}^gF_f)+({}^gf,{}^g\delta\varOmega_\omega)=(\delta\omega,F_f)+(f,\delta\varOmega_\omega).
\label{h2tchern8}
\end{equation}
Consequently, albeit $\mathcal{C}_2$ is not necessarily exact in 
$\Omega_{\mathfrak{v}\mathrm{inv}}{}^*(X)$, its variation $\delta\mathcal{C}_2$ always is. 
These properties indicate that $\mathcal{L}_2$ is the Chern--Simons form of the higher
characteristic class $[\mathcal{C}_2]_{\mathrm{inv}}$. 

Just as the ordinary Chern--Simons action is not invariant 
under the full gauge transformation action, 
the $\CS_2$ action is not invariant under the $\OGau_1(N,\mathfrak{v})$-- action \eqref{7linfdglob}
on $\Conn_2(N,\mathfrak{v})$. 
In fact, \eqref{h2tchern2} implies that  
\begin{equation}
\CS_2({}^g\omega)=\CS_2(\omega)-\kappa_2 Q_2(g)
\label{2tchern3}
\end{equation}
for $g\in\OGau_1(N,\mathfrak{v})$, where the anomaly $Q_2(g)$ is given by 
\begin{equation}
Q_2(g)=\frac{1}{4}\int_N\big[2(d\sigma_g,\varSigma_g)-(\sigma_g,d\varSigma_g)\big].
\label{2tchern4}
\end{equation}
$Q_2(g)$ is in fact simply related to the $\CS_2$ action itself, \pagebreak 
\begin{equation}
Q_2(g)=\kappa_2{}^{-1}\CS_2(\sigma_g),    
\label{2tchern5}
\end{equation}
where we view $\sigma_g,\varSigma_g$ as the components of a flat 2--connection 
(cf. eqs. \eqref{1linfdgloba}, \eqref{1linfdglobb} and \eqref{fFcurv}). 

By \eqref{2tchern4}, the anomaly density is the form $q_2\in\Omega^4(N)$
\begin{equation}
q_2=\frac{1}{4}\big[2(d\sigma_g,\varSigma_g)-(\sigma_g,d\varSigma_g)\big].
\label{d2tchern1}
\end{equation}
As $\sigma_g$ is a 2--connection, $q_2\in\Omega_{\mathfrak{v}}{}^4(N)$. 
Using \eqref{h2tchern2}, \eqref{h2tchern3} and \eqref{h2tchern6}, it is immediately checked that 
$q_2$ is closed. The variation of $q_2$ under 
continuous deformations of the gauge transformation $g$ is instead exact
\begin{equation}
\delta q_2=d(\delta\sigma_g,\varSigma_g).
\label{d2tchern2}
\end{equation}
In analogy to familiar Chern--Simons theory, $Q_2(g)$ is so a topological invariant of $g$,
which we may interpret as a higher winding number of the higher gauge transformation $g$.  

As in the usual Chern--Simons model, 
the fact that the $\CS_2$ action is not $\OGau_1(N,\mathfrak{v})$ invariant prevents 
the a full $\OGau_1(N,\mathfrak{v})$--invariant functional integral quantization of the $\CS_2$ field theory 
unless the pair of the $4$--fold $N$ and the balanced Lie $2$--algebra $\mathfrak{v}$ with invariant form is 
{\it admissible}, that is such that there is a 
positive value of $\kappa_2$ such that $\kappa_2 Q_2(g)\in 2\pi\mathbb{Z}$ for all $g\in\OGau_1(N,\mathfrak{v})$.
Denoting by $\kappa_{2N\mathfrak{v}}$ the smallest value of $\kappa_2$ with such property,
the 1--gauge invariant functional 
integral quantization of the $\CS_2(N,\mathfrak{v})$ theory is in principle 
feasable if the coupling constant $\kappa_2$ is of the form \hphantom{xxxxxxxxxxxxx}
\begin{equation}
\kappa_2=k\kappa_{2N\mathfrak{v}},
\label{2tchern7}
\end{equation}
with $k\in\mathbb{Z}$ an integer. We shall call $k$ {\it level} as in the ordinary theory. 

An important unsolved problem of the theory is the classification of
the admissible pairs $(N,\mathfrak{v})$, assuming that there are any \cite{Soncini:2014ara}. 
For a fixed base manifold $N$, 
a complete understanding of the conditions of admissibility almost certainly 
involves the issue of the integrability of $\mathfrak{v}$ to a semistrict Lie $2$--group $V$,
a rather delicate problem. In the special version of semistrict higher Chern--Simons theory studied in this paper,
this problem is circumvented by the peculiar nature of the underlying Lie 2--algebra $\mathfrak{v}$,
which is constructed from ordinary Lie group data {\it ab initio}.

\vfil\eject

\section{\normalsize \textcolor{blue}{Special higher gauge theory}}\label{sec:specgau}

\hspace{.5cm} 
In this section, we shall study special higher gauge theory in detail. Though this belongs to the realm of
semistrict higher gauge theory, its whole symmetry structure is ultimately encoded in an ordinary
Lie group $G$ with certain properties, making the well--developed tools of standard 
Lie theory available. 

In what follows, 
$G$ is a compact connected Lie group whose the Lie algebra $\mathfrak{g}$ has a non trivial 
center $\mathfrak{z}(\mathfrak{g})$ and is equipped with an invariant symmetric non singular 
bilinear form $(\cdot,\cdot)$ and a choice of a central element $k\in\mathfrak{z}(\mathfrak{g})$ 
such that $(k,k)\not=0$. (See \cite{Cirio:2013xka} for a related set--up.)  
Finally, $N$ is a smooth manifold.



\subsection{\normalsize \textcolor{blue}{The special gauge transformation 2--group}}\label{subsec:spechgs}

\hspace{.5cm}
In this subsection, we shall define and study the special gauge transformation 2--group $\Gau(N,G)$ and its Lie 
2-algebra $\mathfrak{gau}(N,G)$ in detail.
The definition of the content and the operations of $\Gau(N,G)$ 
is to a considerable extent determined by the requirement that $\Gau(N,G)$ admits a 2--group morphism
into the orthogonal gauge transformation 2--group $\OGau(N,\mathfrak{v}_k(\mathfrak{g}))$ for a certain 
balanced Lie 2--algebra $\mathfrak{v}_k(\mathfrak{g})$ with invariant form 
associated with the Lie group $G$, the invariant form $(\cdot,\cdot)$ and the central element $k$. 
The form of $\mathfrak{v}_k(\mathfrak{g})$ in turn is essentially determined by the Lie theoretic 
nature of its constitutive data and by the requirement of being skeletal. 
Because of its ostensible complexity, our construction may appear at first glance 
somewhat arbitrary and lacking motivation. 
It is however rather delicate: our attempts to modify it at several key points preserving
its essential properties have met failure so far, but we cannot rule out the  possibility that 
alternative definitions satisfying the same basic conditions exist. 
Later, we shall show that the structure of $\Gau(N,G)$ subsumes a number of familiar 
construction from ordinary gauge theory.

\vspace{2mm}
 
{\it Special gauge transformations} 

\vspace{2mm}

The set $\Gau_1(N,G)$ of {\it special $G$--1--gauge transformations}
consists of all qua\-druples  $(\gamma,\varsigma_\gamma,\alpha_\gamma,\chi_\gamma)$
with $\gamma\in\Map(N,G)$, $\varsigma_\gamma\in\Omega^2(N,\mathbb{R})$, 
$\alpha_\gamma\in\Omega^0(N,\End(\mathfrak{g}))$, $\chi_\gamma\in\Omega^1(N,\mathfrak{g})$ satisfying
\begin{subequations}
\label{spechgs1-3}
\begin{align}
&(\gamma^{-1}d\gamma,k)=0,
\vphantom{\Big]}
\label{spechgs1}
\\
&(\gamma^{-1}d\gamma,[\gamma^{-1}d\gamma,\gamma^{-1}d\gamma])-6d\varsigma_\gamma=0,
\vphantom{\Big]}
\label{spechgs2}
\\
&(x,\alpha_\gamma(y))+(y,\alpha_\gamma(x))=0,
\vphantom{\Big]}
\label{spechgs3}
\end{align}
\end{subequations}
with $x,y\in\mathfrak{g}$. 

Two special $G$--1--gauge transformations $\beta,\gamma\in\Gau_1(N,G)$ are said to be {\it 2--gauge compatible}
whenever \hphantom{xxxxxxxx}
\begin{equation}
\gamma\beta{}^{-1}=K
\label{spechgs4}
\end{equation}
for some $K\in Z(G)$ seen as a constant element of $\Map(N,G)$ and 
\begin{equation}
\varsigma_\beta=\varsigma_\gamma. 
\label{spechgs5}
\end{equation}
2--compatibility is an equivalence relation in $\Gau_1(N,G)$.

Let $\beta,\gamma\in\Gau_1(N,G)$ be 2--gauge compatible special $G$--1--gauge transformations.
The set $\Gau_2(N,G)(\gamma,\beta)$ of {\it special $G$--2--gauge transformations} $\gamma\Rightarrow \beta$
consists of all triples $(K,\varPhi_K,P_K)$ with $K\in \Map(N,Z(G))$, 
$\varPhi_K\in\Omega^0(N,\End(\mathfrak{g}))$, $P_K\in\Omega^1(N,\mathfrak{g})$ satisfying \hphantom{xxxxxxxx}
\begin{subequations}
\label{spechgs6-7}
\begin{align}
&K^{-1}dK=0,
\vphantom{\Big]}
\label{spechgs6}
\\
&(x,\varPhi_K(y))+(y,\varPhi_K(x))=0
\vphantom{\Big]}
\label{spechgs7}
\end{align}
\end{subequations}
and such that \hphantom{xxxxxxxx}
\begin{subequations}
\label{spechgs8-10}
\begin{align}
&\gamma\beta^{-1}=K,
\vphantom{\Big]}
\label{spechgs8}
\\
&\alpha_\gamma(\pi)-\alpha_\beta(\pi)-\varPhi_K(\pi)=0,
\vphantom{\Big]}
\label{spechgs9}
\\
&\chi_\gamma-\chi_\beta-P_K=0.
\vphantom{\Big]}
\label{spechgs10}
\end{align}
\end{subequations}
We let $\Gau_2(N,G)$ be the set of all special 2--gauge transformations
with arbitrary source and target.

\vspace{2.3mm}
 
{\it Special gauge transformation 2--group} 

\vspace{2.3mm} 

$\Gau(N,G)=(\Gau_1(N,G),\Gau_2(N,G))$ is an infinite dimensional 
strict Lie 2--group, the {\it special $G$--gauge transformation 2--group}. 
The composition and inversion laws and the unit of special $1$--gauge transformations read
\begin{subequations}
\label{spechgs11-22}
\begin{align}
&\beta\diamond\gamma=\beta\gamma,
\vphantom{\Big]}
\label{spechgs11}
\\
&\varsigma_{\beta\diamond\gamma}=\varsigma_\gamma+\varsigma_\beta-(\beta^{-1}d\beta,d\gamma\gamma^{-1}),
\vphantom{\Big]}
\label{spechgs12}
\\
&\alpha_{\beta\diamond\gamma}(\pi)=\alpha_\gamma(\pi)+\gamma^{-1}\alpha_\beta(\gamma\pi\gamma^{-1})\gamma,
\vphantom{\Big]}
\label{spechgs13}
\\
&\chi_{\beta\diamond\gamma}=\chi_\gamma+\gamma^{-1}\chi_\beta\gamma+\alpha_\gamma(\gamma^{-1}\beta^{-1}d\beta\gamma),
\vphantom{\Big]}
\label{spechgs14}
\\
&\gamma^{-1_\diamond}=\gamma^{-1},
\vphantom{\Big]}
\label{spechgs15}
\\
&\varsigma_{\gamma^{-1_\diamond}}=-\varsigma_\gamma,
\vphantom{\Big]}
\label{spechgs16}
\\
&\alpha_{\gamma^{-1_\diamond}}(\pi)=-\gamma\alpha_\gamma(\gamma^{-1}\pi\gamma)\gamma^{-1},
\vphantom{\Big]}
\label{spechgs17}
\\
&\chi_{\gamma^{-1_\diamond}}=\gamma(-\chi_\gamma+\alpha_\gamma(\gamma^{-1}d\gamma))\gamma^{-1},
\vphantom{\Big]}
\label{spechgs18}
\\
&\iota=1,
\vphantom{\Big]}
\label{spechgs19}
\\
&\varsigma_\iota=0,
\vphantom{\Big]}
\label{spechgs20}
\\
&\alpha_\iota(\pi)=0,
\vphantom{\Big]}
\label{spechgs21}
\\
&\chi_\iota=0,
\vphantom{\Big]}
\label{spechgs22}
\end{align}
\end{subequations}
where $\beta,\gamma\in \Gau_1(N,G)$. The horizontal and vertical composition and inversion laws and the units of 
special $2$--gauge transformations are defined by 
\begin{subequations}
\label{spechgs23-37}
\begin{align}
&\varLambda\diamond K=\varLambda K,
\vphantom{\Big]}
\label{spechgs23}
\\
&\varPhi_{\varLambda\diamond K}(\pi)=\varPhi_K(\pi)+\gamma^{-1}\varPhi_\varLambda(\gamma\pi\gamma^{-1})\gamma,
\vphantom{\Big]}
\label{spechgs24}
\\
&P_{\varLambda\diamond K}=P_K+\gamma^{-1}P_\varLambda\gamma+\varPhi_K(\gamma^{-1}\nu^{-1}d\nu\gamma),
\vphantom{\Big]}
\label{spechgs25}
\end{align}
\begin{align}
&K^{-1_\diamond}=K^{-1},
\vphantom{\Big]}
\label{spechgs26}
\\
&\varPhi_{K^{-1_\diamond}}(\pi)=-\gamma\varPhi_K(\gamma^{-1}\pi\gamma)\gamma^{-1},
\vphantom{\Big]}
\label{spechgs27}
\\
&P_{K^{-1_\diamond}}=\gamma(-P_K+\varPhi_K(\gamma^{-1}d\gamma))\gamma^{-1},
\vphantom{\Big]}
\label{spechgs28}
\\
&\varXi\bullet H=\varXi H,
\vphantom{\Big]}
\label{spechgs29}
\\
&\varPhi_{\varXi\bullet H}(\pi)=\varPhi_\varXi(\pi)+\varPhi_H(\pi),
\vphantom{\Big]}
\label{spechgs30}
\\
&P_{\varXi\bullet H}=P_\varXi+P_H, 
\vphantom{\Big]}
\label{spechgs31}
\\
&\varXi^{-1_\bullet}=\varXi^{-1},
\vphantom{\Big]}
\label{spechgs32}
\\
&\varPhi_{\varXi^{-1_\bullet}}(\pi)=-\varPhi_{\varXi}(\pi), 
\vphantom{\Big]}
\label{spechgs33}
\\
&P_{\varXi^{-1_\bullet}}=-P_{\varXi},
\vphantom{\Big]}
\label{spechgs34}
\\
&I_\gamma=1,
\vphantom{\Big]}
\label{spechgs35}
\\
&\varPhi_{I_\gamma}(\pi)=0,
\vphantom{\Big]}
\label{spechgs36}
\\
&P_{I_\gamma}=0,
\vphantom{\Big]}
\label{spechgs37}
\end{align}
\end{subequations}
where $\beta,\gamma,\mu,\nu\in\Gau_1(N,G)$, $K,\varLambda,\varXi,H\in\Gau_2(N,G)$
with $K:\gamma\Rightarrow \beta$, $\varLambda:\nu\Rightarrow \mu$ and $H$,
$\varXi$ composable. 

\vspace{2.3mm}
 
{\it Infinitesimal special gauge transformations}

\vspace{2.3mm} 

Infinitesimal special $1$--gauge transformations are $1$--gauge transformations
in linearized form. Their form can be easily realised expanding \eqref{spechgs1-3}
 around the unit transformation $\iota$ to first order.
The set $\mathfrak{gau}_0(N,G)$ of {\it infinitesimal special $G$--1--gauge transformations}
consists of all quadruples  $(\theta,\dot\varsigma_\theta,\dot\alpha_\theta,\dot\chi_\theta)$
such that $\theta\in\Map(N,\mathfrak{g})$, $\dot\varsigma_\theta\in\Omega^2(N,\mathbb{R})$, 
$\dot\alpha_\theta\in\Omega^0(N,\End(\mathfrak{g}))$, 
$\dot\chi_\theta\in\Omega^1(N,\mathfrak{g})$ and satisfying 
\begin{subequations}
\label{spechgs38-40}
\begin{align}
&(d\theta,k)=0,
\vphantom{\Big]}
\label{spechgs38}
\\
&d\dot\varsigma_\theta=0,
\vphantom{\Big]}
\label{spechgs39}
\\
&(x,\dot\alpha_\theta(y))+(y,\dot\alpha_\theta(x))=0,
\vphantom{\Big]}
\label{spechgs40}
\end{align}
\end{subequations}
with $x,y\in\mathfrak{g}$. 

Infinitesimal special 2--gauge transformations are 2--gauge transformations in linearized form obtained by 
expanding \eqref{0linfdglob} around the unit transformation $I_\iota$ to first order. 
The set $\mathfrak{gau}_1(N,G)$ of {\it infinitesimal special $G$--2--gauge transformations}
consists of all triples $(E,\dot\varPhi_E,\dot P_E)$ such that
$E\in\Map(N,\mathfrak{z}(\mathfrak{g}))$, 
$\dot\varPhi_E\in\Omega^0(N,\End(\mathfrak{g}))$, $\dot P_E\in\Omega^1(N,\mathfrak{g})$ and satisfying
\begin{subequations}
\label{spechgs41-42}
\begin{align}
&dE=0, 
\vphantom{\Big]}
\label{spechgs41}
\\
&(x,\dot\varPhi_E(y))+(y,\dot\varPhi_E(x))=0,
\vphantom{\Big]}
\label{spechgs42}
\end{align}
\end{subequations}
with $x,y\in\mathfrak{g}$.

\vspace{2.3mm}
 
{\it Infinitesimal special gauge transformation Lie 2--algebra}

\vspace{2.3mm} 

$\mathfrak{gau}(N,G)=(\mathfrak{gau}_0(N,G),\mathfrak{gau}_1(N,G))$
is an infinite dimensional strict Lie 2--algebra, the {\it special $G$--gauge transformation 
Lie 2--algebra}. $\mathfrak{gau}(N,G)$ is the Lie 2--algebra of the 
strict Lie 2--group $\Gau(N,G)$, as expected on general grounds. 
The boundary map and the brackets of $\mathfrak{gau}(N,G)$ are given 
by the expressions
\begin{subequations}
\label{spechgs43-47}
\begin{align}
&[\theta,\zeta]_\diamond=[\theta,\zeta],
\vphantom{\Big]}
\label{spechgs43}
\\
&\dot\varsigma_{[\theta,\zeta]_\diamond}=-2(d\theta,d\zeta),
\vphantom{\Big]}
\label{spechgs44}
\\
&\dot\alpha_{[\theta,\zeta]_\diamond}(\pi)=[\theta,\dot\alpha_\zeta(\pi)]-[\zeta,\dot\alpha_\theta(\pi)]
+\dot\alpha_\theta([\zeta,\pi])-\dot\alpha_\zeta([\theta,\pi]),
\vphantom{\Big]}
\label{spechgs45}
\\
&\dot\chi_{[\theta,\zeta]_\diamond}=[\theta,\dot\chi_\zeta]-[\zeta,\dot\chi_\theta]
-\dot\alpha_\theta(d\zeta)+\dot\alpha_\zeta(d\theta),
\vphantom{\Big]}
\label{spechgs46}
\\
&[\theta,\zeta,\eta]_\diamond=0, 
\vphantom{\Big]}
\label{spechgs47}
\end{align}
\end{subequations}
where $\theta,\zeta,\eta\in\mathfrak{gau}_0(N,G)$ and 
\begin{subequations}
\label{spechgs48-54}
\begin{align}
&\partial_\diamond E=E,
\vphantom{\Big]}
\label{spechgs48}
\\
&\dot\varsigma_{\partial_\diamond E}=0,
\vphantom{\Big]}
\label{spechgs49}
\\
&\dot\alpha_{\partial_\diamond E}(\pi)=-\dot\varPhi_E(\pi),
\vphantom{\Big]}
\label{spechgs50}
\end{align}
\begin{align}
&\dot\chi_{\partial_\diamond E}=-\dot P_E,
\vphantom{\Big]}
\label{spechgs51}
\\
&[\theta,E]_\diamond=0, 
\vphantom{\Big]}
\label{spechgs52}
\\
&\dot\varPhi_{[\theta,E]_\diamond}(\pi)=[\theta,\dot\varPhi_E(\pi)]-\dot\varPhi_E([\theta,\pi]),
\vphantom{\Big]}
\label{spechgs53}
\\
&\dot P_{[\theta,E]_\diamond}=[\theta,\dot P_E]+\dot\varPhi_E(d\theta)
\vphantom{\Big]},
\label{spechgs54}
\end{align}
\end{subequations}
where $\theta\in\mathfrak{gau}_0(N,G)$ and $E\in\mathfrak{gau}_1(N,G)$. 

\vspace{2.5mm}
 
{\it The special Lie 2--algebra $\mathfrak{v}_k(\mathfrak{g})$}

\vspace{2.5mm} 

With $\mathfrak{g}$, there is associated a semistrict Lie 2--algebra 
$\mathfrak{v}_k(\mathfrak{g})$, called {\it special}, as follows. $\mathfrak{v}_{k0}(\mathfrak{g})
=\mathfrak{v}_{k1}(\mathfrak{g})=\mathfrak{g}$. 
\begin{subequations}
\label{spechgs55-58}
\begin{align}
&\partial_{\mathfrak{v}_k(\mathfrak{g})}X=0,
\vphantom{\Big]}
\label{spechgs55}
\\
&[x,y]_{\mathfrak{v}_k(\mathfrak{g})}=[x,y],
\vphantom{\Big]}
\label{spechgs56}\\
&[x,X]_{\mathfrak{v}_k(\mathfrak{g})}=[x,X],
\vphantom{\Big]}
\label{spechgs57}
\\
&[x,y,z]_{\mathfrak{v}_k(\mathfrak{g})}=(x,[y,z])k-(x,k)[y,z]-(y,k)[z,x]-(z,k)[x,y],
\vphantom{\Big]}
\label{spechgs58}
\end{align}
\end{subequations}
where in the right hand side $x,y,z\in\mathfrak{v}_{k0}(\mathfrak{g})$ and $X\in\mathfrak{v}_{k1}(\mathfrak{g})$ are
all treated as elements of $\mathfrak{g}$. 
Since the boundary map $\partial_{\mathfrak{v}_k(\mathfrak{g})}$ vanishes,
$\mathfrak{v}_k(\mathfrak{g})$ is skeletal. As every Lie 2--algebra is equivalent to a 
skeletal Lie 2--algebra, the Lie 2--algebras of the above form 
span a broad range of Lie 2--algebra examples. 

As $\dim \mathfrak{v}_{k0}(\mathfrak{g})
=\dim \mathfrak{v}_{k1}(\mathfrak{g})=\dim\mathfrak{g}$, the Lie 2--algebra 
$\mathfrak{v}_k(\mathfrak{g})$ is balanced (cf. app. \ref{sec:linfty}). 
Further, $\mathfrak{v}_k(\mathfrak{g})$ is endowed with an invariant form, 
the pairing $(\cdot,\cdot)$. 

The Lie 2--algebra $\mathfrak{v}_k(\mathfrak{g})$ bears some formal similarity 
to the string Lie 2--algebra $\mathfrak{string}_k(\mathfrak{s})$ of a simple Lie algebra 
$\mathfrak{s}$. 
To see this, let us recall how $\mathfrak{string}_k(\mathfrak{s})$ is defined. 
$\mathfrak{string}_k(\mathfrak{s})$ depends on a parameter $k\in\mathbb{R}$.
The two terms of 
$\mathfrak{string}_k(\mathfrak{s})$ are $\mathfrak{string}_{k0}(\mathfrak{s})=\mathfrak{s}$
and $\mathfrak{string}_{k1}(\mathfrak{s})=\mathbb{R}$ and 
\begin{subequations}
\label{spechgs81--84}
\begin{align}
&\partial_{\mathfrak{string}_k(\mathfrak{s})}X=0,
\vphantom{\Big]}
\label{spechgs81}
\end{align}
\begin{align}
&[x,y]_{\mathfrak{string}_k(\mathfrak{s})}=[x,y],
\vphantom{\Big]}
\label{spechgs82}\\
&[x,X]_{\mathfrak{string}_k(\mathfrak{s})}=0,
\vphantom{\Big]}
\label{spechgs83}
\\
&[x,y,z]_{\mathfrak{string}_k(\mathfrak{s})}=(x,[y,z])k,
\vphantom{\Big]}
\label{spechgs84}
\end{align}
\end{subequations}
where $x,y,z\in\mathfrak{string}_{k0}(\mathfrak{s})$ and $X\in\mathfrak{string}_{k1}(\mathfrak{s})$
and $(\cdot,[\cdot,\cdot])$ is the properly normalized invariant 3--form of $\mathfrak{s}$ associated 
with its Killing form. 
Just as $\mathfrak{v}_k(\mathfrak{g})$, $\mathfrak{string}_k(\mathfrak{s})$ is semistrict and skeletal.

The resemblance of $\mathfrak{v}_k(\mathfrak{g})$ and $\mathfrak{string}_k(\mathfrak{s})$ 
is even more striking for the following reason. 
It is well--known that the three argument bracket of a skeletal semistrict Lie 2--algebra $\mathfrak{v}$
constitutes a 3--cocycle $\mu_{\mathfrak{v}}$ of the Chevalley--Eilenberg complex 
$\mathrm{CE}(\mathfrak{v}_0,\mathfrak{v}_1)$ of the Lie algebra 
$\mathfrak{v}_0$ with values in the $\mathfrak{v}_0$--Lie module $\mathfrak{v}_1$. 
The associated degree 3 cohomology class of $[\mu_{\mathfrak{v}}]
\in H_{\mathrm{CE}}{}^3(\mathfrak{v}_0,\mathfrak{v}_1)$ characterizes 
the equivalence class of $\mathfrak{v}$ in the category $\mathrm{Lie2alg}$ \cite{Baez:2003fs}. 
For $\mathfrak{v}_k(\mathfrak{g})$, we have  
\begin{equation}
\mu_{\mathfrak{v}_k(\mathfrak{g})}=(\pi,[\pi,\pi])k-3(\pi,k)[\pi,\pi].
\label{85}
\end{equation}
The second term in the right hand side is however exact in the Chevalley--Eilenberg 
complex $\mathrm{CE}(\mathfrak{g},\mathfrak{g})$ as 
\begin{equation}
-3(\pi,k)[\pi,\pi]=6\mathcal{Q}_{\mathrm{CE}(\mathfrak{g},\mathfrak{g})}((\pi,k)\pi),
\label{86}
\end{equation}
thus does not contribute to the class $[\mu_{\mathfrak{v}_k(\mathfrak{g})}]$
and in this sense may be dropped. For $\mathfrak{string}_k(\mathfrak{s})$, we have
similarly 
\begin{equation}
\mu_{\mathfrak{string}_k(\mathfrak{s})}=(\pi,[\pi,\pi])k.
\label{87}
\end{equation} 

In ref. \cite{Baez:2005sn}, it is shown that the Lie 2--algebra $\mathfrak{string}_k(\mathfrak{s})$
can be integrated to a Lie 2--group only when the parameter $k$ is integer. More precisely,
in that case there exists an infinite dimensional strict Lie 2--group $P_kS$, whose Lie 2-algebra
$P_k\mathfrak{s}$ is equivalent to $\mathfrak{string}_k(\mathfrak{s})$ in the category
$\mathrm{Lie2alg}$. It is conceivable that a similar property holds for the Lie 2--algebra
$\mathfrak{v}_k(\mathfrak{g})$. Finding a Lie 2--group integrating $\mathfrak{v}_k(\mathfrak{g})$
may help to unveil some of the most subtle aspects of the special version of higher Chern--Simons theory
studied in this paper.

\vspace{2.3mm}
 
{\it The special to orthogonal semistrict higher symmetry 2--group morphism}

\vspace{2.3mm} 

There exists a basic strict 2--group morphism morphism from 
the special $G$--gauge 2--group $\Gau(N,G)$ to 
the orthogonal semistrict gauge transformation 2--group
$\OGau(N,\mathfrak{v}_k(\mathfrak{g}))$ (cf. subsect. \ref{subsec:highgausym}). 
For $\gamma\in\Gau_1(N,G)$, set 
\begin{subequations}
\label{spechgs59-64}
\begin{align}
&g_{\gamma0}(\pi)=\gamma\pi\gamma^{-1}, \hspace{6.9cm}
\vphantom{\Big]}
\label{spechgs59}
\\
&g_{\gamma1}(\varPi)=\gamma\varPi\gamma^{-1},
\vphantom{\Big]}
\label{spechgs60}
\\
&g_{\gamma2}(\pi,\pi)=\gamma(2[\pi,\alpha_\gamma(\pi)]-\alpha_\gamma([\pi,\pi]))\gamma^{-1},
\vphantom{\Big]}
\label{spechgs61}
\\
&\sigma_\gamma\equiv\sigma_{g_\gamma}=\gamma^{-1}d\gamma,
\vphantom{\Big]}
\label{spechgs62}
\\
&\varSigma_\gamma\equiv\varSigma_{g_\gamma}=\varsigma_\gamma k+d\chi_\gamma+[\gamma^{-1}d\gamma,\chi_\gamma], 
\vphantom{\Big]}
\label{spechgs63}
\\
&\tau_\gamma(\pi)\equiv\tau_{g_\gamma}(\pi)
=-(\pi,k)\gamma^{-1}d\gamma+(\pi,\gamma^{-1}d\gamma)k-[\pi,\chi_\gamma]
\vphantom{\Big]}
\label{spechgs64}
\\
&\hspace{4cm}-d\alpha_\gamma(\pi)-[\gamma^{-1}d\gamma,\alpha_\gamma(\pi)]
+\alpha_\gamma([\gamma^{-1}d\gamma,\pi]).
\vphantom{\Big]}
\nonumber
\end{align}
\end{subequations}
Then, $g_\gamma\equiv(g_\gamma,\sigma_\gamma,\varSigma_\gamma,\tau_\gamma)\in\OGau_1(N,\mathfrak{v}_k(\mathfrak{g}))$. 
Next, let $\beta,\gamma\in \Gau_1(N,G)$ and let $K\in\Gau_2(N,G)$ 
with $K:\gamma\Rightarrow\beta$. Set
\begin{subequations}
\label{spechgs65-66}
\begin{align}
&F_K(\pi)=-\gamma\varPhi_K(\pi)\gamma^{-1},
\vphantom{\Big]}
\label{spechgs65}
\\
&A_K\equiv A_{F_K}=-P_K. 
\vphantom{\Big]}
\label{spechgs66}
\end{align}
\end{subequations}
Then, $F_K\equiv(F_K,A_K)\in \OGau_2(N,\mathfrak{v}_k(\mathfrak{g}))$ 
with $F_K:g_\gamma\Rightarrow g_\beta$. 
It can be straightforwardly verified that the mappings $\gamma\rightarrow g_\gamma$ and $K\rightarrow F_K$
define a strict Lie 2--group morphism 
$\mathsans{m}:\Gau(N,G)\rightarrow\OGau(N,\mathfrak{v}_k(\mathfrak{g}))$. 

There is a counterpart of the 2--group morphism $\mathsans{m}$ at the Lie 2--algebra
level. Explicitly, for $\theta\in\mathfrak{gau}_0(N,G)$ set
\begin{subequations}
\label{spechgs67-72}
\begin{align}
&u_{\theta0}(\pi)=[\theta,\pi], 
\vphantom{\Big]}
\label{spechgs67}
\end{align}
\begin{align}
&u_{\theta1}(\varPi)=[\theta,\varPi], 
\vphantom{\Big]}
\label{spechgs68}
\\
&u_{\theta2}(\pi,\pi)=2[\pi,\dot\alpha_\theta(\pi)]-\dot\alpha_\theta([\pi,\pi]), 
\vphantom{\Big]}
\label{spechgs69}
\\
&\dot\sigma_\theta\equiv\dot\sigma_{u_\theta}=d\theta,
\vphantom{\Big]}
\label{spechgs70}
\\
&\dot\varSigma_\theta\equiv\dot\varSigma_{u_\theta}=\dot\varsigma_\theta k+d\dot\chi_\theta, 
\vphantom{\Big]}
\label{spechgs71}
\\
&\dot\tau_\theta(\pi)=\dot\tau_{u_\theta}(\pi)
=-(\pi,k)d\theta+(\pi,d\theta)k-[\pi,\dot\chi_\theta]-d\dot\alpha_\theta(\pi).
\vphantom{\Big]}
\label{spechgs72}
\end{align}
\end{subequations}
Then, $u_\theta\equiv(u_\theta,\dot\sigma_\theta,\dot\varSigma_\theta,\dot\tau_\theta)\in
\mathfrak{ogau}_0(N,\mathfrak{v}_k(\mathfrak{g}))$. 
Likewise, for $E\in\mathfrak{gau}_0(N,G)$ 
\begin{subequations}
\label{spechgs73-74}
\begin{align}
&Q_E(\pi)=-\dot\varPhi_E(\pi)
\vphantom{\Big]}
\label{spechgs73}
\\
&\dot A_E\equiv\dot A_{Q_E}=\dot P_E
\vphantom{\Big]}
\label{spechgs74}
\end{align}
\end{subequations}
Then, $Q_E\equiv(Q_E,\dot A_E)\in \mathfrak{ogau}_1(N,\mathfrak{v}_k(\mathfrak{g}))$.
The maps $\theta\rightarrow u_\theta$ and $E\rightarrow Q_E$
define a strict Lie 2--algebra morphism $\dot{\mathsans{m}}:
\mathfrak{gau}(N,G)\rightarrow\mathfrak{ogau}(N,\mathfrak{v}_k(\mathfrak{g}))$. 

The Lie 2--group morphism $\mathsans{m}$ is generally neither full nor faithful as a 2--functor of strict 
2--categories. Similar remarks apply also to the Lie 2--algebra morphism $\dot{\mathsans{m}}$. 

\vspace{2.5mm}

{\it Homotopic non triviality of special 1--gauge transformations}

\vspace{2.5mm}

If $\gamma\in\Gau_1(N,G)$ is a special 1--gauge transformation, 
its component $\gamma\in\Map(N,G)$ can be viewed as an ordinary gauge transformation. 

Suppose that $L$ is a compact connected Lie group and that the Lie algebra $\mathfrak{l}$ of $L$
is equipped with an invariant symmetric non singular bilinear form $(\cdot,\cdot)$. As it is well--known, 
$L$ is characterized by a canonical closed 3--form $\varTheta_L$
\begin{equation}
\varTheta_L=\frac{1}{48\pi^2}(l^{-1}dl,[l^{-1}dl,l^{-1}dl]),
\label{spechgs78/1}
\end{equation}
where $l^{-1}dl$ denotes the left invariant Maurer--Cartan form of $L$. 
$\varTheta_L$ is insensitive to the center $Z(L)$ of $L$ in the sense that 
\begin{equation}
\varTheta_L=\pi^*\varTheta_{\Ad(L)}
\label{spechgs79/1}
\end{equation}
where $\Ad(L)=L/Z(L)$ is the adjoint Lie group of $L$ and $\pi:L\rightarrow \Ad(L)$ is
the natural projection. $\Ad(L)$ is a semisimple Lie group.
We assume henceforth that the bilinear form $(\cdot,\cdot)$ has been normalized so that 
the cohomology class $[\varTheta_{\Ad(L)}]\in H^3(G,\mathbb{R})$ actually lies in the 
cohomology lattice $H_{\mathbb{Z}}{}^3(G,\mathbb{R})$ 
\footnote{$\vphantom{\dot{\dot{\dot{\dot{x}}}}}$ For a manifold $X$, the cohomology lattice 
$H_{\mathbb{Z}}{}^p(X,\mathbb{R})$ is the image of the natural inclusion 
$H^p(X,\mathbb{Z})\rightarrow H^p(X,\mathbb{R})$. 
$H_{\mathbb{Z}}{}^p(X,\mathbb{R})$ consists of the degree $p$ real cohomology classes 
of $X$ with integer periods. $\vphantom{\ul{\ul{\ul{\ul{g}}}}}$}. 

Every map $\gamma\in\Map(N,G)$ is characterized by the 3--form $w(\gamma)\in\Omega^3(N,\mathbb{R})$ given by
the expression 
\begin{equation}
w(\gamma)=\frac{1}{48\pi^2}(\gamma^{-1}d\gamma,[\gamma^{-1}d\gamma,\gamma^{-1}d\gamma]).
\label{spechgs75}
\end{equation}
If the pairing $(\cdot,\cdot)$ is normalized as assumed in the previous paragraph, 
$w(\gamma)$ is the winding number density of $\gamma$: for any 3--cycle $C\in Z_3(N,\mathbb{Z})$, 
\begin{equation}
W(\gamma,C)=\oint_Cw(\gamma)
\label{spechgs76}
\end{equation}
is the winding number of $\gamma$ on $C$. However, in what follows, it is the winding number density that
plays a basic role. 

For $\gamma\in\Map(N,G)$, $w(\gamma)$ can be expressed in terms of the 3--form $\varTheta_G$ as
\begin{equation}
w(\gamma)=\gamma^*\varTheta_G=(\pi\circ\gamma)^*\varTheta_{\Ad(G)}.
\label{spechgs80}
\end{equation}
Hence, $w(\gamma)$ is a closed 3--form whose cohomology class $[w(\gamma)]\in H^3(N,\mathbb{R})$ lies in the 
cohomology lattice $H_{\mathbb{Z}}{}^3(N,\mathbb{R})$. We have in this way a mapping 
$[w]:\Map(N,G)\rightarrow H_{\mathbb{Z}}{}^3(N,\mathbb{R})$, which we shall call
{\it cohomological winding number map}. Since for $\beta,\gamma\in\Map(N,G)$, one has 
\begin{equation}
w(\beta\gamma)=w(\beta)+w(\gamma)-\frac{1}{8\pi^2}d(\beta^{-1}d\beta,d\gamma\gamma^{-1}),
\label{spechgs79}
\end{equation}
where the third term in the right hand side is exact, $[w]$ is a group morphism. 

Let $\Map_c(N,G)$ be the normal subgroup of $\Map(N,G)$ of the elements $\gamma$ 
homotopic to the unit map $1$ and $\varPi(N,G)=\Map(N,G)/\Map_c(N,G)$ 
be the associated mapping class group. 
For an infinitesimal variation $\delta\gamma$ of $\gamma\in\Map(N,G)$, 
\begin{equation}
\delta w(\gamma)=\frac{1}{16\pi^2}d(\gamma^{-1}\delta\gamma,[\gamma^{-1}d\gamma,\gamma^{-1}d\gamma]).
\label{spechgs78}
\end{equation}
By virtue of the exactness of $\delta w$, the winding number map $[w]$ factors through a group morphism 
$[w]:\varPi(N,G)\rightarrow H_{\mathbb{Z}}{}^3(N,\mathbb{R})$.
The image $\varLambda(N,G)$ of $[w]$ is a (possibly degenerated) 
sublattice of the lattice $H_{\mathbb{Z}}{}^3(N,\mathbb{R})$, which we shall call 
{\it cohomological winding number lattice}. 

We shall call an element $\gamma\in\Map(N,G)$ {\it homotopically non trivial} if it is not 
mapped to the origin of $\varLambda(N,G)$ by the cohomological winding number map. A 1--gauge transformation 
$\gamma\in\Gau_1(N,G)$ is said homotopically 
non trivial if the associated ordinary gauge 
transformation is homotopically non trivial in the above sense.

By \eqref{spechgs2}, for a special 1--gauge transformation $\gamma\in\Gau_1(N,G)$, we have
\begin{equation}
w(\gamma)=\frac{1}{8\pi^2}d\varsigma_\gamma,
\label{spechgs77}
\end{equation}
As $w(\gamma)$ is exact, we have $[w(\gamma)]=0$. 
Therefore, our framework, as it has been formulated up to this point, allows 
only for homotopically trivial 1--gauge transformations $\gamma$.

If we want to include homotopically non trivial special 1-- gauge transformations $\gamma$, 
as it would be presumably required by a fully non perturbative formulation, we must relax 
\eqref{spechgs2} by allowing $\varsigma_\gamma$ to be only locally defined and, 
so, $w(\gamma)$ to be closed rather than merely exact. Formally, this can be done as follows.

Let $\{U_i\}$ be an open covering of $N$ and $\widehat{N}=\coprod_i U_i$ be the disjoint union of 
the $U_i$. A locally defined $p$--form $\alpha$ on $N$ is a $p$--form on $\widehat{N}$.
$\alpha$ is globally defined on $N$ precisely when it is the pull--back of a genuine $p$--form on $N$, 
which we shall denote also by $\alpha$,
under the natural surjection $\widehat{N}\rightarrow N$. We accommodate homotopically non trivial
1--gauge transformations, by letting $\varsigma_\gamma$ to be a locally defined 2--form on $N$ such that 
$d\varsigma_\gamma$ is globally defined on $N$. The homotopically trivial case corresponds 
to the situation where $\varsigma_\gamma$ itself is globally defined on $N$.
To have a well behaved theory, $\varsigma_\gamma$ must however satisfy certain requirements 
which we detail next. 

Let $w_i$, $i=1,\ldots,d$, be closed 3--forms whose cohomology classes $[w_i]\in H^3(N,\mathbb{R})$ 
form a set of generators of the lattice $\varLambda(N,G)$ (and therefore lie in $H_{\mathbb{Z}}{}^3(N,\mathbb{R})$). 
Then, there are group morphism $n_i:\varPi(N,G)\rightarrow \mathbb{Z}$ such that 
\begin{equation}
[w(\bar\gamma)]=\sum_in_i(\bar\gamma)[w_i]
\label{xspechgs2}
\end{equation}
for $\bar\gamma\in\varPi(N,G)$. 
Next, let $\varsigma_i/8\pi^2\in\Omega^2(\widehat{N})$ be fixed primitives of the forms 
$w_i$, 
\begin{equation}
w_i=\frac{1}{8\pi^2}d\varsigma_i.
\label{xspechgs3}
\end{equation}
With any $\bar\gamma\in\varPi(N,G)$, there is then associated a 2--form $\varsigma_{\bar\gamma}\in\Omega^2(\widehat{N})$ 
given by 
\begin{equation}
\varsigma_{\bar\gamma}=\sum_in_i(\bar\gamma)\varsigma_i.
\label{xspechgs4}
\end{equation} 
A group morphism $\varsigma_{-}:\varPi(N,G)\rightarrow \Omega^2(\widehat{N})$ is yielded in this way. 
 
We now require that homotopically non trivial special 1--gauge transformations $\gamma$ satisfy the 
condition that $\varsigma_\gamma-\varsigma_{\bar\gamma}$ 
is a globally defined 2--form, where $\bar\gamma\in\varPi(N,G)$ 
is the mapping class which $\gamma\in\Map(N,G)$ belongs to.
The reason for such a condition will become clear in the study of 
the 1--gauge transformation action on special 2--connections in the next subsection.
All the definitions we have given and results we have found in this subsection extend 
with no modification when homotopically non trivial special 1--gauge transformations $\gamma$ 
are included, since the above theory is completely local and thus insensitive 
to the global definedness of $\varsigma_\gamma$ and the above condition is compatible with the composition 
: we have only to interpret $\varsigma_\gamma$ appropriately. We obtain in this way a strict Lie 2--group
$\overline{\Gau}(N,G)=(\overline{\Gau}_1(N,G),\overline{\Gau}_2(N,G))$ \pagebreak extending $\Gau(N,G)$.
$\Gau_1(N,G)$ is a normal subgroup of $\overline{\Gau}_1(N,G)$
\footnote{$\vphantom{\dot{\dot{\dot{\dot{x}}}}}$ 
If $\beta$, $\gamma$ are homotopically non trivial 1--gauge transformations, then
$\varsigma_{\beta\diamond\gamma}-\varsigma_{\bar\beta\bar\gamma}=
\varsigma_\beta-\varsigma_{\bar\beta}+\varsigma_\gamma$ $-\varsigma_{\bar\gamma}-(\beta^{-1}d\beta,d\gamma\gamma^{-1})$
is globally defined. 
Hence, the restriction imposed on homotopically non trivial 1--gauge transformations
is compatible the composition prescription \eqref{spechgs12}. Similarly, it is also compatible with the 
inversion prescription \eqref{spechgs16}.}. 


The image through the cohomological winding number map $[w]$ of the ordinary gauge transformations
$\gamma\in\Map(N,G)$ coming from 1-gauge transformations $\gamma\in\overline{\Gau}_1(N,G)$ is a generally 
proper sublattice $\varLambda_k(N,G)$ of the winding number lattice 
$\varLambda(N,G)$. Because of the centrality of 
$k$ and \eqref{spechgs80}, a class $[w(\gamma)]\in\varLambda(N,G)$ with $\gamma\in\Map(N,G)$ 
arises from some $\gamma'\in\overline{\Gau}_1(N,G)$ if there is a map $\zeta\in\Map(N,Z(k))$ 
such that $\gamma'=\zeta^{-1}\gamma$ satisfies the condition \eqref{spechgs1}, where 
$Z(k)=\exp(\mathbb{R}k)$ is the $\UU(1)$ subgroup of $G$ generated by $k$. 
To see whether this is the case, we observe to begin with that 
$(\gamma^{-1}d\gamma,k)$ is a closed 1--form. Then, 
\begin{equation}
\zeta(p)=\exp\bigg(\int^p\frac{(\gamma^{-1}d\gamma,k)}{(k,k)}k\bigg)
\label{xspechgs1}
\end{equation}
is $Z(k)$--valued map. By construction $\gamma'=\zeta^{-1}\gamma$ satisfies 
the condition \eqref{spechgs1}. However, as $\zeta$ may be multivalued,
$\gamma'\not\in\Map(N,G)$ in general. $\gamma'\in\Map(N,G)$ only if 
the periods of the 1--form $(\gamma^{-1}d\gamma,k)$ are such 
to yield only elements of the lattice $\ker(\exp|_{\mathbb{R}k})$ in the exponential
\eqref{xspechgs1} when $p$ moves along a closed path, in particular if $(\gamma^{-1}d\gamma,k)$ 
is exact \footnote{$\vphantom{\dot{\dot{\dot{\dot{x}}}}}$ Requiring 
the vanishing of the cohomology $H^1(N,\mathbb{R})$ is going to essentially trivialize 
the 4--dimensional constructions presented in later subsections by implying
the vanishing of the cohomology $H^3(N,\mathbb{R})$ and with it of the whole
winding number lattice $\varLambda(N,G)$.}. Clearly, such a condition may not be verified in general.

\subsection{\normalsize \textcolor{blue}{The special 2--connections}}\label{subsec:specond}

\hspace{.5cm} 
Special 2--connections constitute the basic fields of special 2--gauge theory.
For this reason, the play a central role and \pagebreak 
it is necessary to study their properties in great detail. 

\vspace{2mm}

{\it Special 2--connections and their curvatures}

\vspace{2mm}
A {\it special $G$--2--connection} $\omega$ is simply a $\mathfrak{v}_k(\mathfrak{g})$--2--connection 
(cf. subsects. \ref{subsec:2tchern}, \ref{subsec:spechgs}). Therefore, 
it is a pair $(\omega,\varOmega_\omega)$ with
$\omega\in\Omega^1(N,\mathfrak{g})$, $\varOmega_\omega\in\Omega^2(N,\mathfrak{g})$.
The special 2--connections span a space which we shall denote by $\Conn_2(N,G)$.

The 2--curvature $f$ of a special $G$--2--connection $\omega\in \Conn_2(N,G)$ is the 2--curvature 
$f$ of $\omega$ as a $\mathfrak{v}_k(\mathfrak{g})$--2--connection. Therefore, $f$ 
is a pair $(f,F_f)$ with $f\in\Omega^2(N,\mathfrak{g})$, $F_f\in\Omega^3(N,\mathfrak{g})$.
By combining \eqref{fFcurv} and \eqref{spechgs55-58}, $f$, $F_f$ can be 
com\-puted explicitly. We obtain 
\begin{subequations}
\label{specond1-2}
\begin{align}
&f=d\omega+\frac{1}{2}[\omega,\omega],
\vphantom{\Big]}
\label{specond1}
\\
&F_f=d\varOmega_\omega+[\omega,\varOmega_\omega]-\frac{1}{6~}(\omega,[\omega,\omega])k
+\frac{1}{2}(\omega,k)[\omega,\omega].
\vphantom{\Big]}
\label{specond2}
\end{align}
\end{subequations}
Note that the 2--form curvature component $f$ is given in terms of the 1--form connection  
component $\omega$ by the familiar gauge theoretic expression. 

\vspace{2mm}

{\it Special 1--gauge transformation and connections}

\vspace{2mm}

The special $G$--1--gauge transformation group $\Gau_1(N,G)$ acts on the 
special $G$--2---connection space $\Conn_2(N,G)$.  
For a special 1--gauge transformation $\gamma\in\Gau_1(N,G)$ and a special 
2--connection $\omega\in \Conn_2(N,G)$, the gauge transformed 2--connection ${}^\gamma\omega$
can be computed inserting the relations \eqref{spechgs59-64} into the \eqref{7linfdglob}. 
We find in this way
\begin{subequations}
\label{specond3-4}
\begin{align}
&{}^\gamma\omega\equiv{}^{g_\gamma}\omega=\gamma\omega\gamma^{-1}-d\gamma\gamma^{-1},
\vphantom{\Big]}
\label{specond3}
\\
&{}^\gamma\varOmega_\omega\equiv{}^{g_\gamma}\varOmega_\omega
=\gamma(\varOmega_\omega-D_\omega(\chi_\gamma+\alpha_\gamma(\omega-\gamma^{-1}d\gamma))+\alpha_\gamma(f))\gamma^{-1}
\vphantom{\Big]}
\label{specond4}
\\
&\hspace{5.6cm}-(\varsigma_\gamma-(\omega,\gamma^{-1}d\gamma))k-(\omega,k)d\gamma\gamma^{-1}.
\vphantom{\Big]}
\nonumber
\end{align}
\end{subequations}
From \eqref{spechgs59-64} and \eqref{8linfdglob}, we can compute similarly 
the gauge transformed 2--curvature ${}^\gamma f$. We find \pagebreak 
\begin{subequations}
\label{specond5-6}
\begin{align}
&{}^\gamma f\equiv{}^{g_\gamma}f=\gamma f\gamma^{-1}, 
\vphantom{\Big]}
\label{specond5}
\\
&{}^\gamma F_f\equiv{}^{g_\gamma}F_f
=\gamma(F_f+[\chi_\gamma+\alpha_\gamma(\omega-\gamma^{-1}d\gamma),f]+D_\omega(\alpha_\gamma(f)))\gamma^{-1}
\vphantom{\Big]}
\label{specond6}
\\
&\hspace{6.9cm}+(f,\gamma^{-1}d\gamma)k-(f,k)d\gamma\gamma^{-1}.
\vphantom{\Big]}
\nonumber
\end{align}
\end{subequations}
The $\Gau_1(N,G)$--action is left, as it is straightforward to verify. 

From \eqref{spechgs67-72} and \eqref{7inflinfdglob}, 
the action of an infinitesimal special $G$--1--gauge transformation $\theta\in\mathfrak{gau}_0(N,G)$ on a special 
$G$--2--connection $\omega\in \Conn_2(N,G)$ reads
\begin{subequations}
\label{specond7-8}
\begin{align}
&\delta_\theta\omega\equiv\delta_{u_\theta}\omega=-D_\omega\theta,
\vphantom{\Big]}
\label{specond7}
\\
&\delta_\theta\varOmega_\omega\equiv\delta_{u_\theta}\varOmega_\omega
=[\theta,\varOmega_\omega]-D_\omega(\dot\chi_\theta+\dot\alpha_\theta(\omega))+\dot\alpha_\theta(f)
\vphantom{\Big]}
\label{specond8}
\\
&\hspace{5.6cm}-(\dot\varsigma_\theta-(\omega,d\theta))k-(\omega,k)d\theta.
\vphantom{\Big]}
\nonumber
\end{align}
\end{subequations}
By \eqref{spechgs67-72} and \eqref{7inflinfdglob}, the corresponding action on the 2--curvature $f$ of $\omega$ is  
\begin{subequations}
\label{specond9-10}
\begin{align}
&\delta_\theta f\equiv\delta_{u_\theta}f=[\theta,f],
\vphantom{\Big]}
\label{specond9}
\\
&\delta_\theta F_f\equiv\delta_{u_\theta}F_f
=[\theta,F_f]+[\dot\chi_\theta+\dot\alpha_\theta(\omega),f]+D_\omega(\dot\alpha_\theta(f))
\vphantom{\Big]}
\label{specond10}
\\
&\hspace{6.9cm}+(f,d\theta)k-(f,k)d\theta.
\vphantom{\Big]}
\nonumber
\end{align}
\end{subequations}  

From the above, its is apparent that the 1--form connection component $\omega$ is 
an ordinary gauge field of an ordinary gauge theory with gauge group $G$. 
Special 2--gauge theory can therefore be considered as an ordinary gauge theory with extra 
structure and symmetry.

\vspace{2.33mm}

{\it Action of homotopically non trivial special 
1--gauge transformations} 

\vspace{2.33mm}

The action \eqref{specond3-4} of the group
$\Gau_1(N,G)$ of homotopically trivial special 1--gauge transformations on 
special 2--connection space $\Conn_2(N,G)$ cannot be readily extended 
to one of the full group $\overline{\Gau}_1(N,G)$ of all 1--gauge 
transformations inclusive of the homotopically
non trivial ones (cf. subsect. \ref{subsec:spechgs}). For a transformation $\gamma$ 
of the latter type, \pagebreak $\varsigma_\gamma$ is not a globally defined
2--form, while the global definedness of the 2--form component the gauge transformed 2--connection 
${}^\gamma\omega$ of a special 2--connection $\omega\in\Conn_2(N,G)$  
requires by \eqref{specond4} that $\varsigma_\gamma$ is. This problem can 
be remedied by modifying \eqref{specond4} as follows. 


Although for a homotopically non trivial special 1--gauge 
transformation $\gamma\in\overline{\Gau}_1(N,G)$ the 2--form $\varsigma_\gamma
\in\Omega^2(\widehat{N})$ is not globally defined on $N$, 
the difference $\varsigma_\gamma-\varsigma_{\bar\gamma}\in\Omega^2(N)$ is, where the 2--form 
$\varsigma_{\bar\gamma}$ is defined by \eqref{xspechgs4}. We can thus modify
\eqref{specond4} by replacing $\varsigma_\gamma$  by $\varsigma_\gamma-\varsigma_{\bar\gamma}$ 
in the right hand side and so define a $\overline{\Gau}_1(N,G)$--action on special 
2--connection space $\Conn_2(N,G)$ compatible with the global definedness 
of the 2--form connection component. It can be checked that the $\overline{\Gau}_1(N,G)$--action, 
like the $\Gau_1(N,G)$ one, is left. 

\vspace{2.75mm}

{\it Special 2--gauge transformations as gauge for gauge symmetry}

\vspace{2.75mm}

If $\beta,\gamma\in\Gau_1(N,G)$ are 1--gauge transformations and 
$K\in\Gau_2(N,G)$ is a 2--gauge transformation with $K:\gamma\Rightarrow \beta$, 
then, for a given special 2--connection $\omega\in \Conn_2(N,G)$, 
${}^\beta\omega\not={}^\gamma\omega$ in general. In order to ${}^\beta\omega={}^\gamma\omega$,
the data $\varPhi_K$, $P_K$ characterizing $K$ (cf. subsect. \ref{subsec:spechgs}) must obey the condition 
\begin{equation}
D_\omega(P_K+\varPhi_K(\omega-\gamma^{-1}d\gamma))-\varPhi_K(f)=0,
\label{specond10/1}
\end{equation}
as follows immediately from \eqref{specond3-4} and \eqref{spechgs8-10}. 
When this requirement is fulfilled, $K$ represent a genuine gauge for gauge symmetry. 
Note that the condition depends on the connection's 1--form  component $\omega$.
Condition \eqref{specond10/1} has an infinitesimal counterpart. If $E\in\mathfrak{gau}_1(N,G)$ is an
infinitesimal 2--gauge transformation whose associated infinitesimal 1--gauge transformation
$\partial_\diamond E\in \mathfrak{gau}_0(N,G)$ (cf. eqs. \eqref{spechgs48}--\eqref{spechgs51})
is trivially acting, $\delta_{\partial_\diamond E}\omega=0$, $\delta_{\partial_\diamond E}\varOmega_\omega=0$, then 
the data $\dot\varPhi_E$, $\dot P_E$ characterizing $E$ (cf. subsect. \ref{subsec:spechgs}) 
must obey the condition 
\begin{equation}
D_\omega(\dot P_E+\dot \varPhi_E(\omega))-\dot \varPhi_E(f)=0.
\label{specond10/4}
\end{equation}
We have no geometric interpretation of either \eqref{specond10/1} or \eqref{specond10/4}.

Any special 1--gauge transformation $\gamma\in\Gau_1(N,G)$ such that $\gamma=\kappa$, $\varsigma_\gamma$ $=0$, 
$\alpha_\gamma=0$ and $\chi_\gamma=0$ with $\kappa\in\Map(N,Z(G))$ such that $\kappa^{-1}d\kappa=0$ 
acts trivially on the special 2--connection space $\Conn_2(N,G)$. Indeed, for such a $\gamma$, there is a 
2--gauge transformation $K\in\Gau_2(N,G)$ with $K:\iota\Rightarrow\gamma$ such that $K=\kappa$, $\varPhi_K=0$ 
and $P_K=0$ and thus satisfy\-ing \eqref{specond10/1}.
The 1--gauge transformations $\gamma$ of this form constitute a central subgroup 
$C(N,G)$ of $\Gau_1(N,G)$ isomorphic to the center $Z(G)$ of $G$. 
Hence, the $\Gau_1(N,G)$--action on $\Conn_2(N,G)$ has a kernel containing $C(N,G)$  
and so is not free. This failure of freeness can be at least partly remedied 
by replacing $\Gau_1(N,G)$ by the quotient 
\begin{equation}
\Gau_1{}^*(N,G)=\Gau_1(N,G)/C(N,G),
\label{specond10/2}
\end{equation}
which we shall call reduced 1--gauge transformation group.  As we shall see in sect. \ref{subsec:spe2sympl}, 
$\Gau_1{}^*(N,G)$ plays a basic role in the Hamiltonian analysis of the 1--gauge transformation action.

The Lie algebra $\mathfrak{c}(N,G)$ of $C(N,G)$ consists of those elements $\theta\in\mathfrak{gau}_0(N,G)$ 
such that $\theta=\epsilon$, $\dot\varsigma_\theta=0$, 
$\dot\alpha_\theta=0$ and $\dot\chi_\theta=0$ with $\epsilon\in\Map(N,\mathfrak{z}(\mathfrak{g}))$ such that 
$d\epsilon=0$. Since $C(N,G)$ is a central subgroup of $\Gau_1(N,G)$, $\mathfrak{c}(N,G)$ is a central 
Lie subalgebra of $\mathfrak{gau}_0(N,G)$. The quotient Lie algebra
\begin{equation}
\mathfrak{gau}_0{}^*(N,G)=\mathfrak{gau}_0(N,G)/\mathfrak{c}(N,G).
\label{specond10/3}
\end{equation}
is the Lie algebra of the group $\Gau_1{}^*(N,G)$, the reduced infinitesimal 1--gauge transformation 
Lie algebra. 


\subsection{\normalsize \textcolor{blue}{Special principal 2--bundles}}\label{subsec:spectwobndl}

\hspace{.5cm} 
So far, 
we have tacitly assumed that the geometrical background of our special version of higher gauge 
theory is a trivial principal 2--bundle and that, accordingly, the components of special 2--connections
as well as 1-- and 2--gauge transformations are fields globally defined on the base manifold. \pagebreak 
But we have not defined what 
a special principal 2--bundle and its 2--connections and 1-- and 2--gauge transformations 
actually are in full generality.
The analysis of the global properties of our geometric framework is interesting {\it per se} 
and may shed new light on it. 
Our discussion however will be kept to an elementary level
with no pretence of mathematical rigour. 
To make our approach more understandable, 
we shall emphasize analogies to and differences from ordinary gauge theory. 


In ordinary gauge theory, one assigns a gauge group $G$, a principal $G$--bundle $P$ over a base manifold 
$N$ and a connection $\omega$ of $P$, the gauge field. These are specified by local trivialization matching 
and connection trivializing data with respect to a sufficiently fine open covering $U_i$ of $N$. 
The trivialization matching data are gauge transformations 
$\gamma_{ij}\in\Gau(U_i\cap U_j,G):=\Map(U_i\cap U_j,G)$ with $U_i\cap U_j\not=\emptyset$
obeying the 1--cochain conditions $\gamma_{ii}=\iota$ and $\gamma_{ji}=\gamma_{ij}{}^{-1_\diamond}$ and 
the 1--cocycle condition $\gamma_{jk}\diamond \gamma_{ik}{}^{-1\diamond}\diamond \gamma_{ij}=\iota$ on 
$U_i\cap U_j\cap U_k\not=\emptyset$, where $\diamond$ denotes compositional struc\-ture of gauge transformations. 
The connection trivializing data are connections $\omega_i\in\Conn(U_i,G):=\Omega^1(U_i,\mathfrak{g})$.
As their name suggests, the $\gamma_{ij}$ match the trivializing data $\omega_i$, $\omega_j$ on 
$U_i\cap U_j\not=\emptyset$ via the gauge transformation $\gamma_{ij}$: $\omega_i={}^{\gamma_{ij}}\omega_j$. 
The 1--cocycle condition ensures the consistency of the matching by implying that 
${}^{\gamma_{ij}\,\diamond \,\gamma_{jk}}\omega_k={}^{\gamma_{ik}}\omega_k$ on $U_i\cap U_j\cap U_k\not=\emptyset$. 

In the special version of semistrict higher gauge theory studied in this paper, one should likewise
assign a gauge 2--group $V_k(G)$, a principal $V_k(G)$--2--bundle $Q$ over a base manifold $N$ 
and a special 2--connection $\omega\in \Conn_2(Q)$, the gauge field doublet, where 
$V_k(G)$ is some form of 2--group 
integrating the Lie 2--algebra $\mathfrak{v}_k(\mathfrak{g})$. 
The lack of a simple geometrical model for $Q$ and $\omega$ 
hinders our intuition, but, happily, much as in the ordinary 
case, these objects can be specified by local trivialization 
matching and matching compatibility and 2--connection trivializing data with respect 
to a suitably fine open covering $U_i$ of $N$. 

The trivialization matching data are special 
1--gauge transformations $\gamma_{ij}\in\Gau_1(U_i\cap U_j,G)$ with $U_i\cap U_j\not=\emptyset$.
They 
should obey the 1--cochain conditions \pagebreak 
\begin{subequations}
\label{specond13-14}
\begin{align}
&\gamma_{ii}=\iota,
\vphantom{\Big]}
\label{specond13}
\\
&\gamma_{ji}=\gamma_{ij}{}^{-1_\diamond},
\vphantom{\Big]}
\label{specond14}
\end{align}
\end{subequations}  
as in ordinary gauge theory, where $\diamond$ denotes the compositional structure of 1--gauge transformations 
defined by \eqref{spechgs11-22}. 
However, as is usual in higher gauge theory, the familiar 1--cocycle condition 
$\gamma_{jk}\diamond \gamma_{ik}{}^{-1\diamond}\diamond \gamma_{ij}=\iota$ 
does not hold strictly, but only up to a prescribed 2--gauge 
transformation depending on the underlying covering sets. 
The trivialization matching compatibility data are special 2--gauge transformations 
$K_{ijk}\in\Gau_2(U_i\cap U_j\cap U_k,G)$ with $U_i\cap U_j\cap U_k\not=\emptyset$ such that 
\begin{equation}
K_{ijk}:\gamma_{ij}\diamond \gamma_{jk}\Rightarrow \gamma_{ik}. 
\label{specond11/0}
\end{equation}
It is convenient to impose 2--cochain conditions on the matching compatibility data $K_{ijk}$ as well, namely
\hphantom{xxxxxxxxxxxxxxxxxx} 
\begin{subequations}
\label{specond17--21}
\begin{align}
&K_{iij}=K_{ijj}=I_{\gamma_{ij}}, 
\vphantom{\Big]}
\label{specond17}
\\
&K_{iji}=K_{jij}=I_\iota, 
\vphantom{\Big]}
\label{specond18}
\\
&K_{jik}=I_{\gamma_{ji}}\diamond K_{ijk}{}^{-1_\bullet},
\vphantom{\Big]}
\label{specond19}
\\
&K_{ikj}=K_{ijk}{}^{-1_\bullet}\diamond I_{\gamma_{kj}},
\vphantom{\Big]}
\label{specond20}
\\
&K_{kji}=I_{\gamma_{kj}\diamond \gamma_{ji}}\diamond K_{ijk}{}^{-1_\bullet} \diamond I_{\gamma_{ki}},
\vphantom{\Big]}
\label{specond21}
\end{align}
\end{subequations}  
where $\diamond$ and $\bullet$ denote here respectively the horizontal and vertical compositional 
structures of 2--gauge transformations defined by \eqref{spechgs23-37}. 
Equating as it is natural the two 2--gauge transformations 
$\gamma_{ij}\diamond \gamma_{jk}\diamond \gamma_{kl}\Rightarrow \gamma_{il}$ which
can be built by using $K_{jkl}$, $K_{ikl}$, $K_{ijl}$, $K_{ijk}$ yields the 2--cocycle condition 
\begin{equation}
K_{ikl}\bullet(K_{ijk}\diamond I_{\gamma_{kl}})=K_{ijl}\bullet(I_{\gamma_{ij}}\diamond K_{jkl})
\label{specond12}
\end{equation}
on $U_i\cap U_j\cap U_k\cap U_l\not=\emptyset$.
It is straightforward enough to check that cochain conditions \eqref{specond13-14} and 
\eqref{specond17--21} are mutually compatible and 
consistent with the compatibility condition \eqref{specond11/0} and the cocycle condition
\eqref{specond12} \footnote{$\vphantom{\dot{\dot{\dot{\dot{x}}}}}$ The cochain conditions \eqref{specond13-14},
\eqref{specond17--21} are not mandatory. The theory can be constructed without prescribing them at 
the price of a considerably higher level of complications. To relax \eqref{specond13} in particular,
one must introduce further matching compatibility data, namely 2--gauge transformations 
$\varLambda_i\in\Gau_2(U_i,G)$ such that $\varLambda_i:\gamma_{ii}\Rightarrow 1$.
Equating as reasonable the two 2--gauge transformations $\gamma_{ii}\diamond \gamma_{ij}\Rightarrow \gamma_{ij}$
which can be built using $\varLambda_i$ and $K_{iij}$ and similarly for 
$\gamma_{ij}\diamond \gamma_{jj}\Rightarrow \gamma_{ij}$, we get another set of conditions
namely $\varLambda_i\diamond I_{\gamma_{ij}}=K_{iij}$ and 
$I_{\gamma_{ij}}\diamond \varLambda_j=K_{ijj}$ on the intersections 
$U_i\cap U_j\not=\emptyset$.
Requiring \eqref{specond13-14}, \eqref{specond17--21} is thus useful
making the formal framework more manageable.}. 

The 2--connection trivializing data are special 2--connections $\omega_i\in \Conn_2(U_i,$ $G)$. 
On $U_i\cap U_j\not=\emptyset$, analogously to ordinary gauge theory, 
the data $\omega_i$, $\omega_j$ match 
via the 1--gauge transformation $\gamma_{ij}$, 
\begin{equation}
\omega_i={}^{\gamma_{ij}}\omega_j, 
\label{specond11}
\end{equation}
where the right hand side of the relation is given componentwise by \eqref{specond3-4}. 
The consistency of the matching relations \eqref{specond11} requires that 
\begin{equation}
{}^{\gamma_{ij}\,\diamond \,\gamma_{jk}}\omega_k={}^{\gamma_{ik}}\omega_k 
\label{specond15-16}
\end{equation}  
on $U_i\cap U_j\cap U_k\not=\emptyset$. 
There are two distinct ways of viewing these conditions. If \linebreak we consider the 2--connection trivialization data 
$\omega_i$ as given,  \eqref{specond15-16} is a restriction on the trivialization
matching compatibility data $K_{ijk}$, since \eqref{specond15-16} would obviously hold
if the $K_{ijk}$ were trivial and the cocycle condition $\gamma_{ij}\diamond \gamma_{jk}=\gamma_{ik}$ 
were true. If conversely we consider the matching compatibility data $K_{ijk}$ as given,
\eqref{specond15-16} is a restriction on the allowed 2--connection data $\omega_i$. 
Either way, 2--connection and matching compatibility data cannot be considered as independent from each other.

We shall call the kind of principal $V_k(G)$--2--bundle with 2--connection described by the trivialization 
matching and matching compatibility data $\gamma_{ij}$, $K_{ijk}$ and the 2--connection trivializing data 
$\omega_i$ obeying \pagebreak 
\eqref{specond13-14}, \eqref{specond11/0}, \eqref{specond17--21}, \eqref{specond12}, 
\eqref{specond11} and \eqref{specond15-16} a special principal $G$--2--bundle with 2--connection. 

\vspace{.33mm}
The cochain conditions \eqref{specond13-14}, \eqref{specond17--21}, the matching relation \eqref{specond11/0}
 and the cocycle condition \eqref{specond12} can be made fully explicit using 
the relations \eqref{spechgs11-22}, \eqref{spechgs23-37}. 
We write the trivialization matching and matching compatibility 
data $(\gamma_{ij},\varsigma_{ij},\alpha_{ij},\chi_{ij})$ and 
$(K_{ijk},\varPhi_{ijk},P_{ijk})$ as a shorthand for
$(\gamma_{ij},\varsigma_{\gamma_{ij}},\alpha_{\gamma_{ij}},\chi_{\gamma_{ij}})$ 
and $(K_{ijk},\varPhi_{K_{ijk}}$, $P_{K_{ijk}})$, respectively. 
The cochain conditions \eqref{specond13-14} read as 
\begin{subequations}
\label{specond22-24/1}
\begin{align}
&\gamma_{ji}=\gamma_{ij}{}^{-1},\qquad \gamma_{ii}=1,
\vphantom{\Big]}
\label{specond22}
\\
&\varsigma_{ji}+\varsigma_{ij}=0, 
\vphantom{\Big]}
\label{specond23}
\\
&\alpha_{ji}(\pi)+\gamma_{ij}\alpha_{ij}(\gamma_{ij}{}^{-1}\pi \gamma_{ij})\gamma_{ij}{}^{-1}=0,
\vphantom{\Big]}
\label{specond24}
\\
&\chi_{ji}+\gamma_{ij}\chi_{ij}\gamma_{ij}{}^{-1}-\gamma_{ij}\alpha_{ij}(\gamma_{ij}{}^{-1}d\gamma_{ij})\gamma_{ij}{}^{-1}=0,
\vphantom{\Big]}
\label{specond24/1}
\end{align}
\end{subequations}
while the cochain conditions \eqref{specond17--21} take explicitly the form 
\begin{subequations}
\label{specond29-31}
\begin{align}
&K_{jik}=K_{ikj}=K_{kji}=K_{ijk}{}^{-1},
\quad K_{iij}=K_{jii}=K_{iji}=1,
\vphantom{\Big]}
\label{specond29}
\\
&\varPhi_{jik}(\pi)+\varPhi_{ijk}(\pi)=0, 
\quad \varPhi_{ikj}(\pi)+\gamma_{jk}\varPhi_{ijk}(\gamma_{jk}{}^{-1}\pi \gamma_{jk})\gamma_{jk}{}^{-1}=0, 
\vphantom{\Big]}
\label{specond30}
\\
&\hspace{6cm}\varPhi_{kji}(\pi)+\gamma_{ki}{}^{-1}\varPhi_{ijk}(\gamma_{ki}\pi \gamma_{ki}{}^{-1})\gamma_{ki}=0, 
\vphantom{\Big]}
\nonumber
\\
&P_{jik}+P_{ijk}-\varPhi_{ijk}(\gamma_{ki}d\gamma_{ij}\gamma_{ij}{}^{-1}\gamma_{ki}{}^{-1})=0,
\quad P_{ikj}+\gamma_{jk}P_{ijk}\gamma_{jk}{}^{-1}=0, 
\vphantom{\Big]}
\label{specond31}
\\
&P_{kji}+\gamma_{ki}{}^{-1}P_{ijk}\gamma_{ki} 
\vphantom{\Big]}
\nonumber
\\
&\hspace{2.25cm}
+\gamma_{ki}{}^{-1}\varPhi_{ijk}(d\gamma_{ki}\gamma_{ki}{}^{-1}
+(\gamma_{ji}\gamma_{ik})^{-1}d(\gamma_{ji}\gamma_{ik})-\gamma_{jk}{}^{-1}d\gamma_{jk})\gamma_{ki}=0.
\vphantom{\Big]}
\nonumber
\end{align}
\end{subequations}
With these holding, the matching relation \eqref{specond11/0} reads explicitly as 
\begin{subequations}
\label{specond25-28}
\begin{align}
&\gamma_{jk}\gamma_{ik}{}^{-1}\gamma_{ij}=K_{ijk},
\vphantom{\Big]}
\label{specond25}
\\
&\varsigma_{jk}-\varsigma_{ik}+\varsigma_{ij}-(\gamma_{ij}{}^{-1}d\gamma_{ij},d\gamma_{jk}\gamma_{jk}{}^{-1})=0,
\vphantom{\Big]}
\label{specond26}
\\
&\alpha_{jk}(\pi)-\alpha_{ik}(\pi)+\gamma_{jk}{}^{-1}\alpha_{ij}(\gamma_{jk}\pi \gamma_{jk}{}^{-1})\gamma_{jk}=\varPhi_{ijk}(\pi),
\vphantom{\Big]}
\label{specond27}
\\
&\chi_{jk}-\chi_{ik}+\gamma_{jk}{}^{-1}\chi_{ij}\gamma_{jk}
+\alpha_{jk}(\gamma_{jk}{}^{-1}\gamma_{ij}{}^{-1}d\gamma_{ij}\gamma_{jk})=P_{ijk},
\vphantom{\Big]}
\vphantom{\ul{\ul{\ul{\ul{x}}}}}
\label{specond28}
\end{align}
\end{subequations}
while the cocycle condition \eqref{specond12} yields 
\begin{subequations}
\label{specond32-34}
\begin{align}
&K_{jkl}K_{ikl}{}^{-1}K_{ijl}K_{ijk}{}^{-1}=1,
\vphantom{\Big]}
\label{specond32}
\\
&\varPhi_{jkl}(\pi)-\varPhi_{ikl}(\pi)+\varPhi_{ijl}(\pi)
-\gamma_{kl}{}^{-1}\varPhi_{ijk}(\gamma_{kl}\pi \gamma_{kl}{}^{-1})\gamma_{kl}=0,
\vphantom{\Big]}
\label{specond33}
\\
&P_{jkl}-P_{ikl}+P_{ijl}-\gamma_{kl}{}^{-1}P_{ijk}\gamma_{kl}
+\varPhi_{jkl}(\gamma_{jl}{}^{-1}\gamma_{ij}{}^{-1}d\gamma_{ij}\gamma_{jl})=0. 
\vphantom{\Big]}
\label{specond34}
\end{align}
\end{subequations}
We recall here that the data $(\gamma_{ij},\varsigma_{ij},\alpha_{ij},\chi_{ij})$ 
and $(K_{ijk},\varPhi_{ijk},P_{ijk})$ obey by virtue of their definition the relations \eqref{spechgs1-3}
and \eqref{spechgs6-7}, respectively. 

Let $(\omega_i,\varOmega_i)$ be the 2--connection trivializing data in components.
By \eqref{specond3-4}, the matching relations \eqref{specond11} read as
\begin{subequations}
\label{specond3-4/1}
\begin{align}
&\omega_i=\gamma_{ij}\omega_j\gamma_{ij}{}^{-1}-d\gamma_{ij}\gamma_{ij}{}^{-1},
\vphantom{\Big]}
\label{specond3/1}
\\
&\varOmega_i
=\gamma_{ij}(\varOmega_j-D_{\omega_j}(\chi_{ij}
+\alpha_{ij}(\omega_j-\gamma_{ij}{}^{-1}d\gamma_{ij}))+\alpha_{ij}(f_j))\gamma_{ij}{}^{-1}
\vphantom{\Big]}
\label{specond4/1}
\\
&\hspace{5.6cm}-(\varsigma_{ij}-(\omega_j,\gamma_{ij}{}^{-1}d\gamma_{ij}))k-(\omega_j,k)d\gamma_{ij}\gamma_{ij}{}^{-1},
\vphantom{\Big]}
\nonumber
\end{align}
\end{subequations} 
where $f_j$ is given by \eqref{specond1}. 
The matching consistency condition \eqref{specond15-16} involves both the trivializing data $\omega_i$
and the trivialization matching data $\gamma_{ij}$, $\varPhi_{ijk}$, $P_{ijk}$,  
\begin{equation}
D_{\omega_k}(P_{ijk}+\varPhi_{ijk}(\omega_k-\gamma_{ik}{}^{-1}d\gamma_{ik}))-\varPhi_{ijk}(f_k)=0,
\label{specond35}
\end{equation}
in agreement with relation \eqref{specond10/1}.

In ordinary gauge theory, a globally defined gauge transformation $\eta$ 
is a fiber preserving automorphism of the theory's principal $G$--bundle
$P$ acting on the bundle's connection $\omega$ 
yielding a gauge transformed connection ${}^\eta\omega$. 
If $P$ and $\omega$ are described by the local trivialization matching and connection trivializing data 
$\gamma_{ij}$ and $\omega_i$ with respect to a fine open covering $U_i$ of the base manifold 
$N$, then $\eta$ is described by gauge transformation trivializing data $\eta_i\in \Gau(U_i,G)$ 
such that $\gamma_{ij}\diamond\eta_j=\eta_i\diamond\gamma_{ij}$
on $U_i\cap U_j\not=\emptyset$ and ${}^\eta\omega_i={}^{\eta_i}\omega_i$ on $U_i$.

In the special version of semistrict higher gauge theory we are studying, things 
are more complicated because of the lack of a geometrical model. \pagebreak 
We can characterize a globally defined 
1--gauge transformation $\eta$ as some kind of fiber preserving automorphism of the theory's
principal $V_k(G)$--2--bundle $Q$ acting on the bundle's special 2--connection 
$\omega\in \Conn_2(Q)$ only in terms of local data. 

\vspace{.31mm}
Let $U_i$ be some fine open covering of the base manifold $N$. Then, the 2--bundle $Q$ is specified 
by trivialization matching and matching compatibility data $\gamma_{ij}$
and $K_{ijk}$ obeying \eqref{specond13-14}, \eqref{specond11/0}, \eqref{specond17--21}, \eqref{specond12}.
Further, the allowed 2--connections $\omega$ are described by 2--connection trivializing 
data $\omega_i$ obeying \eqref{specond11} and \eqref{specond15-16}. 

\vspace{.31mm}
The 1--gauge transformation $\eta$ consists of 1--gauge transformation trivializing and 
trivialization matching compatibility data. 
The 1--gauge transformation trivializing data are special 1--gauge transformations $\eta_i\in\Gau_1$ $(U_i,G)$. 
However, in higher gauge theory, the familiar gauge transformation trivialization matching rule 
$\gamma_{ij}\diamond \eta_j=\eta_i\diamond \gamma_{ij}$ does not hold strictly, but only up to a prescribed 2--gauge 
transformation depending on the underlying covering sets. 
The 1--gauge transformation trivialization matching compatibility data are special 2--gauge 
transformations $\varTheta_{ij}\in\Gau_2(U_i\cap U_j,G)$ with $U_i\cap U_j\not=\emptyset$ such that 
\begin{equation}
\varTheta_{ij}:\gamma_{ij}\diamond \eta_j\Rightarrow \eta_i\diamond \gamma_{ij}.
\label{specond36/0}
\end{equation}
These should obey 1--cochain conditions analogous to \eqref{specond17--21}, 
\begin{subequations}
\label{specond38-39}
\begin{align}
&\varTheta_{ii}=I_{\eta_i},
\vphantom{\Big]}
\label{specond38}
\\
&\varTheta_{ji}=I_{\gamma_{ji}}\diamond \varTheta_{ij}{}^{-1_\bullet}\diamond I_{\gamma_{ji}}.
\vphantom{\Big]}
\label{specond39}
\end{align}
\end{subequations}
Equating the two 2--gauge transformations 
$\gamma_{ij}\diamond \gamma_{jk}\diamond\eta_k\Rightarrow \eta_i\diamond \gamma_{ik}$
which can be built using $K_{ijk}$, $\varTheta_{jk}$, $\varTheta_{ik}$, $\varTheta_{ij}$ 
yields the 1--coycle condition 
\begin{equation}
(I_{\eta_i}\diamond K_{ijk})\bullet(\varTheta_{ij}\diamond I_{\gamma_{jk}})\bullet(I_{\gamma_{ij}}\diamond \varTheta_{jk})
=\varTheta_{ik}\bullet (K_{ijk}\diamond I_{\eta_k})
\label{specond37}
\end{equation}
on $U_i\cap U_j\cap U_k\not=\emptyset$.
It is easy to verify that the cochain conditions \eqref{specond38-39}
are consistent with \eqref{specond37}. 
\vspace{1mm}\eject

The 1--gauge transformed special $\vphantom{\dot{\dot{\dot{\dot{x}}}}}$
2--connection ${}^\eta\omega$ is then described by the 
trivializing data given by \hphantom{xxxxxxxxxxxxxxxxxxxxxxxx}
\begin{equation}
{}^\eta\omega_i={}^{\eta_i}\omega_i 
\label{specond36}
\end{equation}
on each $U_i$.  The compatibility of the matching relations 
\eqref{specond11} and the 1--gauge transformation relations \eqref{specond36}
requires that 
\begin{equation}
{}^{\gamma_{ij}\,\diamond \,\eta_j}\omega_j={}^{\eta_i\,\diamond\,\gamma_{ij}}\omega_j
\label{specond40/0}
\end{equation}
on $U_i\cap U_j\not=\emptyset$. 
If we consider the trivialization matching and the 
2--connection trivialization data $\gamma_{ij}$ and $\omega_i$ as given, as we do, then
\eqref{specond40/0} is a restriction on the 1--gauge transformation trivialization matching compatibility 
data $\varTheta_{ij}$, since \eqref{specond40/0} would obviously hold
if the $\varTheta_{ij}$ were trivial and the gauge transform\-ation trivialization matching rule 
$\gamma_{ij}\diamond \eta_j=\eta_i\diamond \gamma_{ij}$ were true. 
Thus, the gauge matching compatibility data  depend on the 
matching and 2--connection data. 

\vspace{.33mm}
We shall call the globally defined 1--gauge transformation described by the 1--gauge transformation trivializing 
and trivialization matching compatibility data $\eta_i$ and $\varTheta_{ij}$ 
satisfying \eqref{specond36/0}, \eqref{specond38-39}, \eqref{specond37}, \eqref{specond36}
and \eqref{specond40/0} a special $G$--1--gauge transformation. 

\vspace{.33mm}
As above, the compatibility relation \eqref{specond36/0}, the cochain conditions \eqref{specond38-39} and the cocycle 
condition \eqref{specond37} can be spelled out rather explicitly using 
the relations \eqref{spechgs11-22}, \eqref{spechgs23-37}. 
We use the shorthands $(\eta_i,\varpi_i,\beta_i,\lambda_i)$ and $(\varTheta_{ij}, \varPsi_{ij},M_{ij})$
for $(\eta_i,\varsigma_{\eta_i},\alpha_{\eta_i},\chi_{\eta_i})$ and $(\varTheta_{ij},\varPhi_{\varTheta_{ij}},P_{\varTheta_{ij}})$,
respectively. The cochain conditions \eqref{specond38-39} take the form
\begin{subequations}
\label{specond40-42}
\begin{align}
&\varTheta_{ji}=\varTheta_{ij}{}^{-1},\qquad \varTheta_{ii}=1,
\vphantom{\Big]}
\label{specond40}
\\
&\varPsi_{ji}(\pi)+\gamma_{ij}\varPsi_{ij}(\gamma_{ij}{}^{-1}\pi\gamma_{ij})\gamma_{ij}{}^{-1}=0,
\vphantom{\Big]}
\label{specond41}
\\
&M_{ji}+\gamma_{ij}M_{ij}\gamma_{ij}{}^{-1}
-\gamma_{ij}\varPsi_{ij}(\gamma_{ij}{}^{-1}\eta_i{}^{-1}d\gamma_{ij}\gamma_{ij}{}^{-1}\eta_i
\gamma_{ij})\gamma_{ij}{}^{-1}=0. 
\vphantom{\Big]}
\label{specond42}
\end{align}
\end{subequations} 
The cocycle condition \eqref{specond37} becomes
\begin{subequations}
\label{specond47-49}
\begin{align}
&\varTheta_{jk}\varTheta_{ik}{}^{-1}\varTheta_{ij}=1,
\vphantom{\Big]}
\label{specond47}
\\
&\varPsi_{jk}(\pi)-\varPsi_{ik}(\pi)+\gamma_{jk}{}^{-1}\varPsi_{ij}(\gamma_{jk}\pi\gamma_{jk}{}^{-1})\gamma_{jk}
\vphantom{\Big]}
\label{specond48}
\\
&\hspace{5.6cm}
+\varPhi_{ijk}(\pi)-\eta_k{}^{-1}\varPhi_{ijk}(\eta_k\pi\eta_k{}^{-1})\eta_k=0,
\vphantom{\Big]}
\nonumber
\\
&M_{jk}-M_{ik}+\gamma_{jk}{}^{-1}M_{ij}\gamma_{jk}
+\varPsi_{jk}(\eta_k{}^{-1}\gamma_{jk}{}^{-1}\gamma_{ij}{}^{-1}d\gamma_{ij}\gamma_{jk}\eta_k)
\vphantom{\Big]}
\label{specond49}
\\
&\hspace{3.75cm}
+P_{ijk}-\eta_k{}^{-1}P_{ijk}\eta_k
+\varPhi_{ijk}(\gamma_{ik}{}^{-1}\eta_i{}^{-1}d\eta_i\gamma_{ik})=0.
\vphantom{\Big]}
\nonumber
\end{align}
\end{subequations}
The compatibility relation  \eqref{specond36/0} reads explicitly as 
\begin{subequations}
\label{specond43-46}
\begin{align}
&\gamma_{ij}\eta_j\gamma_{ij}{}^{-1}\eta_i{}^{-1}=\varTheta_{ij}, \hspace{7.75cm}
\vphantom{\Big]}
\label{specond43}
\\
&\varpi_j-\varpi_i+(\eta_i{}^{-1}d\eta_i,d\gamma_{ij}\gamma_{ij}{}^{-1})-(\gamma_{ij}{}^{-1}d\gamma_{ij},d\eta_j\eta_j{}^{-1})=0,
\vphantom{\Big]}
\label{specond44}
\\
&\beta_j(\pi)-\gamma_{ij}{}^{-1}\beta_i(\gamma_{ij}\pi \gamma_{ij}{}^{-1})\gamma_{ij}-\alpha_{ij}(\pi)
+\eta_j{}^{-1}\alpha_{ij}(\eta_j\pi\eta_j{}^{-1})\eta_j=\varPsi_{ij}(\pi), 
\vphantom{\Big]}
\label{specond45}
\\
&\lambda_j-\gamma_{ij}{}^{-1}\lambda_i\gamma_{ij}-\chi_{ij}+\eta_j{}^{-1}\chi_{ij}\eta_j, 
\vphantom{\Big]}
\label{specond46}
\\
&\hspace{3.05cm}
+\beta_j(\eta_j{}^{-1}\gamma_{ij}{}^{-1}d\gamma_{ij}\eta_j)-\alpha_{ij}(\gamma_{ij}{}^{-1}\eta_i{}^{-1}d\eta_i\gamma_{ij})
=M_{ij}.
\vphantom{\Big]}
\nonumber
\end{align}
\end{subequations}

By \eqref{specond3-4}, the gauge transformation relation \eqref{specond36} reads as
\begin{subequations}
\label{specond3-4/2}
\begin{align}
&{}^\eta\omega_i=\eta_i\omega_i\eta_i{}^{-1}-d\eta_i\eta_i{}^{-1},
\vphantom{\Big]}
\label{specond3/2}
\\
&{}^\eta\varOmega_i
=\eta_i(\varOmega_i-D_{\omega_i}(\lambda_i+\beta_i(\omega_i-\eta_i{}^{-1}d\eta_i))+\beta_i(f_i))\eta_i{}^{-1}
\vphantom{\Big]}
\label{specond4/2}
\\
&\hspace{5.6cm}-(\varpi_i-(\omega_i,\eta_i{}^{-1}d\eta_i))k-(\omega_i,k)d\eta_i\eta_i{}^{-1}.
\vphantom{\Big]}
\nonumber
\end{align}
\end{subequations} 
The 1--gauge transformation consistency condition \eqref{specond40/0} involves the trivializing data $\omega_i$
and $\eta_i$ in addition to the trivialization matching data $\gamma_{ij}$, $\varPsi_{ij}$, $M_{ij}$,
\begin{equation}
D_{\omega_j}(M_{ij}+\varPsi_{ij}(\omega_j-(\gamma_{ij}\eta_j)^{-1}d(\gamma_{ij}\eta_j)))-\varPsi_{ij}(f_j)=0,
\vphantom{\ul{\ul{\ul{\ul{\ul{x}}}}}}
\label{specond50}
\end{equation}
again in accordance to relation \eqref{specond10/1}.

As is well--known, in higher gauge theory, in addition to globally defined 1--gauge transformations, one has also 
globally defined 2--gauge transformations.
A 2--gauge transformation $\varXi$ has two 1--gauge transformations $\eta$, $\eta'$ 
of a principal $V_k(G)$--2--bundle $Q$ with special 2--connection 
$\omega\in \Conn_2(Q)$ as its source and target, $\varXi:\eta\Rightarrow\eta'$. 

Let $U_i$ be a fine covering of the base manifold $N$. Then, 
$Q$ and $\omega$ are described by local trivialization
matching and matching compatibility data $\gamma_{ij}$ and $K_{ijk}$ and 2--connection trivializing 
data $\omega_i$. Further, $\eta$, $\eta'$ are specified 
by 1--gauge transformation trivializing data $\eta_i$, $\eta'{}_i$ and 
trivialization matching compatibility data $\varTheta_{ij}$, $\varTheta'{}_{ij}$.

The 2--gauge transformation $\varXi$ consists of  2--gauge transformation trivializing data. These are 
special 2--gauge transformations $\varXi_i\in\Gau_2(U_i,G)$ with 
\begin{equation}
\varXi_i:\eta_i\Rightarrow \eta'{}_i. 
\label{specond51}
\end{equation}
Compatibility with \eqref{specond36/0} suggests equating the two 2--gauge transformations 
$\gamma_{ij}\diamond\eta_j\Rightarrow \eta'{}_i\diamond\gamma_{ij}$ that can be built using the data 
$\gamma_{ij}$, $\eta_i$, $\varTheta_{ij}$, $\eta'{}_i$, $\varTheta'{}_{ij}$ and $\varXi_i$. This leads to
a cocycle condition, namely 
\begin{equation}
\varTheta'{}_{ij}\bullet(I_{\gamma_{ij}}\diamond \varXi_j)=(\varXi_i\diamond I_{\gamma_{ij}})\bullet \varTheta_{ij}. 
\label{specond52}
\end{equation}

The source and target 1--gauge transformations $\eta$, $\eta'$ of a 2--gauge transformation 
$\varXi:\eta\Rightarrow\eta'$ must have the same action on the 2--connection 
$\omega$ so as to encode gauge for gauge symmetry. This requires that  
\begin{equation}
{}^{\eta_i}\omega_i={}^{\eta'{}_i}\omega_i. 
\label{specond53}
\end{equation}
Since we assume the 2--connection trivializing data $\omega_i$ as well as 
the 1--gauge transformation trivializing data $\eta_i$, $\eta'{}_i$ 
as given,  \eqref{specond53} is a restriction on the 2--gauge transformation trivializing data $\varXi_i$.

We shall call the globally defined 2--gauge transformation described by the 2--gauge transformation trivializing 
data $\varXi_i$ satisfying \eqref{specond51}, \eqref{specond52} 
and \eqref{specond53} a special $G$--2--gauge transformation. 

The relation \eqref{specond51} and the cocycle  condition \eqref{specond52} can be made 
explicit using again the relations \eqref{spechgs11-22}, \eqref{spechgs23-37}. 
We use the shorthands $(\varXi_i, \varLambda_i,N_i)$ for $(\varXi_i,\varPhi_{\varXi_i},P_{\varXi_i})$.
The cocycle condition \eqref{specond52} reads so
\begin{subequations}
\label{specond58-60}
\begin{align}
&\varXi_j\varXi_i{}^{-1}\varTheta'{}_{ij}\varTheta_{ij}{}^{-1}=1,
\vphantom{\Big]}
\label{specond58}
\\
&\varLambda_j(\pi)-\gamma_{ij}{}^{-1}\varLambda_i(\gamma_{ij}\pi\gamma_{ij}{}^{-1})\gamma_{ij}
+\varPsi'{}_{ij}(\pi)-\varPsi_{ij}(\pi)=0, 
\vphantom{\Big]}
\label{specond59}
\\
&N_j-\gamma_{ij}{}^{-1}N_i\gamma_{ij}+\varLambda_j(\eta_j{}^{-1}\gamma_{ij}{}^{-1}d\gamma_{ij}\eta_j)
+M'{}_{ij}-M_{ij}=0.
\vphantom{\Big]}
\label{specond60}
\end{align}
\end{subequations}
Property \eqref{specond51} implies further 
\begin{subequations}
\label{specond54-57}
\begin{align}
&\eta_i\eta'{}_i{}^{-1}=\varXi_i,
\vphantom{\Big]}
\label{specond54}
\\
&\varpi_i-\varpi'{}_i=0,
\vphantom{\Big]}
\label{specond55}
\\
&\beta_i(\pi)-\beta'{}_i(\pi)=\varLambda_i(\pi),
\vphantom{\Big]}
\label{specond56}
\\
&\lambda_i-\lambda'{}_i=N_i.
\vphantom{\Big]}
\label{specond57}
\end{align}
\end{subequations}

The requirement \eqref{specond53} takes the by now familiar form \eqref{specond10/1},
\begin{equation}
D_{\omega_i}(N_i+\varLambda_i(\omega_i-\eta_i{}^{-1}d\eta_i))-\varLambda_i(f_i)=0.
\label{specond61}
\end{equation}

For a principal $V_k(G)$--2--bundle $Q$, globally defined 1-- and 2-- gauge transformations
are the 1-- and 2--cells of an infinite dimensional strict Lie 2--group
$\Gau(N,Q)=(\Gau_1(N,Q),\Gau_2(N,Q))$. The definition of its 
operations and the study its properties is a laborious matter and 
will be tackled elsewhere.

\vspace{2.5mm}

{\it Independence from local data description}

\vspace{2.5mm}

In ordinary gauge theory, two sets of local trivialization matching data $\gamma_{ij}$, $\omega_i$
and $\tilde\gamma_{ij}$, $\tilde\omega_i$ describe the same principal $G$--bundle $P$
with connection if they are related by intertwining gauge transformation data $\epsilon_i$, 
so that $\gamma_{ij}\diamond\epsilon_j=\epsilon_i\diamond\tilde\gamma_{ij}$ 
and $\omega_i={}^{\epsilon_i}\tilde\omega_i$. 

In special gauge theory, \pagebreak 
however, the equivalent trivialization matching data relationship 
$\gamma_{ij}\diamond \epsilon_j=\epsilon_i\diamond \tilde\gamma_{ij}$ does not hold strictly, but only up to a 
prescribed 2--gauge transformation depending on the underlying covering sets. Therefore, we shall say that 
two sets of trivialization matching and matching compatibility and connection trivializing 
data $\gamma_{ij}$, $K_{ijk}$, $\omega_i$ 
and $\tilde\gamma_{ij}$, $\tilde K_{ijk}$, $\tilde\omega_i$ describe the same principal $V_k(G)$--2--bundle $Q$
with 2--connection, if there are intertwining special 1--gauge transformations $\epsilon_i\in\Gau_1(U_i,G)$ 
and 2--gauge transformations $T_{ij}\in\Gau_2(U_i\cap U_j,G)$ such that 
\begin{equation}
T_{ij}:\gamma_{ij}\diamond \epsilon_j\Rightarrow \epsilon_i\diamond \tilde\gamma_{ij}, 
\label{especond36/0}
\end{equation}
obeying 1--cochain conditions \hphantom{xxxxxxxxxxxxx}
\begin{subequations}
\label{especond38-39}
\begin{align}
&T_{ii}=I_{\epsilon_i},
\vphantom{\Big]}
\label{especond38}
\\
&T_{ji}=I_{\gamma_{ji}}\diamond T_{ij}{}^{-1_\bullet}\diamond I_{\tilde\gamma_{ji}}
\vphantom{\Big]}
\label{especond39}
\end{align}
\end{subequations}
and satisfying the further condition
\begin{equation}
(I_{\epsilon_i}\diamond \tilde K_{ijk})\bullet(T_{ij}\diamond I_{\tilde\gamma_{jk}})\bullet(I_{\gamma_{ij}}\diamond T_{jk})
=T_{ik}\bullet (K_{ijk}\diamond I_{\epsilon_k}).
\label{especond37}
\end{equation}
following from equating the two 2--gauge transformations 
$\gamma_{ij}\diamond \gamma_{jk}\diamond\epsilon_k\Rightarrow \epsilon_i\diamond \tilde\gamma_{ik}$
which can be built using $K_{ijk}$, $\tilde K_{ijk}$, $T_{jk}$, $T_{ij}$, $T_{ij}$
and, furthermore, the local 2--connection data $\omega_i$, $\tilde\omega_i$ 
are related as 
\begin{equation}
\omega_i={}^{\epsilon_i}\tilde\omega_i 
\label{especond36}
\end{equation}
and the compatibility condition 
\begin{equation}
{}^{\gamma_{ij}}\omega_j={}^{\epsilon_i\,\diamond\,\tilde\gamma_{ij}}\tilde\omega_j
\label{especond40/0}
\end{equation}
required by the consistency with the matching relation \eqref{specond11} holds. 
Note that \eqref{especond40/0} is a 
restriction on the data $T_{ij}$, since \eqref{especond40/0} would obviously hold 
if the $T_{ij}$ were trivial and the customary rule 
$\gamma_{ij}\diamond \epsilon_j=\epsilon_i\diamond \tilde\gamma_{ij}$ were true. 

\vspace{.33mm}
Relations \eqref{especond36/0}--\eqref{especond37} \pagebreak 
and \eqref{especond36}, \eqref{especond40/0}
can be made fully explicit by expressing them in terms of the components
of $\epsilon_i$ and $T_{ij}$ as we did earlier getting relations formally analogous to 
\eqref{specond22-24/1}--\eqref{specond32-34} and \eqref{specond3-4/1}--\eqref{specond35}.
We leave this straightforward though tedious task to the reader, but we mention 
for reference the following relations
\begin{subequations}
\label{especond41--43}
\begin{align}
&T_{ji}=T_{ij}{}^{-1},\quad T_{ii}=1,
\vphantom{\Big]}
\label{especond41}
\\
&K_{ijk}=T_{jk}T_{ik}{}^{-1}T_{ij}\tilde K_{ijk},
\vphantom{\Big]}
\label{especond42}
\\
&\gamma_{ij}\epsilon_j\tilde\gamma_{ij}{}^{-1}\epsilon_i{}^{-1}=T_{ij},
\vphantom{\Big]}
\label{especond43}
\end{align}
\end{subequations}
where here $\gamma_{ij}$, $K_{ijk}$, $\tilde \gamma_{ij}$, $\tilde K_{ijk}$, 
$\epsilon_i$ and $T_{ij}$ denote the first component of the corresponding
1-- or 2--gauge transformations. 

\vspace{.4mm}
In ordinary gauge theory, when two sets of local data $\gamma_{ij}$, $\omega_i$
and $\tilde\gamma_{ij}$, $\tilde\omega_i$ describe the same principal $G$--bundle $P$ 
with connection and are thus related through intertwining gauge transformation data $\epsilon_i$, 
gauge transformation trivializing data $\eta_i$, $\tilde\eta_i$ are equivalent 
if $\eta_i\diamond \epsilon_i= \epsilon_i\diamond\tilde\eta_i$. Something similar holds in the higher case. 

\vspace{.4mm}
In special gauge theory, however, the gauge transformation trivializing data relationship 
$\eta_i\diamond \epsilon_i=\epsilon_i\diamond\tilde\eta_i$ does not hold strictly, but, again, 
only up to a 2--gauge transformation depending on the underlying covering sets. 
Given two sets of trivialization matching and matching compatibility and connection trivializing 
data $\gamma_{ij}$, $K_{ijk}$, $\omega_i$ and $\tilde\gamma_{ij}$, $\tilde K_{ijk}$, $\tilde\omega_i$  
related by intertwining data $\epsilon_i$, $T_{ij}$ describing the same $V_k(G)$--2--bundle 
$Q$ with 2--connection, we shall say that 
two 1--gauge transformation trivializing and trivialization matching compatibility data 
$\eta_i$, $\varTheta_{ij}$ and $\tilde\eta_i$, $\tilde\varTheta_{ij}$ 
relative to $\gamma_{ij}$, $K_{ijk}$, $\omega_i$ and $\tilde\gamma_{ij}$, $\tilde K_{ijk}$, $\tilde\omega_i$  
respectively describe the same 1--gauge transformation $\eta$ if there are 
intertwining special 2--gauge transformations $X_i\in\Gau_2(U_i,G)$ 
such that 
\begin{equation}
X_i:\eta_i\diamond \epsilon_i\Rightarrow \epsilon_i\diamond\tilde\eta_i
\label{especond51}
\end{equation}
and satisfying the condition \pagebreak 
\begin{equation}
(I_{\epsilon_i}\diamond \tilde \varTheta_{ij})\bullet (T_{ij}\diamond I_{\tilde\eta_j})\bullet(I_{\gamma_{ij}}\diamond X_j)
=(X_i\diamond I_{\tilde\gamma_{ij}})\bullet(I_{\eta_i}\diamond T_{ij})\bullet(\varTheta_{ij}\diamond I_{\epsilon_j})
\label{especond52}
\end{equation}
yielded by equating the two 2--gauge transformations 
$\gamma_{ij}\diamond\eta_j\diamond\epsilon_j\Rightarrow \epsilon_i\diamond\tilde\eta_i\diamond\tilde\gamma_{ij}$ 
that can be built using the data $\gamma_{ij}$, $\eta_i$, $\varTheta_{ij}$, $\tilde\eta_i$, $\tilde\varTheta_{ij}$, 
$\epsilon_i$ and $\varXi_i$ and, moreover,
\begin{equation}
{}^{\eta_i}\omega_i={}^{\epsilon_i\tilde\eta_i}\tilde\omega_i
\label{especond52/1}
\end{equation}
required by the consistency with the 1--gauge transformation action \eqref{specond36}.

Relations \eqref{especond51}, \eqref{especond52}  and \eqref{especond52/1} can also be made fully 
explicit by writing them in terms of the components of $X_i$. 
We shall limit ourselves to furnish for their relevance the following relations, 
\begin{subequations}
\label{especond53-54}
\begin{align}
&\varTheta_{ij}=X_jX_i{}^{-1}\tilde \varTheta_{ij}, 
\vphantom{\Big]}
\label{especond53}
\\&\eta_i\epsilon_i\tilde\eta_i{}^{-1}\epsilon_i{}^{-1}=X_i,
\vphantom{\Big]}
\label{especond54}
\end{align}
\end{subequations}
where $\eta_i$, $\varTheta_{ij}$, $\tilde \eta_i$, $\tilde\varTheta_{ij}$, 
$\epsilon_i$ and $X_i$ denote the first component of the corresponding 1-- or 2--gauge 
transformations. 

Principal $V_k(G)$--2--bundles with 2--connection and 1--gauge transformations 
thereof are characterized by cohomology classes with values in the lattice 
\begin{equation}
\mathfrak{l}(\mathfrak{g})=\ker\big(\exp\big|_{\mathfrak{z}(\mathfrak{g})}\big)
\subset \mathfrak{z}(\mathfrak{g})
\label{cspecond1}
\end{equation}
and the sheaves $\underline{Z(G)}$, $\underline{G/Z(G)}$ of smooth $Z(G)$, $G/Z(G)$-valued functions
\footnote{$\vphantom{\dot{\dot{\dot{\dot{x}}}}}$ We recall a few basic results of cohomology. 
By the definition of $\mathfrak{l}(\mathfrak{g})$, we have an exact sequence of Abelian groups 
\begin{equation}
\xymatrix{0\ar[r]&\mathfrak{l}(\mathfrak{g})\ar[r]^{\iota\,\,}
&\mathfrak{z}(\mathfrak{g})\ar[r]^{\exp\,\,\,}&Z(G)\ar[r]&1}.
\label{cspecond2}
\end{equation}
The exact cohomology sequence associated to this breaks down in segments taking the form 
\begin{align}
&\xymatrix{0\ar[r]&H^n(N,\mathfrak{z}(\mathfrak{g}))/H^n(N,\mathfrak{l}(\mathfrak{g}))
&}
\label{cspecond3}
\\ 
&\hspace{2cm}\xymatrix{&\ar[r]^{\iota\hspace{1.1cm}}
&H^n(N,Z(G))\ar[r]^{\beta\hspace{.4cm}}&\Tor H^{n+1}(N,\mathfrak{l}(\mathfrak{g}))\ar[r]&0},
\nonumber
\end{align}
where $\beta$ 
is the Bockstein morphism and $\Tor A$ is the torsion subgroup of an
Abelian group $A$. We have a further $\vphantom{\dot{\dot{\dot{\dot{x}}}}}$  exact sequence, viz
\begin{equation}
\xymatrix{0\ar[r]&\mathfrak{l}(\mathfrak{g})\ar[r]^{\iota\,\,}
&\underline{\mathfrak{z}(\mathfrak{g})}\ar[r]^{\exp\,\,}&\underline{Z(G)}\ar[r]&1},
\label{cspecond4}
\end{equation}
where $\underline{L}$ denotes the sheaf of smooth functions valued in a Lie group $L$. 
Since $\underline{\mathfrak{z}(\mathfrak{g})}$ is a fine sheaf, the associated 
exact cohomology sequence reduces to the segments 
\begin{equation}
\xymatrix{0\ar[r]&H^n(N,\underline{Z(G)})\ar[r]^{\underline{\beta}\,\,}_{\simeq}
&H^{n+1}(N,\mathfrak{l}(\mathfrak{g}))\ar[r]&0},
\label{cspecond5}
\end{equation}
where $\underline{\beta}$ is again the Bockstein morphism. 
The sequences \eqref{cspecond3} and \eqref{cspecond5} are consistent in the sense that the diagram 
\begin{equation}
\xymatrix@C=2pc@R=.3pc{H^n(N,Z(G))\ar[rd]^{\beta}\ar[dd]_{\iota}&\cr
&\Tor H^{n+1}(N,\mathfrak{l}(\mathfrak{g}))\cr
H^n(N,\underline{Z(G)})\ar[ru]_{\underline{\beta}}&}
\label{cspecond6}
\end{equation}
commutes.}.

Consider a principal $V_k(G)$--2--bundle with connection $Q$ described by 
the trivialization matching and matching compatibility and connection trivializing 
data $\gamma_{ij}$, $K_{ijk}$, $\omega_i$. Relations \eqref{specond29}, \eqref{specond32} show  that
the matching compatibility data $K_{ijk}$ consitute a flat  $Z(G)$--valued \v Cech 2--cocycle. 
By \eqref{especond41}, \eqref{especond42}, this cocycle is defined up to a
flat  $Z(G)$--valued \v Cech 2--coboundary. 
Thus, with $Q$ there is associated a class $K_Q\in H^2(N,\underline{Z(G)})$ whose image in
$H^3(N,\mathfrak{l}(\mathfrak{g}))$ is a torsion class defining a flat  $Z(G)$--gerbe $B_Q$ over $N$. 
Relations \eqref{specond22}, \eqref{specond25} indicate further that the  matching data $\gamma_{ij}$ 
constitute a  $G/Z(G)$--valued \v Cech 1--cocycle. By \eqref{especond43}, 
this cocycle is defined up to a  $G/Z(G)$--conjugation. 
Thus, with $Q$ there is associated a class $\gamma_Q\in H^1(N,\underline{G/Z(G)})$
defining a principal $G/Z(G)$--bundle $P_Q$ over $N$. By \eqref{specond3/1}, 
the $\mathfrak{g}/\mathfrak{z}(\mathfrak{g})$ projection of the 
trivializing data $\omega_i$ are those of a connection $\omega$ of $P_Q$. 
By \eqref{specond25}, the $Z(G)$--gerbe $B_Q$
encodes the obstruction to lifting $P_Q$ to a principal $G$--bundle $\widehat{P}_Q$:
the lift exists provided $B_Q$ is trivial as a smooth gerbe. 
More loosely, \eqref{specond22-24/1}, \eqref{specond29-31} and \eqref{specond32-34} 
state that the 
matching data $(\gamma_{ij},\varsigma_{ij},\alpha_{ij},\chi_{ij})$ and $(K_{ijk},\varPhi_{ijk},P_{ijk})$ 
constitute a kind of non Abelian differential \v Cech 1--cochain and 2--cocycle, respectively,
related according \eqref{specond25-28}. \eqref{specond35} is a requirement the
matching data must obey in order to consistently describe  
the juxtaposition of the 2--connection data $\omega_i, \varOmega_i$ of eq. \eqref{specond3-4/1}. 

Consider next a 1--gauge transformation $\eta$ of principal $V_k(G)$--2--bundle with connection $Q$ 
described by the trivializing and trivialization matching compatibility data 
$\eta_i$, $\varTheta_{ij}$.
Relations \eqref{specond40}, \eqref{specond47} show that the matching compatibility data 
$\varTheta_{ij}$ constitute a flat  $Z(G)$--valued \v Cech 1--cocycle. By \eqref{especond53}
this cocycle is defined up to a flat  $Z(G)$--valued \v Cech 1--coboundary. 
So, with $\eta$ there is associated a class $\varTheta_\eta\in H^1(N,\underline{Z(G)})$ whose image in
$H^2(N,\mathfrak{l}(\mathfrak{g}))$ is a torsion class defining a flat  principal $Z(G)$--bundle $T_\eta$ 
over $N$. Relations \eqref{specond43} indicate further that the data $\eta_i$ describe
an ordinary gauge transformation $\eta$ in the principal $G/Z(G)$--bundle $P_Q$.
\eqref{specond3/2} furnishes the gauge transform ${}^\eta\omega$ of the connection $\omega$
of $P_Q$ encoded by the trivializing data $\omega_i$. 
By \eqref{specond43}, when the $Z(G)$--gerbe $B_Q$ is smoothly trivial and the $G/Z(G)$--bundle 
$P_Q$ is thus liftable to a $G$--bundle $\widehat{P}_Q$, the $Z(G)$--bundle $T_\eta$ 
encodes the obstruction to lifting $\eta$ to a gauge transformation $\widehat{\eta}$ in $\widehat{P}_Q$:
the lift exists provided $T_\eta$ is trivial as a smooth bundle. 
Loosely, \eqref{specond40-42} and \eqref{specond47-49} state that the set of 
matching compatibility data $(\varTheta_{ij}, \varPsi_{ij},M_{ij})$ constitute 
a certain non Abelian differential \v Cech 1--cocycle.
\eqref{specond43-46} are the conditions which mast be obeyed by the date 
$(\eta_i,\varpi_i,\beta_i,\lambda_i)$ and matching data
$(\gamma_{ij},\varsigma_{ij},\alpha_{ij},\chi_{ij})$ for the compatibility
of 1--gauge transformation of the trivializing matching. 
\eqref{specond50} is a requirement the matching consistency data must obey in order to consistently describe  
the 2--connection gauge transform trivializing data $({}^\eta\omega_i,{}^\eta\varOmega_i)$ 
given in eq. \eqref{specond3-4/2}. 

The following important remark is in order. 
When the flat $Z(G)$--gerbe $B_Q$ of a principal $V(G)$--2--bundle $Q$
is trivial and the $G/Z(G)$--bundle $P_Q$ is liftable
to a $G$--bundle $\widehat{P}_Q$, the $G$--valued 1--cocycle $\widehat{\gamma}_{ij}$
describing this latter may fail to satisfy condition \eqref{spechgs1}. In such case,
we do not obtain a description of an equivalent $V(G)$--2--bundle $\widehat{Q}$
characterized by a trivial gerbe $B_{\widehat{Q}}$. 
Assuming this anyway, 
when the flat  principal $Z(G)$--bundle $T_\eta$ of a 1--gauge transformation 
$\eta$ is trivial and the gauge transformation $\eta$ of $P_Q$ is liftable to one $\widehat{\eta}$
of $\widehat{P}_Q$, the $G$--valued 0--cocycle $\widehat{\eta}_i$ may fail to satisfy 
condition \eqref{spechgs1} and thus cannot be seen as a description of an equivalent  
1--gauge transformation $\widehat{\eta}$ of $\widehat{Q}$ 
characterized by a trivial bundle $T_{\widehat{\eta}}$.
This indicates that our topological characterization of $V(G)$--2--bundles
and 1--gauge transformations thereof is presently incomplete.

\vfil\eject

\section{\normalsize \textcolor{blue}{Special Higher Chern--Simons theory}}\label{sec:skhighcs}

\hspace{.5cm} 
In this section, we investigate in depth special 2--Chern--Simons theory, 
a 4--dimensional higher Chern--Simons theory whose symmetry is codified in the special Lie 
2--algebra $\mathfrak{v}_k(\mathfrak{g})$ of a Lie group $G$ with a distinguished central element 
$k\in\mathfrak{z}(\mathfrak{g})$ and having the special gauge transformation 2--group $\Gau(N,G)$ 
as symmetry transformation 2-group (cf. sect. \ref{sec:specgau}). 
We then move to study the the symplectic space of special 2--connections and its reduction.
Finally, using functional integral quantization, we compute the partition function 
of the model. 



\subsection{\normalsize \textcolor{blue}{The special 2--Chern--Simons theory}}\label{subsec:spe2cs}

\hspace{.5cm} 
{\it Special 2--Chern--Simons theory} is a semistrict higher Chern--Simons theory 
having the special Lie 2--algebra $\mathfrak{v}_k(\mathfrak{g})$ as symmetry 
algebra. 
%
%
%
%
The geometric background of the theory is a trivial special principal $G$--2--bundle $Q$
on a closed 4--dimensional base manifold $N$. Special 2--connection $\omega\in \Conn_2(Q)$ are thus 
globally defined fields.
Using the definition \eqref{spechgs55-58} of the Lie 2--algebra $\mathfrak{v}_k(\mathfrak{g})$
in the general expression \eqref{2tchern1} of the 2--Chern--Simons action $\CS_2$, we obtain readily 
the special 2--Chern--Simons action,  
\begin{align}
&\CS_2(\omega)
=\kappa_2\int_N\Big[\Big(d\omega+\frac{1}{2}[\omega,\omega],\varOmega_\omega\Big)
-\frac{1}{6}(\omega,k)(\omega,[\omega,\omega])\Big].
\vphantom{\Big]}
\label{spec2cs1} 
\end{align}
Taking advantage from the richer structure of the Lie 2--algebra $\mathfrak{v}_k(\mathfrak{g})$, 
we shall add to this a coupling term to a background closed 3--form $H$
\begin{equation}
\varDelta\!\CS_2(\omega;H)=8\pi^2\kappa_2\int_N (\omega,k)H.
\label{spec2cs2}
\end{equation}
The resulting total action is therefore
\begin{align}
\overline{\CS}_2(\omega;H)&
=\kappa_2\int_N\Big\{\Big(d\omega+\frac{1}{2}[\omega,\omega],\varOmega_\omega\Big)
\vphantom{\Big]}
\label{spec2cs3}
\\
&\hspace{4cm}+(\omega,k)\Big[-\frac{1}{6}(\omega,[\omega,\omega])+8\pi^2H\Big]\Big\}.
\vphantom{\Big]}
\nonumber 
\end{align}
Because of the background $H$, the field equations are no longer given by the flatness conditions
\eqref{2tchern2}, but they take the more general form
\begin{subequations}
\label{spec2cs4-5}
\begin{align}
&f=0,
\vphantom{\Big]}
\label{spec2cs4}
\\
&F_f+8\pi^2Hk=0.
\vphantom{\Big]}
\label{spec2cs5}
\end{align}
\end{subequations}
The 2--curvature components $f$, $F_f$ are here given explicitly by \eqref{specond1-2}. 
It is readily verified using the Bianchi identity \eqref{FBianchi} that the closedness of the 3--form $H$ 
is a necessary condition for the integrability of these equations.

Projecting the field equations on the line of $\mathfrak{g}$ spanned by $k$, we get
\begin{subequations}
\label{spec2cs6-7}
\begin{align}
&d(\omega,k)=0,
\vphantom{\Big]}
\label{spec2cs6}
\\
&d(\varOmega_\omega,k)+8\pi^2(k,k)(\cs_1+H)=0,
\vphantom{\Big]}
\label{spec2cs7}
\end{align}
\end{subequations}
where $\cs_1$ denotes 
the ordinary Chern--Simons form of the 1--form connection component $\omega$
\hphantom{xxxxxxxxxxxxxxxxxxxxxxxx}
\begin{equation}
\cs_1=\frac{1}{8\pi^2}\Big(\omega,d\omega+\frac{1}{3}[\omega,\omega]\Big).
\label{spec2cs8/-1}
\end{equation}
Thus, $(\omega,k)$ is a closed 1--form. Further, 
$\cs_1$ is a closed 3--form cohomologous to $-H$ via $(\varOmega_\omega,k)$. 

Combining  \eqref{spec2cs5} and \eqref{spec2cs7}, we find the following relation
\begin{equation}
\frac{1}{8\pi^2(k,k)}(F_f,k)=\cs_1 +\text{exact terms}. 
\label{spec2cs8/0}
\end{equation}
This relation furnishes an interesting 
interpretation of the de Rham cohomology class $[(F_f,k)/8\pi^2(k,k)]\in H^3(N,\mathbb{R})$.

The special 2--Chern--Simons action $\overline{\CS}_2(\omega;H)$ is invariant under homotopically trivial special
1--gauge transformations $\gamma\in\Gau_1(N,G)$, while $\overline{\CS}_2(\omega;H)$ 
transforms by a simple shift of the background 3--form $H$ under homotopically non trivial 
1--gauge transformations $\gamma\in\overline{\Gau}_1(N,G)$. We have indeed
\begin{equation}
\overline{\CS}_2({}^\gamma\omega;H)=\overline{\CS}_2(\omega;H+w(\bar\gamma)),
\label{spec2cs8}
\end{equation}
where the 1--gauge transformation $\bar\gamma$ is defined in subsect. \ref{subsec:specond} 
and $w(\bar\gamma)$ is the winding number density of $\bar\gamma$ 
given by \eqref{spechgs75}. 

The 2--Chern--Simons action $\overline{\CS}_2(\omega;H)$ has also Abelian 
on shell symmetries when the first field equation \eqref{spec2cs4} holds. These 
symmetries show up and play a distinguished role in the functional integral quantization of 
the model we shall carry out in subsect. \ref{subsec:funct2cs}. 

In the quantization of ordinary Chern--Simons theory, the invariance 
of the Boltzmann exponential $\exp(i\CS_1(\omega))$ under homotopically non trivial gauge 
transformations entails level quantization. Apparently, our special 2--Chern--Simons theory does not 
enjoy an analogous property. By \eqref{spec2cs3} and \eqref{spec2cs8}, full invariance of the 
exponential $\exp(i\CS_2(\omega))$ under homotopically non trivial 1--gauge transformations 
does not obtain unless there is some mechanism that localizes the 1--form 2--connection component 
$(\omega,k)$ on a lattice of integer period 1-forms. It is not immediately clear what such mechanism
might be, if indeed it does exist at all. 
For this reason, 
we shall thus treat $\kappa_2$ as a continuous parameter.


%
%

\subsection{\normalsize \textcolor{blue}{The special 2--connection symplectic space 
and its reduction
}}\label{subsec:spe2sympl}

\hspace{.5cm} 
It is well--known that there is a intimate relationship between ordinary 3--dimensional Chern--Simons 
theory and 2--dimensional topological gauge theory and that the rich geometric structure of the latter 
is based among other things on the existence of a natural symplectic structure on connection space
with respect to which the gauge transformation action is Hamiltonian, with the curvature map acting as 
moment map. It is natural to wonder whether there is a similar relationship between our 4--dimensional 
special 2--Chern--Simons theory and some kind of special 3--dimensional topological gauge theory 
characterized by a symplectic structure on 2--connection space with respect to which the 1--gauge 
transformation action is Hamiltonian with moment map closely related to the 2--curvature map. 
The answer is affirmative and, though there still are open issues, it is already possible 
to delineate some features of the 3--dimensional theory.


Consider a special principal $G$--2--bundle $P$ on a closed 3--dimensional base manifold $M$. 
For simplicity, we assume $P$ to be trivial so that the components of its special 2--connections 
are globally defined forms. 

The object of our study will be the space $\Conn_2(M,G)$ of the special $G$--2--connections 
$\omega$. Unlike what one may naively expect based on the analogy to 
ordinary gauge theory, $\Conn_2(M,G)$ is {\it not} an affine space modelled on the vector space 
$T\Conn_2(M,G)=\Omega^1(M,\mathfrak{g})\oplus\Omega^2(M,\mathfrak{g})$ with a linear 
action of the spe\-cial 1--gauge transformation group $\Gau_1(M,G)$. 
In fact, by \eqref{specond3-4}, for a 1--gauge transformation 
$\gamma\in\Gau_1(M,G)$, the 2--form component ${}^\gamma\varOmega_\omega$ of the gauge transformed 
2--connection $({}^\gamma\omega,{}^\gamma\varOmega_\omega)$ of a 
2--connection $(\omega,\varOmega_\omega)$ depends quadratically 
on the 1--form component $\omega$ of the latter, 
so that the difference of two 2--connections $(\omega,\varOmega_\omega)$, 
$(\omega',\varOmega_\omega{}')$ cannot transform linearly
under $\gamma$. 

\vspace{2.4mm}

{\it The symplectic structure}

\vspace{2.4mm}

Just as the space of ordinary connections in two dimensions is symplectic, 
the 2--connection space $\Conn_2(M,G)$ carries a natural symplectic structure, 
\begin{equation}
\varSigma_{M,G}=\int_M(\delta\omega,\delta\varOmega_\omega). 
\label{specmu1}
\end{equation}
A Poisson bracket structure on $\Conn_2(M,G)$ is thus defined. This can be written compactly as follows.
For $a\in\Omega^p(M,\mathfrak{g})$, $A\in\Omega^{3-p}(M,\mathfrak{g})$ with $0\leq p\leq 3$, set
\begin{equation}
\langle a, A\rangle=\int_M(a,A).
\label{specmu2}
\end{equation}
Then, the Poisson brackets read \hphantom{xxxxxxxxx} 
\begin{equation}
\{\langle\omega,\varXi_\xi\rangle,\langle\xi,\varOmega_\omega\rangle\}=\langle\xi,\varXi_\xi\rangle
\label{specmu3}
\end{equation}
with $\xi\in\Omega^1(M,\mathfrak{g})$, $\varXi_\xi\in\Omega^2(M,\mathfrak{g})$.

Denoting by $V_\gamma:\Conn_2(M,G)\rightarrow\Conn_2(M,G)$ the action \eqref{specond3-4}
of a special 1--gauge transformation $\gamma\in\Gau_1(M,G)$ on $\Conn_2(M,G)$, 
we have
\begin{equation}
V_\gamma{}^*\varSigma_{M,G}=\varSigma_{M,G}.
\label{specmu4}
\end{equation}
$\varSigma_{M,G}$ is therefore gauge invariant. 

\vspace{2.4mm}

{\it Hamiltonianity of 1--gauge transformation and the moment map}

\vspace{2.4mm}

It is natural to wonder if the special 1--gauge transformation action $V_\cdot$ 
is Hamiltonian, that is if there exists an {\it equivariant moment map}
$\mu:\Conn_2(M,G)\rightarrow \mathfrak{gau}_0(M,G)^\vee$ generating the action at the infinitesimal level. 
Before attempting the construction of $\mu$, we have to recall that by convention
$\mu$ is supposed to generate infinitesimally the right counterpart $\bar V_\cdot$
of the left action $V_\cdot$, which is defined by $\bar V_\gamma=V_{\gamma^{-1_\diamond}}$
with $\gamma\in \Gau_1(M,G)$. 
Through a straightforward calculation  analogous to that yielding \eqref{specond7-8}, we find 
\begin{subequations}
\label{specmu5-6}
\begin{align}
&\bar\delta_\theta\omega=D_\omega\theta,
\vphantom{\Big]}
\label{specmu5}
\\
&\bar\delta_\theta\varOmega_\omega
=-[\theta,\varOmega_\omega]+D_\omega(\dot\chi_\theta+\dot\alpha_\theta(\omega))-\dot\alpha_\theta(f)
\vphantom{\Big]}
\label{specmu6}
\\
&\hspace{5cm}+(\dot\varsigma_\theta-(\omega,d\theta))k+(\omega,k)d\theta
\vphantom{\Big]}
\nonumber
\end{align}
\end{subequations}
with $\theta\in\mathfrak{gau}_0(M,G)$. 
The moment map $\mu$ should thus be the Ham\-iltonian function for the 
infinitesimal right special 1--gauge transformation action $\bar\delta$,
\begin{subequations}
\label{specmu7-8}
\begin{align}
&\{\mu(\theta),\langle\omega,\varXi_\xi\rangle\}=\langle \bar\delta_\theta\omega,\varXi_\xi\rangle,
\vphantom{\Big]}
\label{specmu7}
\\
&\{\mu(\theta),\langle\xi,\varOmega_\omega\rangle\}=\langle\xi,\bar\delta_\theta\varOmega_\omega\rangle,
\vphantom{\Big]}
\label{specmu8}
\end{align}
\end{subequations}
and be equivariant, that is 
\begin{equation}
\{\mu(\theta),\mu(\zeta)\}=\mu([\theta,\zeta]_\diamond)
\label{specmu9}
\end{equation}
for $\theta,\zeta\in\mathfrak{gau}_0(M,G)$, the Lie bracket $[\theta,\zeta]_\diamond$
being defined in \eqref{spechgs43}--\eqref{spechgs46}.$\vphantom{\ul{\ul{\ul{x}}}}$

A moment map $\mu:\Conn_2(M,G)\rightarrow \mathfrak{gau}_0(M,G)^\vee$ with such properties is 
\begin{align}
&\mu(\omega)(\theta)
=\int_M\Big[(F_f,\theta^\perp)+(f,\dot\chi_{\theta^\perp}+\dot\alpha_{\theta^\perp}(\omega)
\label{specmu10}
\\
&\hspace{5cm}+(\omega,\theta^\perp)k
-(\omega,k)\theta^\perp)+(\omega,k)\dot\varsigma_{\theta^\perp}\Big],
\vphantom{\Big]}
\nonumber
\end{align} 
where the 2--curvature components $f$, $F_f$ are given by \eqref{specond1}, \eqref{specond2} and 
the mapping ${}^\perp:\mathfrak{gau}_0(M,G)\rightarrow \mathfrak{gau}_0(M,G)$ is defined by 
\begin{subequations}
\label{specmu11-14}
\begin{align}
&\theta^\perp=\theta-\frac{(\theta,k)k}{(k,k)},
\vphantom{\Big]}
\label{specmu11}
\\
&\dot\varsigma_{\theta^\perp}=\dot\varsigma_\theta,
\vphantom{\Big]}
\label{specmu12}
\\
&\dot\alpha_{\theta^\perp}(\pi)=\dot\alpha_\theta(\pi),
\vphantom{\Big]}
\label{specmu13}
\\
&\dot\chi_{\theta^\perp}=\dot\chi_\theta.
\vphantom{\Big]}
\label{specmu14}
\end{align}
\end{subequations}
${}^\perp$ is in fact a Lie algebra morphism, as it is immediately verified from
\eqref{spechgs38-40} and \eqref{spechgs43}--\eqref{spechgs46}.
The key property of the morphism is that 
\begin{equation}
(\theta^\perp,k)=0.
\label{specmu15}
\end{equation}
This can be used to simplify the expression of $\mu(\theta)$ once the explicit expressions
\eqref{specond1-2}) of $f$, $F_f$ are plugged in into \eqref{specmu10}.

\vspace{2.4mm}

{\it Hamiltonian reduction} 

\vspace{2.4mm}

It is not possible to carry out the Hamiltonian reduction of the Hamiltonian symplectic $\Gau_1(M,G)$--manifold 
$(\Conn_2(M,G),\varSigma_{M,G})$ via the moment map $\mu$ directly. The reason for that is that 
the 1--gauge transformation group $\Gau_1(M,G)$
does not act freely on $\Conn_2(M,G)$. As was demonstrated  earlier in subsect. \ref{subsec:specond}, $\Gau_1(M,G)$
contains a finite dimensional central subgroup $C(M,G)\simeq Z(G)$ acting trivially on $\Conn_2(M,G)$. To carry 
out the reduction, it is therefore appropriate to mod out $C(M,G)$ by replacing $\Gau_1(M,G)$ with its reduced form 
$\Gau_1{}^*(M,G)=\Gau_1(M,G)/C(M,G)$ (cf. eq. \eqref{specond10/2}).
Upon doing so, \linebreak 
the Hamiltonian reduction can be consistently performed provided  the moment 
map $\mu:\Conn_2(M,G)\rightarrow\mathfrak{gau}_0(M,G)^\vee$ is projectable to a moment map 
$\mu^*:\Conn_2(M,G)$ $\rightarrow\mathfrak{gau}_0{}^*(M,G)^\vee$, where 
$\mathfrak{gau}_0{}^*(N,G)=\mathfrak{gau}_0(N,G)/\mathfrak{c}(N,G)$ is the reduced form of the 
Lie algebra $\mathfrak{gau}_0(M,G)$ (cf. eq. \eqref{specond10/3}). To this end, it is 
sufficient that $\mu(\theta)=0$ for all $\theta\in\mathfrak{c}(M,G)$, a property that can be
straightforwardly verified by inspecting \eqref{specmu10}. 

\vspace{.3mm}
The next step of the Hamiltonian reduction procedure is the study the conditions under which 
the subspace $\mu^{*-1}(0)$ of $\Conn_2(M,G)$ is a closed embedded submanifold. For this to be the case, it is 
sufficient that $0$ is a regular value of $\mu^*$, 
that is that the differential $\delta\mu^*(\omega):T_\omega\Conn_2(M,G)\rightarrow 
\mathfrak{gau}_0{}^*(M,G)^\vee$ is surjec\-tive for every $\omega\in \mu^{*-1}(0)$. 
A simple computations gives
\begin{equation}
\delta\mu^*(\omega)(\theta)=\int_M\Big[(\delta\omega,\bar\delta_\theta\varOmega_\omega)
-(\delta\varOmega_\omega,\bar\delta_\theta\omega)\big]
\label{specmu15/1}
\end{equation}
for arbitrary $\theta\in\mathfrak{gau}_0(M,G)$, 
where $\bar\delta_\theta\omega$, $\bar\delta_\theta\varOmega_\omega$ are the infinitesimal 1--gauge 
variations of $\omega$, $\varOmega_\omega$ given by \eqref{specmu5}, \eqref{specmu6}. From
\eqref{specmu15/1}, it appears that 
$\delta\mu^*(\omega)$ is surjective provided $\delta\mu^*(\omega)(\theta)=0$ only
for $\theta\in\mathfrak{c}(M,G)$, that is if the 2--connection $\omega$ is $\Gau_1{}^*(M,G)$--irreducible.
From now on, we assume that this is the case. If this requirement failed to be satisfied, 
the offending reducible 2--connections $\omega$ would have to be removed by hand from $\mu^{*-1}(0)$. 

\vspace{.25mm}
From \eqref{specmu10}, recalling the \eqref{spechgs38-40}, it is found that 
the vanishing locus $\mu^{*-1}(0)$ of $\mu^*$ in $\Conn_2(M,G)$ consists of the 2--connections 
$\omega$ 
obeying the equations
\begin{subequations}
\label{specmu16-18}
\begin{align}
&(\omega-dak,k)=0, 
\vphantom{\Big]}
\label{specmu16}
\\
&f=0,
\vphantom{\Big]}
\label{specmu17}
\\
&F_f+8\pi^2 Hk=0
\vphantom{\Big]}
\label{specmu18}
\end{align}
\end{subequations}
for some $a\in\Omega^1(M,\mathbb{R})$ and $H\in\Omega^3(M,\mathbb{R})$
Eq. \eqref{specmu18} implies that $H$ is closed as it is readily verified
using the Bianchi identities \eqref{fFcurv}. \pagebreak 
By the assumed absence $\Gau_1{}^*(M,G)$--reducible 2--connections $\omega\in\mu^{*-1}(0)$,
the quotient $\matheul{C}_{M,G}=\mu^{*-1}(0)//\Gau_1{}^*(M,G)$ is a smooth manifold endowed with a symplectic structure 
$\varSigma_{{\eul C}_{M,G}}$ induced by $\varSigma_{M,G}$, as established by the classic Weinstein--Marsden theorem.
The problem with what we have done is that we have applied results known to be valid in a finite dimensional
set--up to an essentially infinite dimensional problem. More work is required for a sounder understanding
of this issue.

\vspace{2.4mm}

{\it $H$--flat 2--connections and their moduli space}

\vspace{2.4mm}

Eqs. \eqref{specmu17}, \eqref{specmu18} coincide with \eqref{spec2cs4},
\eqref{spec2cs5}. \eqref{specmu17} implies that $(d\omega,k)=0$, 
as in \eqref{spec2cs6}.
Eq. \eqref{specmu16} is compatible with but in general stronger than this relation. 
If $H^1(M,\mathbb{R})=0$, eq. \eqref{specmu16} is subsumed 
by eq. \eqref{specmu17} and can thus be dropped. We assume this to be the case henceforth.

For $H\in\Omega^3(M,\mathbb{R})$ a closed 3--form, 
we shall call a special $G$--2--connection $\omega\in \Conn_2(M,G)$ whose 2--curvature $f$
satisfies \eqref{specmu17}, \eqref{specmu18} {\it $H$--flat} and we shall denote 
by $\mathsans{F}_H(M,G)$ the subspace of $\mu^{*-1}(0)$ of such  2--connections. 
$\mu^{*-1}(0)$ is clearly the union of the subspaces $\mathsans{F}_H(M,G)$ for all possible
closed 3--forms $H$. So, it may be useful to study $\mathsans{F}_H(M,G)$ for fixed $H$. 

\vspace{.25mm}
By \eqref{specond5-6}, the $H$--flat 2--connection space 
$\mathsans{F}_H(M,G)$ is invariant under the 
special 1--gauge transformation group
$\Gau_1(M,G)$ and so also under its reduced form $\Gau_1{}^*(M,G)$. 
The moduli space ${\eul F}_H(M,G)=\mathsans{F}_H(M,G)/\Gau_1{}^*(M,G)$ is hence defined.
Again, ${\eul F}_H(M,G)$ is not a well--behaved space because of the possible existence of 
reducible special 2-connections, fixed points of the $\Gau_1{}^*(M,G)$--action on $\mathsans{F}_H(M,G)$. 

$\mathsans{F}_H(M,G)$ depends only on the de Rham cohomology class $[H]\in H^3(M,\mathbb{R})$
of the closed 3--form $H$. Indeed, it is not difficult to check that if $B\in\Omega^2(M,\mathbb{R})$ is a 2--form,
$\mathsans{F}_{H+dB}(M,G)={}^{\gamma_B}\mathsans{F}_H(M,G)$, where $\gamma_B\in\Gau_1(M,G)$ is 
the 1--gau\-ge transformation specified by the data $\gamma_B=1$, $\varsigma_{\gamma_B}=-dB$,
$\alpha_{\gamma_B}(\pi)=0$ and $\chi_{\gamma_B}=0$. Since $\mathsans{F}_H(M,G)$ is 1--gauge invariant, 
$\mathsans{F}_{H+dB}(M,G)=\mathsans{F}_H(M,G)$. Consequently, also the moduli space ${\eul F}_H(M,G)$
depends only on the class $[H]$. So, we shall denote $\mathsans{F}_H(M,G)$ 
and ${\eul F}_H(M,G)$ by $\mathsans{F}_{[H]}(M,G)$ and ${\eul F}_{[H]}(M,G)$, respectively, 
if it is necessary to emphasize this property. 

\vspace{.4mm}
At least at the formal level, it is possible to describe the local geometry of ${\eul F}_H(M,G)$
as follows. We assume temporarily that $M$ has general dimension and revert to 3 dimensions later. 
Let $\omega\in \mathsans{F}_H(M,G)$ be a fixed $H$--flat special 
2--connection. We define a cochain complex $(\mathsans{C}_\omega , \delta_\omega )$
depending on $\omega$ as follows. The complex has just three non vanishing terms at degree
$0$, $1$, $2$, 
\begin{equation}
\xymatrix{0\ar[r]&\mathsans{C}_\omega {}^0\ar[r]^{\delta_\omega\,\,}
&\mathsans{C}_\omega {}^1\ar[r]^{\delta_\omega\,\,}&\mathsans{C}_\omega {}^2\ar[r]&0}.
\label{specmu25}
\end{equation}
$\mathsans{C}_\omega {}^0$ is simply the Lie algebra $\mathfrak{gau}_0(M,G)$. 
$\mathsans{C}_\omega {}^1$ is the tangent space $T_\omega \Conn_2(M,G)$ 
of $\Conn_2(M,G)$ at $\omega$. Finally, 
$\mathsans{C}_\omega {}^2$ is the tangent space $T_{0,-8\pi^2Hk}\Curv_{2\omega} (M,G)$
at $(0,-8\pi^2Hk)$ of $\Curv_{2\omega} (M,G)$, where $\Curv_{2\omega} (M,G)$
is the space of solutions $a$ of the 2--Bianchi relations, that is, explicitly,  
of the pairs $(a,A_a)$ with $a\in\Omega^2(M,\mathfrak{g})$, $A_a\in\Omega^3(M,\mathfrak{g})$
obeying 
\begin{subequations}
\label{specmu19-20}
\begin{align}
&D_\omega a=0,
\vphantom{\Big]}
\label{specmu19}
\\
&D_\omega A_a-[a,\varOmega_\omega]+\frac{1}{2}([\omega,\omega],a)k-(\omega,k)[\omega,a]
-\frac{1}{2}(a,k)[\omega,\omega]=0.
\vphantom{\Big]}
\label{specmu20}
\end{align}
\end{subequations}
If $\omega$ were a general 2--connection and $a=f$,  these would be by \eqref{spechgs55-58}
the Bian\-chi identities \eqref{fFBianchi} for the special Lie 2--algebra $\mathfrak{v}_k(\mathfrak{g})$. 
The coboundary $\delta_\omega $ opera\-tor acts as follows. For $\theta\in\mathsans{C}_\omega {}^0$,
$\delta_\omega \theta=(\bar\delta_\theta\omega,\bar\delta_\theta\varOmega_\omega)$ with $\bar\delta_\theta$ 
the infinitesimal special 1--gauge transformation operator \eqref{specmu5-6} here with $f=0$. Hence,  
\begin{subequations}
\label{specmu21-22}
\begin{align}
&\delta_\omega \theta_1=D_\omega\theta, 
\vphantom{\Big]}
\label{specmu21}
\\
&\delta_\omega \theta_2
=-[\theta,\varOmega_\omega]+D_\omega(\dot\chi_\theta+\dot\alpha_\theta(\omega))
+(\dot\varsigma_\theta-(\omega,d\theta))k+(\omega,k)d\theta,
\vphantom{\Big]}
\label{specmu22}
\end{align}
\end{subequations}
where in the left hand side the indices $1, 2,\ldots$ denote form degree. \pagebreak 
For $\beta\in \mathsans{C}_\omega {}^1$ with 1-- and 2--form components $\beta$, $\dot\varOmega_\beta$, 
$\delta_\omega \beta=(\delta_\beta f,\delta_\beta F_f)$
with $\delta_\beta$ denoting infinitesimal variation of $(\omega,\varOmega_\omega)$ 
of the amount $(\beta,\dot\varOmega_\beta)$, 
\begin{subequations}
\label{specmu23-24}
\begin{align}
&\delta_\omega \beta_2=D_\omega\beta,
\vphantom{\Big]}
\label{specmu23}
\\
&\delta_\omega \beta_3
=D_\omega\dot\varOmega_\beta+[\beta,\varOmega_\omega]
\vphantom{\Big]}
\label{specmu24}
\\
&\hspace{3cm}
-\frac{1}{2}(\beta,[\omega,\omega])k+\frac{1}{2}(\beta,k)[\omega,\omega]+(\omega,k)[\omega,\beta].
\vphantom{\Big]}
\nonumber
\end{align}
\end{subequations}
It is now straightforward to show that 
\begin{equation}
\delta_\omega{}^2=0. \vphantom{\Bigg]}
\label{specmu26}
\end{equation}
The verification relies crucially on the $H$--flatness of the 2--connection $\omega$, eqs. 
\eqref{specmu17}, \eqref{specmu18}. All this holds in any dimensions. In 3 dimensions, 
which is the case we are concerned with,
some of the above properties get trivial by dimensional reason, e. g. eq. \eqref{specmu20}. 

Next, we describe the cohomology of the complex $(\mathsans{C}_\omega,\delta_\omega )$.
The cohomology space $H^0(\mathsans{C}_\omega,\delta_\omega)$ is just the Lie subalgebra $\mathfrak{inv}(\omega)$ 
of $\mathfrak{gau}_0(M,G)$ of the infinitesimal special 1--gauge transformation $\theta$
leaving the 2--connection $\omega$ invariant. $H^0(\mathsans{C}_\omega,\delta_\omega)$  contains the Lie algebra
$\mathfrak{c}(M,G)\simeq \mathfrak{z}(\mathfrak{g})$ of $C(N,G)$ as subalgebra. 
As $\mathfrak{inv}(\omega)$ transforms according to the adjoint action
of $\Gau_1(M,G)$ on $\mathfrak{gau}_0(M,G)$ under the 1--gauge transformation action of $\Gau_1(M,G)$ on 
$\mathsans{F}_H(M,G)$, the deviation of $H^0(\mathsans{C}_\omega,\delta_\omega)$ from $\mathfrak{c}(M,G)$
measures how singular the moduli space ${\eul F}_H(M,G)$ at 
$[\omega]$ is. $H^1(\mathsans{C}_\omega,\delta_\omega)$ is the vector space of tangent vectors 
to $\mathsans{F}_H(M,G)$ at $\omega$ modulo those vectors that 
are given by infinitesimal 1--gauge transformations. It can thus 
be identified with the tangent space to the moduli space ${\eul F}_H(M,G)$ at a regular point $[\omega]$. 
$H^2(\mathsans{C}_\omega,\delta_\omega)$ describes the deformations of the solutions of the Bianchi relations 
\eqref{specmu19-20} modulo those of the form resulting from deformations of 2--connections. 
It would be interesting to find out under what conditions these cohomology spaces are 
finite dimensional, since this would indicate the finite dimensionality of the moduli space
${\eul F}_H(M,G)$. 


\subsection{\normalsize \textcolor{blue}{Functional integral quantization 
}}
\label{subsec:funct2cs}

\hspace{.5cm} 
In this subsection, we shall attempt the functional integral quantization 
of special 2--Chern--Simons theory. This will unveil the kind of
 topological quantum field theory the model is.

The partition function of special $G$--2--Chern--Simons theory is
\begin{equation}
Z_{\mathrm{s2CS}}(H)
=\frac{1}{V} \int D\omega D\varOmega_\omega \exp(i\overline{\CS}_2(\omega,\varOmega_\omega;H)),
\label{funct2cs1}
\end{equation}
where $\overline{\CS}_2(\omega,\varOmega_\omega;H)$ is the special 2--Chern--Simons action 
given in \eqref{spec2cs3} and $V$ is the carefully defined functional volume of the gauge 
modulo gauge for gauge symmetry. The functional integration is extended to the whole space 
$\Conn_2(N,G)$ of special 2--connections. Upon fixing a Riemannian metric $g$ on $N$, 
the functional measures $D\omega$, $D\varOmega_\omega$ are those induced by 
the tangent space Hilbert norms
\begin{subequations}
\label{efunct2cs1-2}
\begin{align}
&\Vert\delta\omega\Vert^2=\frac{1}{(k,k)}\int_N(\delta\omega,*\delta\omega),
\vphantom{\Big]}
\label{efunct2cs1}
\\
&\Vert\delta\varOmega_\omega\Vert^2=\frac{1}{(k,k)}\int_N(\delta\varOmega_\omega,*\delta\varOmega_\omega),
\vphantom{\Big]}
\label{efunct2cs2}
\end{align}
\end{subequations}
where $*$ is the Hodge star operator of $g$ 
\footnote{$\vphantom{\dot{\dot{\dot{\dot{x}}}}}$ \label{fn:funct}
We recall a few basic facts about functional integration stating in this way also our conventions. 
If $\mathcal{F}$ is a real Hilbert manifold, then, for any $f\in\mathcal{F}$, the tangent space 
$T_f\mathcal{F}$ is a Hilbert space. A functional measure $Df$ on $\mathcal{F}$ is defined by assigning
a smoothly varying functional measure $D\delta f_f$ on the tangent space $T_f\mathcal{F}$ for 
each $f\in\mathcal{F}$ according to the following rules. 

If $\mathcal{H}$ is a real Hilbert space with Hilbert inner product $\langle\cdot,\cdot\rangle$
the associated functional measure $D\phi$ on $\mathcal{H}$ is defined as the translation invariant measure
normalized so that 
\begin{equation}
\int_{\mathcal{H}}D\phi\exp(-\Vert\phi\Vert^2/2)=1.
\label{ffunct2cs1}
\end{equation}

The functional determinant $\Det(\varDelta)$ of a positive selfadjoint 
linear operator $\varDelta:\mathcal{H}\rightarrow\mathcal{H}$ is
\begin{equation}
(\Det(\varDelta))^{-1/2}=\int_{\mathcal{H}}D\phi\exp(-\langle\phi,\varDelta\phi\rangle/2).
\label{ffunct2cs2}
\end{equation}

The functional Dirac delta function $\delta(\phi)$ on $\mathcal{H}$ is normalized so that
\begin{equation}
\int_{\mathcal{H}}D\phi\,F(\phi)\delta(\phi)=F(0),
\label{ffunct2cs3}
\end{equation}
for any function $F:\mathcal{H}\rightarrow \mathbb{R}$. $\vphantom{\dot{\dot{\dot{x}}}}$

A linear invertible mapping $T:\mathcal{H}'\rightarrow\mathcal{H}$ of Hilbert spaces
induces a change of functional integration variables $\phi=T\phi'$. Its Jacobian $J_T$ satisfies 
\begin{equation}
\int_{\mathcal{H}}D\phi\,F(\phi)=J_T\int_{\mathcal{H}'}D\phi'\,F(T\phi')
\label{ffunct2cs4}
\end{equation}
for any function $F:\mathcal{H}\rightarrow \mathbb{R}$. $J_T$ is given by 
\begin{equation}
J_T=(\Det(T^+T))^{1/2}.
\label{ffunct2cs5}
\end{equation}
with the determinant defined according to \eqref{ffunct2cs2}.

When a Hilbert space $\mathcal{H}$ is decomposable as an orthogonal direct sum of a collection of Hilbert spaces
$\mathcal{H}_\alpha$, $\mathcal{H}=\bigoplus_\alpha \mathcal{H}_\alpha$, the functional measure $D\phi$ of $\mathcal{H}$ 
factorizes accordingly in the product of the functional measures $D\phi_\alpha$ of $\mathcal{H}_\alpha$, 
\begin{equation}
D\phi=\text{$\prod_\alpha$} D\phi_\alpha.
\label{ffunct2cs6}
\end{equation}
The other properties of functional integration are formal consequences of the above. 
}. \pagebreak 
The normalization factor $(k,k)$ is conventional and may be dropped. Here, we assume that 
$(k,k)>0$; it is always possible to have this condition fulfilled by reversing the overall sign of 
the form $(\cdot,\cdot)$. 

\vspace{.25mm}
The measures $D\omega$, $D\varOmega_\omega$ are invariant under the $\overline{\Gau}_1(N,G)$--action 
\eqref{specond3-4}, as it is straightforward to check. Because of the special from of \eqref{specond4}
the translation invariance of the tangent space functional measures 
is crucial for this property to hold.  

\vspace{.25mm}
To perform the 
integration in \eqref{funct2cs1}, we use a suitable orthogonal decomposition 
of the 2--connection components $\omega$, $\varOmega_\omega$ with respect to 
the invariant bilinear form $(\cdot,\cdot)$ of $\mathfrak{g}$.
Explicitly, this reads
\begin{subequations}
\label{funct2cs3-4}
\begin{align}
&\omega=\omega_{0\,}k+\omega_s, 
\vphantom{\Big]}
\label{funct2cs3}
\\
&\varOmega_\omega=\varOmega_{\omega0\,}k+\varOmega_{\omega s}
\vphantom{\Big]}
\label{funct2cs4}
\end{align}
\end{subequations}
with $\omega_0\in\Omega^1(M,\mathbb{R})$, $\omega_s\in\Omega^1(N,(\mathbb{R}k)^\perp)$,
$\varOmega_{\omega0}\in\Omega^2(M,\mathbb{R})$, $\varOmega_{\omega s}\in\Omega^2(N,(\mathbb{R}k)^\perp)$, 
where $\Omega^p(N,(\mathbb{R}k)^\perp)$ is the subspace of 
$\Omega^p(N,\mathfrak{g})$ spanned by the $p$--forms $\varUpsilon_s$ 
val\-ued in the Lie subalgebra $(\mathbb{R}k)^\perp$ 
of $\mathfrak{g}$, i. e. satisfying $(\varUpsilon_s,k)=0$. 
The tangent space Hilbert norms of $\omega$, $\varOmega_\omega$ of eqs. \eqref{efunct2cs1-2} 
induce compatible norms for the components $\omega_0$, $\omega_s$, $\varOmega_{\omega 0}$, $\varOmega_{\omega s}$,
in terms of which their functional measures $D\omega_0$, $D\omega_s$, $D\varOmega_{\omega 0}$, $D\varOmega_{\omega s}$ 
can be built. 
The Jacobian relating the combined measures $D\omega D\varOmega_\omega$ and
$D\omega_0D\omega_sD\varOmega_{\omega 0}D\varOmega_{\omega s}$ is easily seen to equal $1$. 

\vspace{.33mm}
To carry out the functional integration, it is useful to employ the following change of functional variables
\begin{subequations}
\label{funct2cs2}
\begin{align}
&\tilde\varOmega_{\omega 0}=\varOmega_{\omega 0},
\vphantom{\Big]}
\label{funct2cs2a}
\\
&\tilde\varOmega_{\omega s}=\varOmega_{\omega s}-(k,k)\omega_0\omega_s
\vphantom{\Big]}
\label{funct2cs2b}
\end{align}
\end{subequations}
corresponding to the redefinition $\tilde\varOmega_\omega=\varOmega_\omega-(\omega,k)\omega$
of the 2--form connection component $\varOmega_\omega$.
The functional Jacobian of the transformation is equal to $1$, as it is immediately verified. 

\vspace{.33mm}
For future reference, it is useful to have the expressions of the orthogonal 
components ${}^\gamma\omega_0$, ${}^\gamma\omega_s$ and 
${}^\gamma\tilde\varOmega_{\omega0}$, ${}^\gamma\tilde\varOmega_{\omega s}$ 
of the gauge transformed  2--connection components ${}^\gamma\omega$, ${}^\gamma\varOmega_\omega$
for a possibly homotopically non trivial special 1--gauge transformation $\gamma\in\overline{\Gau}_1(N,G)$.
From \eqref{specond3-4}, on account of \eqref{spechgs1}, we obtain the following expressions
\begin{subequations}
\label{funct2cs7-8}
\begin{align}
&{}^\gamma\omega_0=\omega_0,
\vphantom{\Big]}
\label{funct2cs7}
\\
&{}^\gamma\omega_s=\gamma\omega_s\gamma^{-1}-d\gamma\gamma^{-1},
\vphantom{\Big]}
\label{funct2cs8}
\\
&{}^\gamma\tilde\varOmega_{\omega 0}
=\tilde\varOmega_{\omega 0}-d(\chi_{\gamma 0}+\alpha_{\gamma 0}(
\omega_s-\gamma^{-1}d\gamma))
\vphantom{\Big]}
\nonumber
\\
&\hspace{5.3cm}+\alpha_{\gamma 0}(
f_s)-\varsigma_\gamma+\varsigma_{\bar\gamma}+(\omega_s,\gamma^{-1}d\gamma),
\vphantom{\Big]}
\label{funct2cs7/1}
\\
&{}^\gamma\tilde\varOmega_{\omega s}
=\gamma(\tilde\varOmega_{\omega s}-D_{\omega_s}(\chi_{\gamma s}+\alpha_{\gamma s}(\omega_{0\,}k+
\omega_s-\gamma^{-1}d\gamma))
\vphantom{\Big]}
\nonumber
\\
&\hspace{7.3cm}+\alpha_{\gamma s}(d\omega_{0\,}k+f_s))\gamma^{-1},
\vphantom{\Big]}
\label{funct2cs8/1}
\end{align}
\end{subequations}
where we employ the orthogonal decompositions
$\alpha_\gamma(\pi)=\alpha_{\gamma 0}(\pi)k+\alpha_{\gamma s}(\pi)$
and $\chi_\gamma=\chi_{\gamma 0}k+\chi_{\gamma s}$ with $\alpha_{\gamma 0}(\pi)\in\Omega^0(N,\mathbb{R}[1])$, 
$\alpha_{\gamma s}(\pi)\in\Omega^0(N,(\mathbb{R}k)^\perp[1])$ and $\chi_{\gamma 0}\in\Omega^1(N,\mathbb{R})$,
$\chi_{\gamma s}\in\Omega^1(N,(\mathbb{R}k)^\perp)$ 
\footnote{$\vphantom{\dot{\dot{\dot{\dot{x}}}}}$ 
If we write an element $x\in\mathfrak{g}$ as 
$x=x_{0\,}k+x_s$ with $x_s\in (\mathbb{R}k)^\perp$, we have 
$\alpha_{\gamma 0}(x_{0\,}k+x_s)=-(x_s,\alpha_{\gamma s}(k))$ and $(x_s,\alpha_{\gamma s}(y_s))
+(x_s,\alpha_{\gamma s}(y_s))=0$ 
because of the 
requirement \eqref{spechgs3}.} and $f_s$ is the curvature of  $\omega_s$ given by \eqref{specond1}. 
Thus, $\omega_0$ is gauge invariant while $\omega_s$ behaves as a genuine gauge field.
The in\-terpretation of the way $\tilde\varOmega_{\omega 0}$ and $\tilde\varOmega_{\omega s}$ transform
is less transparent. 
The functional measures $D\omega_0$, $D\omega_s$, $D\varOmega_{\omega 0}$, $D\varOmega_{\omega s}$
are invariant under the $\overline{\Gau}_1(N,G)$--action 
\eqref{funct2cs7-8} again by the translation invariance of the 
tangent space measures.

Writing the  special 2--Chern--Simons action $\overline{\CS}_2(\omega,\varOmega_\omega;H)$
in terms of the orthogonal components fields $\omega_0$, $\omega_s$,
$\tilde\varOmega_{\omega0}$, $\tilde\varOmega_{\omega s}$,  
the partition function $Z_{\mathrm{s2CS}}(H)$ of the special 2--Chern--Simons theory takes then the form 
\begin{align}
&Z_{\mathrm{s2CS}}(H)
=\frac{1}{V}\int D\omega_0 \,D\omega_s \,D\tilde\varOmega_{\omega0} \,D\tilde\varOmega_{\omega s} 
\vphantom{\Big]}
\label{funct2cs9}
\\
&\hspace{1cm}\exp\bigg\{i\kappa_2(k,k)\int_N\bigg[
\frac{1}{(k,k)}(f_s,\tilde\varOmega_{\omega s})
+\omega_0\big(d\tilde\varOmega_{\omega0}+8\pi^2(\cs_{1s}+H)\big)\bigg]\bigg\},
\vphantom{\Big]}
\nonumber
\end{align}
where $f_s$ and $\cs_{1s}$ are given by \eqref{specond1} and \eqref{spec2cs8/-1} in terms of $\omega_s$. 

For fixed $H$, the partition function $Z_{\mathrm{s2CS}}(H)$ depends only 
on the de Rham cohomology class $[H]\in H^3(N,\mathbb{R})$ 
of $H$, as the $\tilde\varOmega_{\omega0}$ integration enforces the constraint $d\omega_0=0$. We shall thus denote 
the partition function as $Z_{\mathrm{s2CS}}([H])$ rather than $Z_{\mathrm{s2CS}}(H)$
to emphasize this property.

Under a homotopically non trivial special 1--gauge transformation $\gamma\in\overline{\Gau}_1$ $(N,G)$,
the action $\overline{\CS}_2(\omega,\varOmega_\omega;H)$ is not invariant, as $H$ gets shifted by 
the winding number $w(\bar\gamma)$ density by \eqref{spec2cs8}. So, the partition function $Z_{\mathrm{s2CS}}([H])$ 
satisfies 
\begin{equation}
Z_{\mathrm{s2CS}}([H]+h)=Z_{\mathrm{s2CS}}([H])
\vphantom{\bigg]}
\label{funct2cs10}
\end{equation}
for any $h\in \varLambda_k(N,G)$, where $\varLambda_k(N,G)\subset H_{\mathbb{Z}}{}^3(N,\mathbb{R})$ 
denotes the cohomological winding number lattice 
defined earlier in subsect. \ref{subsec:spechgs}. 
Therefore, $Z_{\mathrm{s2CS}}([H])$ depends not just on the cohomology class 
$[H]\in H^3(N,\mathbb{R})$ of $H$, but on the equivalence class of 
$[H]_w\in H^3(N,\mathbb{R})/\varLambda_k(N,G)$ 
of $[H]$. The \pagebreak (possibly degenerated) 
torus $H^3(N,\mathbb{R})/\varLambda_k(N,G)$ 
is thus the effective background field space of the model. 
We shall henceforth denote the partition function $Z_{\mathrm{s2CS}}([H]_w)$ 
instead than $Z_{\mathrm{s2CS}}([H])$ to emphasize this property.

Performing the $\tilde\varOmega_{\omega s}$ integration in \eqref{funct2cs9}, 
the partition function $Z_{\mathrm{s2CS}}([H]_w)$ takes the form 
\begin{align}
&Z_{\mathrm{s2CS}}([H]_w)
=(|\kappa_2|(k,k))^{-D_{2s}}\frac{1}{V}\int \,D\omega_s \,\delta(f_s)
\vphantom{\Big]}
\label{funct2cs11}
\\
&\hspace{2.75cm}\int  D\omega_0\,D\tilde\varOmega_{\omega 0}
\exp\bigg[i\kappa_2(k,k)\int_N\omega_0\big(d\tilde\varOmega_{\omega 0}+8\pi^2(\cs_{1s}+H)\big)\bigg],
\vphantom{\Big]}
\nonumber
\end{align}
where $D_{2s}$ is the formal dimension of $\Omega^2(N,(\mathbb{R}k)^\perp)$. 

Next, we consider the $\omega_0$, $\tilde\varOmega_{\omega 0}$ integrations in \eqref{funct2cs11}. 
We notice that when the connection com\-ponent 
$\omega_s$ is flat, $f_s=0$, as it is in the above functional integral by virtue  of the delta function
$\delta(f_s)$, 
the ordinary Chern--Simons 3--form $\cs_{1s}$ is closed because $d\cs_{1s}=(f_s,f_s)=0$. 
Since $H$ is also closed, the integrand turns out to be independent from the 2--connection components
$\omega_{0\mathrm{ex}}$, and $\tilde\varOmega_{\omega 0 \mathrm{ex}}$, $\tilde\varOmega_{\omega 0\mathrm{h}}$ 
\footnote{$\vphantom{\dot{\dot{\dot{\dot{x}}}}}$ Let $X$ be a closed manifold with Riemannian metric $h$. 
In Hodge--de Rham theory, the space $\Omega^p(X,\mathbb{R})$
of $p$--forms on $X$ is a Hilbert space with the norm 
\begin{equation}
\Vert\alpha\Vert^2=\int_X\alpha*\alpha,
\label{ffunct2cs7}
\end{equation}
where $*$ is the Hodge operator associated with $h$.
The de Rham operator $d:\Omega^p(X,\mathbb{R})\rightarrow \Omega^{p+1}(X,\mathbb{R})$ has thus
an adjoint operator $d^+:\Omega^{p+1}(X,\mathbb{R})\rightarrow \Omega^p(X,\mathbb{R})$. 
Both $d$ and $d^+$ are nilpotent, $d^2=0$, $d^{+2}=0$. 
A form $\alpha\in\Omega^p(X,\mathbb{R})$ is said closed if $d\alpha=0$, exact if 
$\alpha=d\beta$ for some $\beta\in\Omega^{p-1}(X,\mathbb{R})$, coclosed if $d^+\alpha=0$, coexact if 
$\alpha=d^+\beta$ for some $\beta\in\Omega^{p+1}(X,\mathbb{R})$ and harmonic if $d\alpha=d^+\alpha=0$.
The corresponding subspaces of $\Omega^p(X,\mathbb{R})$ will be 
denoted below by $\Omega_{\mathrm{cl}}{}^p(N,\mathbb{R})$,
$\Omega_{\mathrm{ex}}{}^p(N,\mathbb{R})$, $\Omega_{\mathrm{cocl}}{}^p(N,\mathbb{R})$,
$\Omega_{\mathrm{coex}}{}^p(N,\mathbb{R})$ and $\Omega_{\mathrm{h}}{}^p(N,\mathbb{R})$, respectively. 

The Hodge Laplacian is the operator $\varDelta=d^+d+dd^+:\Omega^p(X,\mathbb{R})\rightarrow \Omega^p(X,\mathbb{R})$.
It turns out that $\Omega_{\mathrm{h}}{}^p(N,\mathbb{R})=\ker\varDelta$. 
%
%

The space of $p$--forms $\Omega^p(X,\mathbb{R})$ enjoys the orthogonal direct sum decomposition 
\begin{equation}
\Omega^p(X,\mathbb{R})=\Omega_{\mathrm{ex}}{}^p(N,\mathbb{R})\oplus\Omega_{\mathrm{h}}{}^p(N,\mathbb{R})
\oplus\Omega_{\mathrm{coex}}{}^p(N,\mathbb{R}).
\label{ffunct2cs8}
\end{equation}
Correspondingly, $\vphantom{\dot{\dot{\dot{\dot{x}}}}}$any $p$--form $\alpha$ can be expressed uniquely as the sum 
$\alpha=\alpha_{\mathrm{ex}}+\alpha_{\mathrm{h}}+\alpha_{\mathrm{coex}}$ of 
its exact, harmonic and coexact components.
In particular, $\Omega_{\mathrm{cl}}{}^p(N,\mathbb{R})
=\Omega_{\mathrm{ex}}{}^p(N,\mathbb{R})\oplus\Omega_{\mathrm{h}}{}^p(N,\mathbb{R})$ and 
$\Omega_{\mathrm{cocl}}{}^p(N,\mathbb{R})=\Omega_{\mathrm{h}}{}^p(N,\mathbb{R})\oplus\Omega_{\mathrm{coex}}{}^p(N,\mathbb{R})$
and $\alpha_{\mathrm{coex}}=0$ for $\alpha\in\Omega_{\mathrm{cl}}{}^p(N,\mathbb{R})$ and 
$\alpha_{\mathrm{ex}}=0$ for $\alpha\in\Omega_{\mathrm{cocl}}{}^p(N,\mathbb{R})$. Moreover,
$\Omega_{\mathrm{h}}{}^p(N,\mathbb{R})=\Omega_{\mathrm{cl}}{}^p(N,\mathbb{R})\cap\Omega_{\mathrm{cocl}}{}^p(N,\mathbb{R})$
and $\alpha_{\mathrm{ex}}=\alpha_{\mathrm{coex}}=0$ for $\alpha\in \Omega_{\mathrm{h}}{}^p(N,\mathbb{R})$. 
}, as it is not difficult to check. This reflects an on shell 
Abelian \pagebreak gauge symmetry of the theory  \hphantom{xxxxxxxxxxxxxx}
\begin{subequations}
\label{efunct2cs12-13}
\begin{align}
&\omega_0\rightarrow \omega_0+d\phi,
\vphantom{\Big]}
\label{efunct2cs12}
\\
&\tilde\varOmega_{\omega 0}\rightarrow \tilde\varOmega_{\omega 0}+d\varPhi+Z,
\vphantom{\Big]}
\label{efunct2cs13}
\end{align}
\end{subequations}
where $\phi\in\Omega^0(N,\mathbb{R})$, $\varPhi\in\Omega^1(N,\mathbb{R})$, $Z\in\Omega_{\mathrm{h}}{}^2(N,\mathbb{R})$
(cf. subsect. \ref{subsec:spe2cs}). This gauge symmetry in turn enjoys 
a gauge for gauge symmetry
\begin{subequations}
\label{efunct2cs14-15}
\begin{align}
&\phi\rightarrow \phi +z,
\vphantom{\Big]}
\label{efunct2cs14}
\\
&\varPhi=\varPhi+d\varPsi +W,
\vphantom{\Big]}
\label{efunct2cs15}
\end{align}
\end{subequations}
where $z\in\Omega_{\mathrm{h}}{}^0(N,\mathbb{R})$, $\varPsi\in\Omega^0(N,\mathbb{R})$, 
$W\in\Omega_{\mathrm{h}}{}^1(N,\mathbb{R})$. 
Finally, we have a gauge for gauge for gauge symmetry \hphantom{xxxxxx}
\begin{equation}
\varPsi\rightarrow\varPsi+U,
\label{efunct2cs16}
\end{equation}
where $U\in\Omega_{\mathrm{h}}{}^0(N,\mathbb{R})$. 
The effective functional volume of the above Abelian gauge symmetry is so given by 
$V_0=V_1V_2{}^{-1}V_3$, where $V_1$, $V_2$, $V_3$ are the geometric functional 
volumes of the symmetries \eqref{efunct2cs12-13},
\eqref{efunct2cs14-15} and \eqref{efunct2cs16}, respectively. 
The alternating exponents of the factors are due to the gauge for gauge symmetry reducing
the effectively acting gauge symmetry and consequently its volume. 
$V_1$, $V_2$, $V_3$  are given by 
$V_1=\Vol\Omega^0(N,\mathbb{R})\Vol\Omega^1(N,\mathbb{R})\Vol\Omega_{\mathrm{h}}{}^2(N,\mathbb{R})$,
$V_2=\Vol\Omega_{\mathrm{h}}{}^0(N,\mathbb{R})\Vol\Omega^0(N,\mathbb{R})\Vol\Omega_{\mathrm{h}}{}^1(N,\mathbb{R})$,
$V_3=\Vol\Omega_{\mathrm{h}}{}^0(N,\mathbb{R})$. Hence, we have 
$V_0=\Vol\Omega^1(N,\mathbb{R})\Vol\Omega_{\mathrm{h}}{}^2(N,\mathbb{R})/\Vol\Omega_{\mathrm{h}}{}^1(N,\mathbb{R})$.
By virtue of the factorization $\Vol\Omega^1(N,\mathbb{R})=\Vol\Omega_{\mathrm{ex}}{}^1(N,\mathbb{R})
\Vol\Omega_{\mathrm{h}}{}^1(N,\mathbb{R})\Vol\Omega_{\mathrm{coex}}{}^1(N,\mathbb{R})$ and the relation 
$\Vol\Omega_{\mathrm{ex}}{}^2(N,\mathbb{R})=\Det(d^+d\big|_{\Omega_{\mathrm{coex}}{}^1(N,\mathbb{R})})^{1/2}
\Vol\Omega_{\mathrm{coex}}{}^1(N,\mathbb{R})$, we find finally \pagebreak 
\begin{equation}
V_0=\frac{\Vol\Omega_{\mathrm{ex}}{}^1(N,\mathbb{R})\Vol\Omega_{\mathrm{ex}}{}^2(N,\mathbb{R})\Vol\Omega_{\mathrm{h}}{}^2(N,\mathbb{R})}
{\Det(d^+d\big|_{\Omega_{\mathrm{coex}}{}^1(N,\mathbb{R})})^{1/2}}.
\label{efunct2cs17}
\end{equation}

Since the integrand in \eqref{funct2cs11} is independent from the 2--connection components
$\omega_{0\mathrm{ex}}$, and $\tilde\varOmega_{\omega 0 \mathrm{ex}}$, $\tilde\varOmega_{\omega 0\mathrm{h}}$, 
the functional integration on these latter is trivial and its contribution is a divergent factor
cancelled out by the factor $V_0$ of the gauge volume $V$. Letting $V'=V/V_0$ be  the volume of the gauge 
symmetry other than the one considered above, \eqref{funct2cs11}  reduces into 
\begin{align}
&Z_{\mathrm{s2CS}}([H]_w)
=(|\kappa_2|(k,k))^{-D_{2s}}\Det(d^+d\big|_{\Omega_{\mathrm{coex}}{}^1(N,\mathbb{R})})^{1/2}\frac{1}{V'}\int D\omega_s\,\delta(f_s)
\vphantom{\Big]}
\label{efunct2cs18}
\\
&\hspace{.5cm}
\int D\omega_{0\mathrm{h}}D\omega_{0\mathrm{coex}}\,D\tilde\varOmega_{\omega 0 \mathrm{coex}}\,
\vphantom{\Big]}
\nonumber
\\
&\hspace{3.5cm}
\exp\bigg\{i\kappa_2(k,k)\int_N\big[\omega_{0\mathrm{coex}}d\tilde\varOmega_{\omega 0 \mathrm{coex}}
+8\pi^2\omega_{0\mathrm{h}}(\cs_{1s}+H)_{\mathrm{h}}\big]\bigg\}.
\vphantom{\Big]}
\nonumber
\end{align}
The $\omega_{0\mathrm{coex}}$, $\tilde\varOmega_{\omega 0 \mathrm{coex}}$ integration can be carried out
by means of the change of variable $\tilde\varOmega_{\omega 0 \mathrm{coex}}=d^+*\varUpsilon_{\mathrm{coex}}$
with $\varUpsilon_{\mathrm{coex}}\in \Omega_{\mathrm{coex}}{}^1(N,\mathbb{R})$. This produces a Jacobian
$\Det(dd^+\big|_{\Omega_{\mathrm{ex}}{}^3(N,\mathbb{R})})^{1/2}=\Det(d^+d\big|_{\Omega_{\mathrm{coex}}{}^1(N,\mathbb{R})})^{1/2}$.
The integral under consideration then takes the form 
\begin{align}
&\int D\omega_{0\mathrm{coex}}\,D\tilde\varOmega_{\omega 0 \mathrm{coex}}\,
\exp\bigg(i\kappa_2(k,k)\int_N\omega_{0\mathrm{coex}}d\tilde\varOmega_{\omega 0 \mathrm{coex}}\bigg)
\vphantom{\Big]}
\label{efunct2cs19}
\\
&~=\Det(d^+d\big|_{\Omega_{\mathrm{coex}}{}^1(N,\mathbb{R})})^{1/2}
\int D\omega_{0\mathrm{coex}}\,D\,\varUpsilon_{\mathrm{coex}}
\exp\bigg(i\kappa_2(k,k)\int_N\omega_{0\mathrm{coex}}*d^+d\varUpsilon_{\mathrm{coex}}\bigg)
\vphantom{\Big]}
\nonumber
\\
&~=(|\kappa_2|(k,k))^{-B_{1\mathrm{coex}}}\Det(d^+d\big|_{\Omega_{\mathrm{coex}}{}^1(N,\mathbb{R})})^{-1/2},
\vphantom{\Big]}
\nonumber
\end{align}
where $B_{1\mathrm{coex}}$ is the formal dimension of $\Omega_{\mathrm{coex}}{}^1(N,\mathbb{R})$.
Using \eqref{efunct2cs19} in \eqref{efunct2cs18}, we obtain then
\begin{align}
&Z_{\mathrm{s2CS}}([H]_w)
=(|\kappa_2|(k,k))^{-D_{2s}-B_{1\mathrm{coex}}}\frac{1}{V'}\int D\omega_s\,\delta(f_s)
\vphantom{\Big]}
\label{efunct2cs20}
\\
&\hspace{4cm}
\int D\omega_{0\mathrm{h}}
\exp\bigg[i8\pi^2\kappa_2(k,k)\int_N\omega_{0\mathrm{h}}(\cs_{1s}+H)_{\mathrm{h}}\bigg].
\vphantom{\Big]}
\nonumber
\end{align}

Next, we tackle the $\omega_{0h}$ integration. 
Let $\zeta_{1a}$, $a=1,\ldots,b_1(N)$ be a basis of $\Omega_{\mathrm{h}}{}^1(N,\mathbb{R})$, which, for convenience, we shall 
assume to be constituted by forms with integer periods. $\omega_{0h}$ can be expanded as \hphantom{xxxxxxxxxxx}
\begin{equation}
\omega_{0h}=\sum_{a=1}^{b_1(N)}t^a\zeta_{1a}, 
\label{funct2cs16}
\end{equation}
where the $t^a$ are real numerical variables. We now use these $t^a$ as new integration variables.
The Jacobian of the transformation is $[\deg G_1/(2\pi)^{b_1(N)}]^{1/2}$, where $G_1$ is 
the Gramian matrix of the basis $\zeta_{1a}$ 
\begin{equation}
G_{1ab}=\int_N \zeta_{1a}*\zeta_{1b}.
\label{funct2cs15}
\end{equation}
Proceeding in this way, noting that, for a closed 3--form $\varTheta$, 
 $\int_N \zeta_{1a}\varTheta_{\mathrm{h}}=\int_{\PD(\zeta_{1a})}\varTheta$,  
where $\PD(\alpha)$ denotes a codimension $p$ closed submanifold of $N$ 
Poincar\'e dual of a closed $p$--form $\alpha$, we find 
\begin{align}
&\int D\omega_{0\mathrm{h}}
\exp\bigg[i8\pi^2\kappa_2(k,k)\int_N\omega_{0\mathrm{h}}(\cs_{1s}+H)_{\mathrm{h}}\bigg]
\vphantom{\Big]}
\label{efunct2cs21}
\\
&\hspace{1.5cm}=(\deg G_1)^{1/2}(2(2\pi)^{3/2}|\kappa_2|(k,k))^{-b_1(N)}\prod_{k=a}^{b_1(N)}
\delta\bigg(\int_{\PD(\zeta_{1a})}(\cs_{1s}+H)\bigg).
\vphantom{\Big]}
\nonumber
\end{align}
Inserting \eqref{efunct2cs21} into \eqref{efunct2cs20}, we find
\begin{align}
&Z_{\mathrm{s2CS}}([H]_w)
=(|\kappa_2|(k,k))^{-D_{2s}-B_{1\mathrm{cocl}}}\frac{(\deg G_1)^{1/2}}{(2(2\pi)^{3/2})^{b_1(N)}}
\vphantom{\Big]}
\label{funct2cs17}
\\
&\hspace{4cm}
\frac{1}{V'}\int D\omega_s\,\delta(f_s)\prod_{k=a}^{b_1(N)}
\delta\bigg(\int_{\PD(\zeta_{1a})}(\cs_{1s}+H)\bigg).
\vphantom{\Big]}
\nonumber
\end{align}
where $B_{1\mathrm{cocl}}=B_{1\mathrm{coex}}+b_1(N)$ is the formal dimension of $\Omega_{\mathrm{cocl}}{}^1(N,\mathbb{R})$.  

We now tackle the problem of the $\omega_s$ integration. 
The gauge symmetry left over after the integrating out $\tilde\varOmega_{\omega s}$,
$\omega_0$, $\tilde\varOmega_{\omega0}$ is given by the gauge transformations
\eqref{funct2cs8}. The corresponding gauge group $\Map_k(N,G)$ is formed by the maps
$\gamma\in\Map(N,G)$ satisfying the condition \eqref{spechgs1}. 
We remind that homotopically non trivial gauge transformations are included. 
The trivial ones form an invariant subgroup $\Map_{kc}(N,G)$ of $\Map_k(N,G)$.

We first fix the homotopically trivial gauge symmetry. \pagebreak 
On general grounds, under the action of $\Map_{kc}(N,G)$ the space of connections 
$\omega_s$ decomposes in pairwise disjoint layers, the elements of each of which
have conjugate stabilizer subgroups. 
The stabilizer subgroup $S_{c\omega_s}\subset\Map_{kc}(N,G)$ of a connection $\omega_s$ 
is formed by the gauge transformations $\gamma$ such that 
${}^\gamma\omega_s=\omega_s$.  
Since, the layers with larger stabilizer subgroups have larger codimensions, 
only the main layer with the smallest stabilizer subgroup contributes effectively 
to the functional integral. We shall therefore tacitly assume that the integration 
is restricted to this latter. 

The volume of the homotopically trivial gauge group $\Map_{kc}(N,G)$ is comput\-ed by 
a functional measure $D\gamma$ defined with respect to a biinvariant tangent space Hilbert norm
on $\Map_{kc}(N,G)$ and consequently itself biinvariant.
Because the possible non triviality of the stabilizer subgroup,
the homotopically trivial gauge symmetry effective volume $V'$ in \eqref{funct2cs17} is given by 
\begin{equation}
V'=\frac{\Vol\Map_{kc}(N,G)}{\Vol S_{c\omega^{{}_0}{}_s}},
\label{efunct2cs22}
\end{equation}
where $\omega^{{}_0}{}_s$ is a reference connection. 
Since the stabilizer subgroups of the connections $\omega_s$ are 
mutually conjugate, $\Vol S_{c\omega_s}$ is independent from $\omega_s$
and so the choice of $\omega^{{}_0}{}_s$  is immaterial. 

To fix the homotopically trivial $\Map_{kc}(N,G)$ gauge symmetry, we impose a standard Lorenz--like gauge fixing condition 
\begin{equation}
D_{\omega^{{}_0}{}_s}{}^+(\omega_s-\omega^{{}_0}{}_s)=0.
\label{funct2cs18}
\end{equation}
The corresponding Faddeev--Popov functional is given by 
\begin{equation}
\varDelta_{\mathrm{FP}}(\omega_s)^{-1}=\int_{\Map_{kc}(N,G)} D\gamma\,
\delta(D_{\omega^{{}_0}{}_s}{}^+({}^\gamma\omega_s -\omega^{{}_0}{}_s)).
\label{funct2cs19}
\end{equation}
Standard gauge theoretic techniques furnish 
\begin{equation}
\varDelta_{\mathrm{FP}}(\omega_s)=\frac{\Det'(D_{\omega^{{}_0}{}_s}{}^+D_{\omega_s{}'})}{\Vol S_{c\omega^{{}_0}{}_s}},
\label{funct2cs20}
\end{equation}
where $\det'(D_{\omega^{{}_0}{}_s}{}^+D_{\omega_s})$ is the functional determinant of the 
Faddeev--Popov \pagebreak 
operator $D_{\omega^{{}_0}{}_s}{}^+D_{\omega_s}$ with the zero modes removed
and $\omega_s{}'$ is the unique 
connection satisfying \eqref{funct2cs18} such that $\omega_s{}'={}^\gamma\omega_s$ for some $\gamma\in\Map_{kc}(N,G)$. 
The above expression is only formal. If the functional integration prescriptions listed 
in fn. \ref{fn:funct} were rigorously applied, the functional determinant in the numerator would
be $[\Det'((D_{\omega^{{}_0}{}_s}{}^+D_{\omega_s{}'})^+D_{\omega^{{}_0}{}_s}{}^+D_{\omega_s{}'})]^{1/2}$, but an 
object like this cannot be soundly defined in quantum field theory, unless $\omega_s{}'=\omega^{{}_0}{}_s$, in which case 
it reduces simply to $\Det'(D_{\omega^{{}_0}{}_s}{}^+D_{\omega^{{}_0}{}_s})$. Instead, it is reasonable to split 
the differential operator as $D_{\omega^{{}_0}{}_s}{}^+D_{\omega_s{}'}
=D_{\omega^{{}_0}{}_s}{}^+D_{\omega^{{}_0}{}_s}+D_{\omega^{{}_0}{}_s}{}^+\ad(\omega_s{}'-\omega^{{}_0}{}_s))$ 
and view the two terms respectively as 
the kinetic and gauge coupling terms of the Faddeev--Popov ghost/antighost system, 
respectively. Reinterpreting the determinant along these lines
leads to the following more precise expression
\begin{equation}
\Det'(D_{\omega^{{}_0}{}_s}{}^+D_{\omega_s{}'})
=\Det'(D_{\omega^{{}_0}{}_s}{}^+D_{\omega^{{}_0}{}_s})W_{\mathrm{FP}}(\omega_s{}'-\omega^{{}_0}{}_s),
\label{efunct2cs23}
\end{equation}
where $W_{\mathrm{FP}}(a_s)$ is the normalized generating functional of 
the ghost/antighost current correlation functions. Therefore, 
\eqref{funct2cs20} is to be replaced by the more precise expression
\begin{equation}
\varDelta_{\mathrm{FP}}(\omega_s)=\frac{\Det'(D_{\omega^{{}_0}{}_s}{}^+D_{\omega^{{}_0}{}_s})
W_{\mathrm{FP}}(\omega_s{}'-\omega^{{}_0}{}_s)}{\Vol S_{c\omega^{{}_0}{}_s}}. 
\label{efunct2cs25}
\end{equation}

According to standard Faddeev--Popov theory, the gauge fixing is achieved by inserting into the functional 
integral of the partition function the term $\varDelta_{\mathrm{FP}}(\omega_s)$
$\delta(D_{\omega^{{}_0}{}_s}{}^+(\omega_s-\omega^{{}_0}{}_s))$
and then replacing $\omega_s{}'$ with $\omega_s$. In this way, one finds
\begin{align}
&Z_{\mathrm{s2CS}}([H]_w)
=(|\kappa_2|(k,k))^{-D_{2s}-B_{1\mathrm{cocl}}}\frac{(\deg G_1)^{1/2}}{(2(2\pi)^{3/2})^{b_1(N)}}
\Det'(D_{\omega^{{}_0}{}_s}{}^+D_{\omega^{{}_0}{}_s})
\vphantom{\Big]}
\label{funct2cs21}
\\
&\hspace{.5cm}
\int D\omega_s \,\delta(f_s)\delta(D_{\omega^{{}_0}{}_s}{}^+(\omega_s-\omega^{{}_0}{}_s))
W_{\mathrm{FP}}(\omega_s-\omega^{{}_0}{}_s)
\prod_{a=1}^{b_1(N)}\delta\bigg(\int_{\PD(\zeta_{1a})}(\cs_{1s}+H)\bigg).
\vphantom{\Big]}
\nonumber
\end{align} 
We now interpret the result that we have obtained.

In \eqref{funct2cs21}, only the homotopically trivial $\Map_{kc}(N,G)$ 
symmetry has been fixed. The homotopically non trivial symmetry associated with 
the mapping class group $\Map_k(N,G)/\Map_{kc}(N,G)$ conversely has not. 
Hence, the $\omega_s$ integration is restricted by the gauge fixing to the $\Map_{kc}(N,G)$--orbit space
$\matheul{M}^0{}_k(N,G)$ of the space of flat 1--form connections. To complete the gauge fixing, 
one should express the partition function as an integral over the $\Map_k(N,G)$--orbit space
$\matheul{M}_k(N,G)$  by factoring out the volume of the residual 
unfixed $\Map_k(N,G)/\Map_{kc}(N,G)$ symmetry. 

As far as we can see, the fixing of the leftover gauge symmetry cannot be done through 
the familiar field theoretic techniques used so far. We notice that $\matheul{M}^0{}_k(N,G)$ 
is a covering space of $\matheul{M}_k(N,G)$, the deck transformation group
being precisely 
$\Map_k(N,G)/\Map_{kc}(N,G)$. The fixing thus requires the determination of 
and the restriction of integration to a fundamental region 
of the deck group 
in $\matheul{M}^0{}_k(N,G)$, a task  
necessitating a suitable prior parametrization of 
$\matheul{M}^0{}_k(N,G)$ itself. Let us assume that this somehow has been done.  

In \eqref{funct2cs21}, there appears the ordinary Chern--Simons 3--form $\cs_{1s}$. 
When $\omega_s$ is flat, as it is in the present case, $\cs_{1s}$ is closed and 
so defines a degree $3$ de Rham cohomology class 
$[\cs_{1s}]\in H^3(N,\mathbb{R})$. 
Furthermore, if $\gamma\in\Map_k(N,G)$ is a homotopically non trivial 
gauge transformation, $\cs_1({}^\gamma\omega_s)=\cs_{1s}+w(\gamma)+$ exact terms, $w(\gamma)$ being the winding
number density of $\gamma$ given in \eqref{spechgs75}. Since the cohomology class $[w(\gamma)]$ 
lies in $\varLambda_k(N,G)$, the class $[\cs_{1s}]_w
\in H^3(N,\mathbb{R})/\varLambda_k(N,G)$ is defined in a fully gauge invariant 
manner. Hence, $[\cs_{1s}]_w$ 
is defined on the orbit space $\matheul{M}_k(N,G)$.

If $\omega_s(t)$ is a smooth path in the space of flat 1--form connections $\omega_s$, 
the derivative $d\cs_1(\omega_s(t))/dt$ is exact for every $t$. Hence, the de Rham cohomology class
$[\cs_1(\omega_s(t))]$ is independent from $t$. This property is obviously preserved by all 
gauge transformation. The class $[\cs_{1s}]_w$ 
is therefore constant on each path connected component of $\matheul{M}_k(N,G)$.

From what observed above, upon inspection of \eqref{funct2cs21}, it appears 
that the partition functions $Z_{\mathrm{s2CS}}([H]_w)$ localizes on the locus in $\matheul{M}_k(N,G)$
of the connection components $\omega_s$ such that  $[\cs_{1s}+H]_w=0$. 
The non vanishing of the partition function thus detects the possible values
of the class $[\cs_{1s}]_w$ and thus the path connected component structure 
of $\matheul{M}_k(N,G)$.

Is the quantum field theory we have described topological, that is independ\-ent 
from the background metric $g$ on $N$? It should. The only potential source of $g$--dependence 
of the partition function 
comes from the normalization of the functional measures and the gauge fixing insertions.
However, the $g$--dependence of the measure amounts to a mere $g$ dependent Jacobian 
factor that gets absorbed in correlators of gauge invariant 
operators. Further, the $g$--dependence of the gauge fixing insertions
should amount to BRST exact contribution which have no effects on those
correlators. A more thorough analysis of these matters would anyway be welcome.



\vfil\eject

\vfil\eject

\appendix

\section{\normalsize \textcolor{blue}{Lie $2$--group and $2$--algebra theory}}\label{sec:2tLinftytheor}

\hspace{.5cm} 
In the following appendices, we collect various results on $2$--groups
and Lie $2$--algebras and their automorphisms disseminated 
 in the mathematical literature in order to define our terminology and notation and for 
reference throughout in the text. A good introduction to these matters
tailored for higher gauge theoretic applications  
is provided in \cite{Baez:2010ya}.


\subsection{\normalsize \textcolor{blue}{Strict $2$--groups}}\label{sec:twogr}

\hspace{.5cm} 
The theory of strict $2$--groups is formulated most elegantly in the 
language of higher category theory \cite{Baez5}. Here, we shall limit ourselves to providing
the basic definitions and properties.

We provide now the definition of strict $2$--group. 

\hspace{.5cm} A strict $2$--group (in delooped form)
consists of the following set of data: 
\begin{enumerate}

\item a set of $1$-cells $V_1$;

\item a composition law of $1$--cells $\circ: V_1\times V_1\rightarrow V_1$;

\item a inversion law of $1$--cells ${}^{-1_\circ}: V_1\rightarrow V_1$;

\item a distinguished unit $1$--cell $1\in V_1$;

\item for each pair of $1$--cells $a,b\in V_1$, a set of $2$--cells $V_2(a,b)$;

\item for each quadruple of $1$--cells $a,b,c,d\in V_1$, a horizontal composition law of $2$--cells
$\circ:V_2(a,c)\times V_2(b,d)\rightarrow V_2(b\circ a,d\circ c)$;

\item for each pair of $1$--cells $a,b\in V_1$, 
a horizontal inversion law of $2$--cells ${}^{-1_\circ}: V_2(a,b)\rightarrow V_2(a^{-1_\circ},b^{-1_\circ})$;

\item for each triple of $1$--cells $a,b,c\in V_1$, a vertical composition law of $2$--cells
$\bfdot:V_2(a,b)\times V_2(b,c)\rightarrow V_2(a,c)$;

\item for each pair of $1$--cells $a,b\in V_1$, 
a vertical inversion law of $2$--cells ${}^{-1_\bfdot}\!\!: V_2(a,b)\rightarrow V_2(b,a)$;

\item for each $1$--cell $a$, a distinguished 
unit $2$--cell $1_a\in V_2(a,a)$.
\end{enumerate}
 These are required to satisfy the following axioms. 
\begin{subequations}
\label{twogr1}
\begin{align}
&(c\circ b)\circ a=c\circ(b\circ a),
\vphantom{\Big]}
\label{twogr1a}
\\
&a^{-1_\circ}\circ a=a\circ a^{-1_\circ}=1,
\vphantom{\Big]}
\label{twogr1b}
\\
&a\circ 1=1\circ a=a,
\vphantom{\Big]}
\label{twogr1c}
\\
&(C\circ B)\circ A=C\circ(B\circ A),
\vphantom{\Big]}
\label{twogr1d}
\\
&A^{-1_\circ}\circ A=A\circ A^{-1_\circ}=1_1,
\vphantom{\Big]}
\label{twogr1e}
\\
&A\circ 1_1=1_1\circ A=A,
\vphantom{\Big]}
\label{twogr1f}
\\
&(C\bfdot B)\bfdot A=C\bfdot(B\bfdot A),
\vphantom{\Big]}
\label{twogr1g}
\\
&A^{-1_\bfdot}\!\bfdot A=1_a,\qquad A\bfdot A^{-1_\bfdot}=1_b,
\vphantom{\Big]}
\label{twogr1h}
\\
&A\bfdot 1_a=1_b\bfdot A=A,
\vphantom{\Big]}
\label{twogr1i}
\\
&(D\bfdot C)\circ(B\bfdot A)=(D\circ B)\bfdot(C\circ A).
\vphantom{\Big]}
\label{twogr1j}
\end{align}
\end{subequations}%
Here and in the following, $a,b,c,\dots\in V_1$, $A,B,C,\dots\in V_2$, where 
$V_2$ denotes the set of all $2$-cells. For clarity, we often denote $A\in V_2(a,b)$
as $A:a\Rightarrow b$. 
All identities involving the vertical composition and inversion hold whenever defined.  
Relation \eqref{twogr1j} is called interchange law. 
In the following, we shall denote a $2$--group such as the above as $V$ or $(V_1,V_2)$
or $(V_1,V_2,\circ,{}^{-1_\circ},\bfdot,{}^{-1_\bfdot},1_-)$ to emphasize the underlying structure.

$V$ is in fact a one--object
strict $2$--category in which all $1$--morphisms are invertible  and all $2$--morphisms
are both horizontal and vertical invertible, a one--object strict $2$--groupoid. 

If $(V_1,V_2,\circ,{}^{-1_\circ},\bfdot,{}^{-1_\bfdot},1_-)$ \pagebreak 
is a strict $2$--group, then 
$(V_1,\circ,{}^{-1_\circ},1)$ is an ordinary group and $(V_1,V_2,\bfdot,{}^{-1_\bfdot},1_-)$ is a groupoid. 
Viewing this as a category $V$ having $V_1$, $V_2$  as its collection of objects and morphisms, 
$\circ:V\times V\rightarrow V$ and ${}^{-1_\circ}:V\rightarrow V$ are both functors and 
$V$ turns out to be a strict monoidal category in which every morphism is invertible 
and every object has a strict inverse.

\subsection{\normalsize \textcolor{blue}{Lie $2$--algebras}}\label{sec:linfty}

\hspace{.5cm} 
A Lie $2$--algebra consists of the following set of data: 
\begin{enumerate}

\item a pair of vector spaces on the same field
$\mathfrak{v}_0,\mathfrak{v}_1$;

\item a linear map $\partial:\mathfrak{v}_1\rightarrow\mathfrak{v}_0$;

\item a linear map $[\cdot,\cdot]:\mathfrak{v}_0\wedge \mathfrak{v}_0\rightarrow \mathfrak{v}_0$;

\item a linear map $[\cdot,\cdot]:\mathfrak{v}_0\otimes \mathfrak{v}_1\rightarrow \mathfrak{v}_1$;

\item a linear map 
$[\cdot,\cdot,\cdot]:\mathfrak{v}_0\wedge \mathfrak{v}_0\wedge \mathfrak{v}_0\rightarrow \mathfrak{v}_1$
\footnote{$\vphantom{\bigg[}$ We denote by $[\cdot,\cdot]$ both 
$2$--argument brackets. It will be clear from context which is which.}.

\end{enumerate}
These are required to satisfy the following axioms:
\begin{subequations}
\label{2tlinalg}
\begin{align}&[\pi,\partial\varPi]-\partial[\pi,\varPi]=0,
\vphantom{\Big]}
\label{2tlinalga}
\\
&[\partial \varPi,\varPi]=0, 
\vphantom{\Big]}
\label{2tlinalgb}
\\
&3[\pi,[\pi,\pi]]-\partial[\pi,\pi,\pi]=0,
\vphantom{\Big]}
\label{2tlinalgc}
\\
&2[\pi,[\pi,\varPi]]-[[\pi,\pi],\varPi]-[\pi,\pi,\partial\varPi]=0,
\vphantom{\Big]}
\label{2tlinalgd}
\\
&4[\pi,[\pi,\pi,\pi]]-6[\pi,\pi,[\pi,\pi]]=0.
\vphantom{\Big]}
\label{2tlinalge}
\end{align}
\end{subequations}
where $\pi$ and $\Pi$ are given by
\begin{subequations}
\label{gammaCdef}
\begin{align}
\pi&=\pi^a\otimes e_a,
\vphantom{\Big]}
\label{gammadef}
\\
\varPi&=\varPi^\alpha\otimes E_\alpha,
\vphantom{\ul{\ul{\ul{\ul{\ul{\ul{g}}}}}}}
\vphantom{\Big]}
\label{Cdef}
\end{align}
\end{subequations}
$\{e_a\}$, $\{E_\alpha\}$ being bases of $\mathfrak{v}_0$, $\mathfrak{v}_1$
and $\{\pi^a\}$, $\{\varPi^\alpha\}$ being the bases of $\mathfrak{v}_0{}^\vee[1]$, 
$\mathfrak{v}_1{}^\vee[2]$ dual to $\{e_a\}$, $\{E_\alpha\}$, respectively. 
Here, $\mathfrak{v}_0{}^\vee[1]$ and $\mathfrak{v}_1{}^\vee[2]$ are
the $1$ and $2$ step degree shifted duals of $\mathfrak{v}_0$, $\mathfrak{v}_1$
assumed to have degree $0$.
We shall denote a Lie $2$--algebra such as the above 
by $\mathfrak{v}$ or, more explicitly, by $(\mathfrak{v}_0,\mathfrak{v}_1)$ or 
$(\mathfrak{v}_0,\mathfrak{v}_1,\partial,[\cdot,\cdot],[\cdot,\cdot,\cdot])$
to emphasize its underlying structure. 

Similarly to ordinary Lie algebras, the  Chevalley--Eilenberg algebra 
$\mathrm{CE}(\mathfrak{v})$ of $\mathfrak{v}$ is the graded commutative algebra 
$S(\mathfrak{v}_0{}^\vee[1]\oplus\mathfrak{v}_1{}^\vee[2])\simeq \bigwedge^*\mathfrak{v}_0{}^\vee
\otimes\bigvee^*\mathfrak{v}_1{}^\vee$ 
generated by $\mathfrak{v}_0{}^\vee[1]\oplus\mathfrak{v}_1{}^\vee[2]$.
The Chevalley--Eilenberg differential $\mathcal{Q}_{\mathrm{CE}(\mathfrak{v})}$ 
is the degree $1$ differential defined by 
\begin{subequations}
\label{2tlinalgQ}
\begin{align}
\mathcal{Q}_{\mathrm{CE}(\mathfrak{v})}\pi&=-\frac{1}{2}[\pi,\pi]+\partial \varPi,
\vphantom{\Big]}
\label{2tlinalgQa}
\\
\mathcal{Q}_{\mathrm{CE}(\mathfrak{v})}\varPi&=-[\pi,\varPi]+\frac{1}{6}[\pi,\pi,\pi].
\vphantom{\Big]}
\label{2tlinalgQb}
\end{align}
\end{subequations}
$\mathcal{Q}_{\mathrm{CE}(\mathfrak{v})}$ turns out to be nilpotent,
\begin{equation}
\mathcal{Q}_{\mathrm{CE}(\mathfrak{v})}{}^2=0,
\label{wQ2=0}
\end{equation}
in virtue of the relations \eqref{2tlinalg}.
$(\mathrm{CE}(\mathfrak{v}),\mathcal{Q}_{\mathrm{CE}(\mathfrak{v})})$ is a so cochain complex. 
The associated Chevalley--Eilenberg cohomology
$H_{CE}{}^*(\mathfrak{v})$ is the Lie $2$--algebra cohomology of $\mathfrak{v}$
generalizing ordinary Lie algebra cohomology. 

A Lie $2$--algebra $\mathfrak{v}$ is said balanced if $\dim\mathfrak{v}_0=\dim\mathfrak{v}_1$.
For any non balanced Lie $2$--algebra $\mathfrak{v}$, there exists a
balanced Lie $2$--algebra $\mathfrak{v}^\sim$ minimally extending $\mathfrak{v}$. 

Let $\mathfrak{v}$ be a balanced Lie $2$--algebra. An invariant form 
on $\mathfrak{v}$ is a non singular bilinear mapping 
$(\cdot,\cdot):\mathfrak{v}_0\times \mathfrak{v}_1\rightarrow \mathbb{R}$
enjoying the following properties.
\begin{subequations}
\label{linfty1,2,3}
\begin{align}
&(\partial X,Y)-(\partial Y,X)=0, 
\vphantom{\Big]}
\label{linfty1}
\\
&([\pi,x],X)+(x,[\pi,X])=0,
\vphantom{\Big]}
\label{linfty2}
\\
&(x,[\pi,\pi,y])+(y,[\pi,\pi,x])=0,
\vphantom{\Big]}
\label{linfty3}
\end{align}
\end{subequations}
for any $x,y\in\mathfrak{v}_0$, $X,Y\in \mathfrak{v}_1$.

\subsection{\normalsize \textcolor{blue}{The Lie $2$--algebra
automorphism group }}\label{sec:linftyauto}

Let $\mathfrak{v}$ be a Lie $2$--algebra. A Lie $2$--algebra
$1$--automorphism of $\mathfrak{v}$ consists of the following data:
\begin{enumerate}

\item a vector space automorphism  $\phi_0:\mathfrak{v}_0\rightarrow\mathfrak{v}_0$;

\item a vector space automorphism $\phi_1:\mathfrak{v}_1\rightarrow\mathfrak{v}_1$; 

\item a vector space morphism $\phi_2:\mathfrak{v}_0\wedge\mathfrak{v}_0\rightarrow\mathfrak{v}_1$. 

\end{enumerate}
These are required to satisfy the following relations:
\begin{subequations}
\label{mor2tlinalg}
\begin{align}
&\phi_0(\partial \varPi)-\partial\phi_1(\varPi)=0,
\vphantom{\Big]}
\label{mor2tlinalga}
\\
&\phi_0([\pi,\pi])-[\phi_0(\pi),\phi_0(\pi)]-\partial\phi_2(\pi,\pi)=0,
\vphantom{\Big]}
\label{mor2tlinalgb}
\\
&\phi_1([\pi,\varPi])-[\phi_0(\pi),\phi_1(\varPi)]-\phi_2(\pi,\partial \varPi)=0,
\vphantom{\Big]}
\label{mor2tlinalgc}
\\
&3[\phi_0(\pi),\phi_2(\pi,\pi)]+3\phi_2(\pi,[\pi,\pi])
\vphantom{\Big]}
\label{mor2tlinalgd}
\\
&\hspace{3cm}+[\phi_0(\pi),\phi_0(\pi),\phi_0(\pi)]-\phi_1([\pi,\pi,\pi])=0.  
\vphantom{\Big]}
\nonumber
\end{align}
\end{subequations}
In the following, we shall denote a $1$--morphism 
such as the above one by $\phi$ or, more explicitly, by 
$(\phi_0,\phi_1,\phi_2)$ to emphasize its constituent components. 
We shall denote the set of all $1$--automorphisms of $\mathfrak{v}$ 
by $\Aut_1(\mathfrak{v})$.

For any two Lie $2$--algebra $1$--automorphisms $\phi,\psi$, 
a {\it Lie $2$--algebra $2$--auto\-morphism} from $\phi$ to $\psi$ 
consists of a single datum:
\begin{enumerate}

\item a linear map $\varPhi:\mathfrak{v}_0\rightarrow\mathfrak{v}_1$. 

\end{enumerate}
This must satisfy the following relations
\begin{subequations}
\label{mor0tlinalg}
\begin{align}
&\phi_0(\pi)-\psi_0(\pi)-\partial\varPhi(\pi)=0,
\vphantom{\Big]}
\label{mor0tlinalga}
\\
&\phi_1(\varPi)-\psi_1(\varPi)-\varPhi(\partial \varPi)=0,
\vphantom{\Big]}
\label{mor0tlinalgb}
\\
&\phi_2(\pi,\pi)-\psi_2(\pi,\pi)+[\phi_0(\pi)+\psi_0(\pi),\varPhi(\pi)]
-\varPhi([\pi,\pi])=0.
\vphantom{\Big]}
\label{mor0tlinalgc}
\end{align}
\end{subequations}  %
We shall write a $2$--automorphism such as this as $\varPhi$ or as $\varPhi:\phi\Rightarrow\psi$
to emphasize its source and target.  
We shall denote the set of all $2$--automorphisms 
$\varPhi:\phi\Rightarrow\psi$ by $\Aut_2(\mathfrak{v})(\phi,\psi)$
and the set of all $2$--automorphisms $\varPhi$ by $\Aut_2(\mathfrak{v})$. 

$\Aut_1(\mathfrak{v})$, $\Aut_2(\mathfrak{v})$ are the sets of $1$-- and $2$--cells
of a strict $2$--group $\Aut(\mathfrak{v})$ for the operations and units 
\begin{subequations}
\label{mor3tlinalg}
\begin{align}
&\psi\circ \phi_0(\pi)=\psi_0\phi_0(\pi),
\vphantom{\Big]}
\label{mor3tlinalga}
\\
&\psi\circ \phi_1(\varPi)=\psi_1\phi_1(\varPi),
\vphantom{\Big]}
\label{mor3tlinalgb}
\\
&\psi\circ \phi_2(\pi,\pi)=\psi_1\phi_2(\pi,\pi)+\psi_2(\phi_0(\pi),\phi_0(\pi)),
\vphantom{\Big]}
\label{mor3tlinalgc}
\\
&\phi^{-1_\circ}{}_0(\pi)=\phi_0{}^{-1}(\pi), \hspace{4.05cm}
\vphantom{\Big]}
\label{mor3/2tlinalgd}
\\
&\phi^{-1_\circ}{}_1(\varPi)=\phi_1{}^{-1}(\varPi), 
\vphantom{\Big]}
\label{mor3/2tlinalge}
\\
&\phi^{-1_\circ}{}_2(\pi,\pi)=-\phi_1{}^{-1}\phi_2(\phi_0{}^{-1}(\pi),\phi_0{}^{-1}(\pi)).
\vphantom{\Big]}
\label{mor3/2tlinalgf}
\\
&\id_0(\pi)=\pi,
\vphantom{\Big]}
\label{mor3tlinalgg}
\\
&\id_1(\varPi)=\varPi,
\vphantom{\Big]}
\label{mor3tlinalgh}
\\
&\id_2(\pi,\pi)=0,
\vphantom{\Big]}
\label{mor3tlinalgi}
\\
&\varPsi\circ \varPhi(\pi)=\varPsi\lambda_0(\pi)+\psi_1\varPhi(\pi)=\varPsi\mu_0(\pi)+\phi_1\varPhi(\pi),
\vphantom{\Big]}
\label{mor4tlinalga}
\\
&\varPhi^{-1_\circ}(\pi)=-\lambda_1{}^{-1}\varPhi\mu_0{}^{-1}(\pi)=-\mu_1{}^{-1}\varPhi\lambda_0{}^{-1}(\pi),
\vphantom{\Big]}
\label{mor4/1tlinalgb}
\\
&\varLambda\bfdot \varTheta(\pi)=\varTheta(\pi)+\varLambda(\pi),
\vphantom{\Big]}
\label{mor4tlinalgb}
\\
&\varTheta^{-1_\bfdot}(\pi)=-\varTheta(\pi),
\vphantom{\Big]}
\label{mor4/1tlinalgd}
\\
&\mathrm{Id}_\phi(\pi)=0.
\vphantom{\Big]}
\label{mor4tlinalgc}
\end{align}
\end{subequations} 
where $\varPhi:\lambda\Rightarrow\mu$, $\varPsi:\phi\Rightarrow\psi$, 
$\varTheta:\rho\Rightarrow\sigma$, $\varLambda:\sigma\Rightarrow\tau$.

Let $\mathfrak{v}$ be a balanced Lie $2$--algebra equipped with an invariant form 
$(\cdot,\cdot)$. 
A $1$--automorphism $\phi\in\Aut_1(\mathfrak{v})$ is said orthogonal if 
\begin{subequations}
\label{linfty4,5}
\begin{align}
&(\phi_0(x),\phi_1(X))=(x,X),
\vphantom{\Big]}
\label{linfty4}
\end{align}
\begin{align}
&(\phi_0(x),\phi_2(y,z))+(\phi_0(z),\phi_2(y,x))=0,
\vphantom{\Big]}
\label{linfty5}
\end{align}
\end{subequations}
for any $x,y,z\in\mathfrak{v}_0$, $X\in \mathfrak{v}_1$. 
We shall denote by $\OAut_1(\mathfrak{v})$ the set of all orthogonal elements
$\phi\in\Aut_1(\mathfrak{v})$. 

A $2$--automorphism $\varPhi\in\Aut_2(\mathfrak{v})(\phi,\psi)$, $\phi,\psi\in\Aut_1(\mathfrak{v})$
being two $1$--auto\-morphism,
is said orthogonal if both $\phi,\,\psi$ are. For any $\phi,\psi\in\OAut_1(\mathfrak{v})$,
we shall set $\OAut_2(\mathfrak{v})(\phi,\psi)=\Aut_2(\mathfrak{v})(\phi,\psi)$. We further set
$\OAut_2(\mathfrak{v})=\bigcup_{\phi,\psi\in\OAut_1(\mathfrak{v})}$ $\Aut_2(\mathfrak{v})(\phi,\psi)$. 

The following theorem holds true. 
$\OAut(\mathfrak{v})=(\OAut_1(\mathfrak{v}),\OAut_2(\mathfrak{v}))$ 
is a Lie $2$--subgroup of the strict Lie $2$--group
$\Aut(\mathfrak{v})=(\Aut_1(\mathfrak{v}),\Aut_2(\mathfrak{v}))$, 
by which we mean that $\OAut(\mathfrak{v})$ is closed under all operations of the strict $2$--group  
$\Aut(\mathfrak{v})$ (cf. app. \ref{sec:linftyauto}).

{\it The derivation Lie $2$--Lie algebra}

Let $\mathfrak{v}$ be a Lie $2$--algebra.
The derivation strict Lie $2$--Lie algebra $\mathfrak{aut}(\mathfrak{v})$ 
of $\mathfrak{v}$ is described as follows. 

An element of $\alpha$
of $\mathfrak{aut}_0(\mathfrak{v})$, a $1$--derivation,  consists of three mappings.
\begin{enumerate}

\item a vector space morphism $\alpha_0:\mathfrak{v}_0\rightarrow\mathfrak{v}_0$;

\item a vector space morphism $\alpha_1:\mathfrak{v}_1\rightarrow\mathfrak{v}_1$;

\item a vector space morphism $\alpha_2:\mathfrak{v}_0\wedge\mathfrak{v}_0\rightarrow\mathfrak{v}_1$.

\end{enumerate} 
These must satisfy the following relations:
\begin{subequations}
\label{mor5tlinalg}
\begin{align}
&\alpha_0(\partial \varPi)-\partial\alpha_1(\varPi)=0,
\vphantom{\Big]}
\label{mor5tlinalga}
\\
&\alpha_0([\pi,\pi])-[\alpha_0(\pi),\pi]-[\pi,\alpha_0(\pi)]-\partial\alpha_2(\pi,\pi)=0,
\hspace{1cm}
\vphantom{\Big]}
\label{mor5tlinalgb}
\\
&\alpha_1([\pi,\varPi])-[\alpha_0(\pi),\varPi]-[\pi,\alpha_1(\varPi)]-\alpha_2(\pi,\partial \varPi)=0,
\vphantom{\Big]}
\label{mor5tlinalgc}
\\
&3[\pi,\alpha_2(\pi,\pi)]+3\alpha_2(\pi,[\pi,\pi])
\vphantom{\Big]}
\label{mor5tlinalgd}
\\
&\hspace{4cm}+3[\pi,\pi,\alpha_0(\pi)]-\alpha_1([\pi,\pi,\pi])=0.
\vphantom{\Big]}
\nonumber
\end{align}
\end{subequations}

An element of $\varGamma$ of 
$\mathfrak{aut}_1(\mathfrak{v})$, a $2$--derivation, consists of a single mapping. 
\begin{enumerate}

\item a vector space morphism $\varGamma:\mathfrak{v}_0\rightarrow\mathfrak{v}_1$.

\end{enumerate} 
No restrictions are imposed on it. 

The boundary map and the brackets of $\mathfrak{aut}(\mathfrak{v})$ 
are given by the expressions 
\begin{subequations}
\label{mor7tlinalg}
\begin{align}
&\partial_\circ \varGamma_0(\pi)=-\partial\varGamma(\pi), \hspace{6.1cm}
\vphantom{\Big]}
\label{mor7tlinalgx}
\\
&\partial_\circ \varGamma_1(\varPi)=-\varGamma(\partial \varPi),
\vphantom{\Big]}
\label{mor7tlinalgy}
\\
&\partial_\circ \varGamma_2(\pi,\pi)=2[\pi,\varGamma(\pi)]-\varGamma([\pi,\pi]),
\vphantom{\Big]}
\label{mor7tlinalgz}
\\
&[\alpha,\beta]_{\circ 0}(\pi)=\alpha_0\beta_0(\pi)-\beta_0\alpha_0(\pi), 
\vphantom{\Big]}
\label{mor7tlinalga}
\\
&[\alpha,\beta]_{\circ 1}(\varPi)=\alpha_1\beta_1(\varPi)-\beta_1\alpha_1(\varPi),
\vphantom{\Big]}
\label{mor7tlinalgb}
\\
&[\alpha,\beta]_{\circ 2}(\pi,\pi)=\alpha_1\beta_2(\pi,\pi)+2\alpha_2(\beta_0(\pi),\pi)
\vphantom{\Big]}
\label{mor7tlinalgc}
\\
&\qquad\qquad\qquad\qquad\qquad -\beta_1\alpha_2(\pi,\pi)-2\beta_2(\alpha_0(\pi),\pi),
\vphantom{\Big]}
\nonumber
\\
&[\alpha,\varGamma]_\circ (\pi)=\alpha_1\varGamma(\pi)-\varGamma\alpha_0(\pi),
\vphantom{\Big]}
\label{mor7tlinalgv}
\\
&[\alpha,\beta,\gamma]_\circ (\pi)=0.
\vphantom{\Big]}
\label{mor7tlinalgw}
\end{align}
\end{subequations}
Relations  \eqref{mor5tlinalg} ensure that the basic relations \eqref{2tlinalg}
are satisfied by the above boundary and brackets.

A $1$--derivation $\alpha\in\mathfrak{aut}_0(\mathfrak{v})$ 
is said orthogonal if 
\begin{subequations}
\label{linfty8,9}
\begin{align}
&(\alpha_0(x),X)+(x,\alpha_1(X))=0,
\vphantom{\Big]}
\label{linfty8}
\\
&(x,\alpha_2(y,z))+(z,\alpha_2(y,x))=0,
\vphantom{\Big]}
\label{linfty9}
\end{align}
\end{subequations}
for any $x,y,z\in\mathfrak{v}_0$, $X\in \mathfrak{v}_1$. 
We shall denote by $\mathfrak{oaut}_0(\mathfrak{v})$ the subset of all orthogonal elements
$\alpha\in\mathfrak{aut}_0(\mathfrak{v})$.

A $2$--derivation $\varGamma\in\mathfrak{aut}_1(\mathfrak{v})$ 
is said orthogonal if, for $x,y,z\in\mathfrak{v}_0$, $X\in\mathfrak{v}_1$,  
\begin{subequations}
\label{linfty10,11}
\begin{align}
&(\partial\varGamma(x),X)+(x,\varGamma(\partial X))=0,
\vphantom{\Big]}
\label{linfty10}
\\
&(y,[x,\varGamma(z)]+[z,\varGamma(x)])+(x,\varGamma([y,z]))+(z,\varGamma([y,x]))=0. 
\vphantom{\Big]}
\label{linfty11}
\end{align}
\end{subequations}
We shall denote by $\mathfrak{oaut}_1(\mathfrak{v})$ the subset of all orthogonal elements
$\varGamma\in\mathfrak{aut}_1(\mathfrak{v})$.

The following theorem holds true. 
$\mathfrak{oaut}(\mathfrak{v})=(\mathfrak{oaut}_0(\mathfrak{v}),\mathfrak{oaut}_1(\mathfrak{v}))$ 
is a strict Lie $2$--subalgebra of 
$\mathfrak{aut}(\mathfrak{v})=(\mathfrak{aut}_0(\mathfrak{v}),\mathfrak{aut}_1(\mathfrak{v}))$, 
by which we mean that $\mathfrak{oaut}(\mathfrak{v})$
is closed under all operations of the strict Lie $2$--algebra  
$\mathfrak{aut}(\mathfrak{v})$.

For any Lie $2$--algebra $\mathfrak{v}$ with invariant form, $\OAut(\mathfrak{v})$ 
is a strict Lie $2$--group having precisely $\mathfrak{oaut}(\mathfrak{v})$ as its associated strict 
Lie $2$--algebra. 

\vfil\eject

\end{document}